\documentclass{aa}
\usepackage{graphicx}
\usepackage{txfonts}
\usepackage{natbib}
\bibpunct{(}{)}{;}{a}{}{,}

\newcommand{\thetabold}{\mbox{\boldmath$\theta$}}

\begin{document}

\title{Bayesian inversion of Stokes profiles}
\author{A. Asensio Ramos\inst{1}
  \and M. J. Mart\'{\i}nez Gonz\'alez\inst{2} 
  \and J. A. Rubi\~no-Mart\'in\inst{1}}

\offprints{aasensio@iac.es}

\institute{
 Instituto de Astrof\'{\i}sica de Canarias, 38205, La Laguna, Tenerife, Spain
 \and
LERMA, Observatoire 
de Paris-Meudon, 5 place Jules Janssen, 92195, Meudon, France \\
\email{aasensio@iac.es}
}
\date{Received ; Accepted}
\titlerunning{Bayesian inversion of Stokes profiles}
\authorrunning{Asensio Ramos et al.}

\abstract
{Inversion techniques are the most powerful methods to obtain information about the thermodynamical and magnetic 
properties of solar and stellar atmospheres. In the last years, we have witnessed the development
of highly sophisticated inversion codes that are now widely applied to
spectro-polarimetric observations. The majority of these inversion codes are based on the optimization of 
a complicated non-linear merit function. The large experience gained over the years have facilitated the
recovery of the model that best fits a given observation. However, and except for the recently developed
inversion codes based on database search algorithms together with the application of Principal Component Analysis, no 
reliable and statistically well-defined confidence intervals
can be obtained for the parameters inferred from the inversions.}
{A correct estimation of the confidence intervals for all the parameters that describe the model is mandatory.
Additionally, it is fundamental to apply efficient techniques to assess the ability of models to reproduce
the observations and to what extent the models have to be refined or can be simplified.}
{Bayesian techniques are applied to analyze the performance of the model to fit a given observed Stokes
vector. The posterior distribution, that takes into account both the information about the priors and the
likelihood, is efficiently sampled using a Markov Chain Monte Carlo method. For simplicity, we focus on
the Milne-Eddington approximate solution of the radiative transfer equation and we only take into account
the generation of polarization through the Zeeman effect. However, the method is
extremely general and other more complex forward models can be applied, even allowing for the presence of atomic
polarization.}
{We illustrate the ability of the method with the aid of different problems, from academic to more realistic 
examples. We show that the information provided by the posterior distribution turns out to be fundamental
to understand and determine the amount of information available in the Stokes profiles in these particular cases.}
{}

\maketitle


\section{Introduction}
One of the most important breakthroughs in the interpretation of spectro-polarimetric observations
has been the development and systematic application of inversion techniques 
\citep[see e.g.,][and references therein]{bellotrubio_spw4_06}. They have allowed
to extract as much information as possible from the observed Stokes profiles. A 
model that is assumed to be successful in describing the astrophysical plasma that we are
observing is proposed by the physicist. This model is defined by a set of parameters, usually associated with interesting
physical quantities. It can happen that these physical parameters are not direct observables that can be
obtained directly from the Stokes profiles. Then, inversion techniques try to adjust the parameters that
characterize the selected model so that the emergent Stokes profiles reproduce, as good as possible,
the observed profiles.

The initial steps in the development of inversion codes were limited by the computational
time. These inversion techniques often need the application of time-consuming non-linear
optimization methods. For this reason, the first generation of inversion codes either used very simple
models to reproduce the observed Stokes profiles or introduced some additional
physical ingredients in the inversion scheme so that the complications were highly reduced 
\citep[e.g.,][]{auer_heasly_house77,keller90}. This is the reason why simple Milne-Eddington 
atmospheres \citep[ME,][]{auer_heasly_house77,landi_landolfi04} have been widely applied for 
the retrieval of some kind of average magnetic field vector
in the line formation region. Although the assumptions in which ME atmospheres are based
may not be exactly fulfilled in the solar atmosphere, they have been extensively used. The reason is
their inherent simplicity and the fact that there is an analytic expression for the emergent Stokes 
profiles in terms of the physical parameters. 

A great leap forward was the development of inversion codes based on the concept of response
functions \citep{sir92}. They have facilitated the inversion of Stokes profiles so that it is now
possible to infer the vertical stratification of the thermodynamical and magnetic properties
of the atmosphere if the information is present on the Stokes profiles. The presence of vertical 
variations along the line-of-sight of the physical
properties are of importance for explaining the strong asymmetries observed in sunspots
and faculae \citep{illing_sunspot75,sanchez_almeida89}.

The development of such powerful and computationally efficient inversion codes has led to an
extensive number of applications to a large variety of solar atmospheric structures 
\citep[e.g.,][]{westendorp_nature97,sanchez_almeida_misma97,lites98,shibu03,bommier07}.
In spite of the success, it is important to be very cautious when using an inversion code. They
essentially carry out the optimization of a given merit function with respect to a set of 
parameters. It is fundamental, however, to be cautious with several fundamental points. First,
the number of free parameters cannot be as large as desired. The reason is that
the amount of information available in the observed Stokes profiles might not be enough to constrain the value
of many of these parameters. Reasons for this can be assigned to the presence of noise that mask
the line profile dependence on certain parameters or the fact the Stokes parameters are completely
insensitive to a parameter due to the intrinsic line formation process. Second, the parameters
that we use to describe a given model might not be completely independent so that there exists
(possibly nonlinear) combinations of these parameters that give rise to exactly the same
emergent Stokes profiles. Among these degeneracies, we can find the well-known ambiguity associated with
the projection of the magnetic field vector on the plane of the sky, the degeneracy between the filling
factor and the longitudinal component of the magnetic field strength (the magnetic flux density) in the
weak-field regime and less-known degeneracies between thermodynamical and magnetic parameters
for magnetic field structures organized in small scales \citep{martinez_gonzalez06}. 
Third, since the optimization problem is usually solved with the aid of gradient descent
methods like the Levenberg-Marquardt scheme, the solution given by the inversion code might not be that
corresponding to the global minimum (in case a global minimum is present).

Recently, several
works have faced this problem from a different point of view. On the one hand, 
\cite{asensio_ramos06} has introduced the usage of
model selection algorithms for the interpretation of spectro-polarimetric observations. Given 
a set of possible models that can be used to describe
the observations, these algorithms help us select the most probable one using a
quantitative approach.
These algorithms, based on the Occam's Razor, favor models that
better fit the observations with a reduced set of parameters, while disfavoring too complicated models
even if they match the observations or those that badly fit the observables. On the other 
hand, \cite{asensio_dimension07} have
applied algorithms based on geometrical considerations to estimate the intrinsic amount of information present
in the Stokes profiles. The intrinsic dimension of the manifold in which the observables lie
can be associated with the number of independent free parameters that can be used when
proposing a model to describe the observations. They have also shown that the amount of information 
present in an observed dataset increases monotonically with the number of spectral lines included.
Also of interest is the work carried out by \cite{socas_navarro_inversion04} for estimating
the level of detail of the stratification of atmospheric parameters one can obtain from 
selected spectral lines.

An important advance in the development of inversion codes was the application of database search
algorithms in conjunction with Principal Component
Analysis (PCA) to the inversion of Stokes profiles \citep{rees_PCA00,arturo_casini02,skumanich02,casini05}.
For the moment, this inversion technique has been applied only to simplified ME atmospheres and to
microstructured magnetic atmospheres \citep[MISMA;][]{socasnavarro02} but nothing (except
for a computational problem) avoids using more complicated models. 
It is based on the direct comparison between the observed Stokes profiles and all the possible ones that can
be built by varying the parameters that describe the model atmosphere. This comparison is not done
with the profiles themselves, but with the coefficients of the projection of both the observed and theoretical
profiles into a given basis. The key point of the PCA inversion is that this basis set is obtained from the
synthetic Stokes profiles themselves. Consequently, it already encodes valuable information about the 
line formation mechanism.
The fact that we compare the observed Stokes profile with the whole database allows us to reach the global
minimum, instead of getting stuck in local minima.
A subproduct of using a 
database is that it is possible to define an error bar and give uncertainties into the inferred
physical parameters.


This paper addresses the development of an inversion scheme that allows to characterize the 
probability distribution
of parameters of the model that better fit the observed Stokes profiles. To this end, we adopt a Bayesian
approach to infer the most probable values of the parameters and to extract their confidence levels.

\section{Bayesian Inversion of Stokes Profiles}
Our aim is to develop an inversion code that can obtain all the physical information present in the 
observed Stokes profiles and that can give us detailed statistical information. This statistical
information allows us to estimate a real error bar for each parameter and whether a parameter of a given 
model is constrained by the observables or not. This information turns out to be fundamental 
so that one can trust the value of the inferred parameters and properly analyze the observations.

\subsection{Forward modeling}
The Bayesian formalism is extremely general and can be applied to any model that explains
a given set of observations. We are interested in the Stokes profiles emerging from a given
atmosphere. Let $\mathbf{S}=(I,Q,U,V)^\dagger$ be the Stokes vector ($\dagger$ indicating transpose). The vectorial 
radiative transfer equation describes the variation along a given ray of the Stokes vector $\mathbf{S}$ depending on
the absorption and emission properties of the medium:
\begin{equation}
\frac{d\mathbf{S}}{ds} = \mathbf{\epsilon} - \mathbf{K} \mathbf{S},
\label{eq:radiative_transfer}
\end{equation}
where $\mathbf{\epsilon}=(\epsilon_I,\epsilon_Q,\epsilon_U,\epsilon_V)^\dagger$ is the emission vector 
and $\mathbf{K}$ is the propagation matrix:
\begin{equation}
\mathbf{K} = 
\left( \begin{array}{cccc}
\eta_I & \eta_Q & \eta_U & \eta_V \\
\eta_Q & \eta_I & \rho_V & -\rho_U \\
\eta_U & -\rho_V & \eta_I & \rho_Q \\
\eta_V & \rho_U & -\rho_Q & \eta_I
\end{array} \right).
\end{equation}
In principle, once $\mathbf{\epsilon}$ and $\mathbf{K}$ are known for all the points along the considered ray, it is 
possible to solve Eq. (\ref{eq:radiative_transfer}) and obtain the synthetic emergent Stokes parameters. However,
for simplicity, we will focus on the Milne-Eddington approximation, although we plan to apply Bayesian inversion
techniques to other more complex problems. Of interest is the case of inversion under local thermodynamic 
equilibrium \citep[LTE;][]{sir92} in which strong degeneracies may be present \citep[see][]{martinez_gonzalez06} and
the case of scattering polarization and the Hanle effect with many (and even unknown) degeneracies 
\citep[][]{house77,casini_judge99,trujillo99,trujillo01,casini05}. In the ME approximation 
\citep[see, e.g.,][]{auer_heasly_house77,landi_landolfi04}, we assume that the ratio between the line absorption
coefficient and the continuum absorption coefficient does not vary with depth in the atmosphere and that the 
line source function has a linear dependence on the optical depth along the line-of-sight. Furthermore, we assume that the
magnetic field vector $\mathbf{B}$ and the bulk velocity are constant with depth. 

Here we focus on the Zeeman effect as the mechanism that generates and modifies the polarization state of the
atmosphere. In this case, the elements of the propagation matrix and of the emission vector can be easily calculated
\cite[e.g.,][]{landi_landolfi04}. These elements depend on the strength of the magnetic field and on the
specific orientation of the field vector with respect to the line-of-sight. After these assumptions, the well-known 
Milne-Eddington analytical solution of the radiative transfer equation can be applied.
In the code, the effect of the magnetic field
on the energy levels can be treated under the simple linear Zeeman regime, under the more general incomplete Paschen-Back
regime or even hyperfine structure can be included.

\subsection{Posterior probability}
The interest to extract all the information available in the observations has led to the systematic
application of methods based on the Bayesian approach. A myriad of problems can be tackled under this framework that has
strong theoretical roots. We present the fundamental ideas of the formalism, although much more detailed
information can be found in several monographs \citep[see e.g.,][]{neal93}.
Let us assume a model $M$ that is used to describe a given dataset $D$. In our case, the model $M$ is the
Milne-Eddington approximation. It is parameterized in terms of the 
vector of physical quantities $\thetabold$ that contains the usual ME 
parameters: doppler width of the line in wavelength units ($\Delta \lambda_\mathrm{dopp}$), wavelength 
shift due to a macroscopic bulk velocity ($v_\mathrm{mac}$),
gradient of the source function ($\beta$), ratio between the line and continuum absorption coefficients ($\eta_0$), 
line damping parameter ($a$) and magnetic field vector parameterized by its modulus, inclination and azimuth
with respect to a given reference direction ($B$, $\theta_B$ and $\phi_B$, respectively).
It is customary to have some initial information about the
physical parameters. For instance, an estimation of the range of variation of the physical parameters
might be available, although sometimes it can be a very rough one (for instance, a limitation to
positive or negative values). This information is incorporated into a prior distribution $p(\thetabold)$. When
the information contained into the data $D$ is incorporated in the problem, our state of knowledge of the
parameters change according to the Bayes theorem:
\begin{equation}
p(\thetabold|D) \propto p(\thetabold) p(D|\thetabold).
\label{eq:bayes_theorem}
\end{equation}
The posterior distribution $p(\thetabold|D)$ represents our state of knowledge of the parameters once the
information of the dataset has been taken into account. The term $p(D|\thetabold)$ is the so-called
likelihood function and gives information about how well a particular set of parameters predicts 
the observed data. The Bayes 
theorem states that whether a model $M$ becomes plausible after the data $D$ has been taken into account
depends on how plausible the model was before taking into account the data and how well the model
predicts the data. The simplicity of the Bayes theorem hides all its potential and this kind of
reasoning has led to a variety of applications: it has been widely used in cosmological
analyses \citep[e.g.,][]{lewis02,rubino_martin03,rebolo04}, gravitational wave analyses \citep[e.g.,][]{cornish05}, 
gravitational lensing \citep[e.g.,][]{brewer_lensing07}, oscillation of solar-like stars 
\citep[e.g.,][]{brewer_oscillations07} and many more. 
A powerful inversion code can be
built based on the Bayes theorem. Once the posterior distribution $p(\thetabold|D)$ is known, the
position of the maximum value gives the most probable combination of parameters that fit the
data. Not only this, but we can also analyze the confidence of the parameters. Consequently, degeneracies,
ambiguities and the rest of problems that arise in typical inversion codes (except for those based
on PCA) can be investigated in great detail.

Let us analyze in detail the terms appearing in the right hand side of in Eq. (\ref{eq:bayes_theorem}). As
mentioned above, the prior distribution contains all the information that we know about the parameters
without taking into account the observed data. In the most simple case, we can assume that all the 
parameters are statistically independent, so that the prior distribution can be written as:
\begin{equation}
p(\thetabold) = \prod_{i=1}^{N_\mathrm{par}} p(\theta_i),
\end{equation}
where the $\{\theta_i\}$ are the parameters included in the model and $N_\mathrm{par}$ is the number of
such parameters. Unless physical information is available, we typically only know the range of
variation of the parameters, so that we can write:
\begin{equation}
p(\theta_i) = H(\theta_i,\theta_i^\mathrm{min},\theta_i^\mathrm{max}),
\end{equation}
where $H(x,a,b)$ is the top-hat function:
\begin{equation}
H(x,a,b) = \left\{
\begin{array} {clc}
\frac{1}{b-a} & & a < x < b \\
0 & & \mathrm{otherwise}
\end{array}
\right.
\end{equation}
As an example, consider the prior of a uniform magnetic field vector $\mathbf{B}$. In order to
guarantee such an uniform magnetic field vector, we have to sample uniformly the volume element
$dV=r^2\,dr\,d(\cos \theta_B) \, d\phi_B$. We can assume that the magnetic 
field strength cannot be larger than $B_\mathrm{max}$,
so that its prior is a top-hat function that is non-zero in the interval $[0,B_\mathrm{max}^3]$.
When focusing on the solar atmosphere, a reasonable choice is $B_\mathrm{max} \approx 4000$ G so that all
the physically relevant cases can be covered (quiet Sun, sunspots, faculae, etc.). The inclination and the 
azimuth have to be obviously limited to the ranges $[0,\pi]$ and $[0,2\pi]$, respectively. If we neglect any
correlation between the magnetic field strength, inclination and azimuth, the final prior on the
magnetic field vector is given by:
\begin{equation}
p(\mathbf{B}) = H(B^3,0,B_\mathrm{max}^3) H(\cos \theta_B,-1,1) H(\phi_B,0,2\pi).
\end{equation}
Interestingly, it is possible to include correlations between the parameters. As an example, let us assume that the 
stronger the magnetic field, the more vertical it is. Additionally, weaker fields can be found with all kinds of
inclinations. A simple prior distribution that fulfills the previous assumptions is:
\begin{eqnarray}
p(\mathbf{B}) &=& \frac{1}{C} \left\{ 1+\exp \left[ -(B-B_\mathrm{max})^2 /\sigma_\mathrm{B}^2 \right] \exp \left[ -\theta_B^2/\sigma_\theta^2 \right] \right\} \nonumber \\
& \times & H(\phi_B,0,2\pi),
\end{eqnarray}
where $\sigma_\mathrm{B}^2$ and $\sigma_\theta^2$ control the shape of the prior and $C$ is a normalization constant.

The second term, $p(D|\thetabold)$, termed the likelihood, measures the probability that a model determined
by a set of parameters $\thetabold$ fits a given observation $D$. For simplifying the notation, it is advantageous to
particularize to the case of the inversion of Stokes profiles. In spite of this particularization, the method
remains still very general. The data $D$ that we are facing consists on a set of four vectorial quantities, i.e., the
wavelength dependence of the four Stokes parameters. The number of wavelength points in each Stokes profile is
indicated by $N_\lambda$. The value of the Stokes parameter $i=0,1,2,3$ (that we
associate with the more usual notation I, Q, U and V) at a wavelength $\lambda_j$ is represented by the
quantity $S^\mathrm{obs}_i(\lambda_j)$. When these Stokes parameters are observed with a spectro-polarimeter attached to
a telescope, they contain a certain level of noise. If these observational errors are independent and have
a Gaussian distribution, their distributions can be described with their standard deviations 
$\sigma_i(\lambda_j)$, i.e., the noise
level for each Stokes parameter at wavelength $\lambda_j$. Strictly speaking, the noise in the observed
Stokes profiles should be poissonian because it comes mainly from photon noise. However, for consistency
with other works, we choose the noise to be normally distributed, which will be a good
approximation if the number of photons is high enough. Typically, we will deal with wavelength-independent
noise, so that only the four quantities $\sigma_i$ are needed. 
Let $S^\mathrm{syn}_i(\lambda_j)$ be the
Stokes parameters that emerge when the forward problem is solved in a given model $M$ parameterized by the
vector of parameters $\thetabold$. Taking into account the previous definitions and assumptions, the likelihood
function is defined as \cite[e.g.,][]{mackay03}:
\begin{equation}
p(D|\thetabold) \propto e^{ -\frac{1}{2} \chi^2 },
\label{eq:likelihood}
\end{equation}
where we have introduced the usual merit function $\chi^2$:
\begin{equation}
\chi^2 = \frac{1}{4 N_\lambda} \sum_{i=1}^4 \sum_{j=1}^{N_\lambda} \left( \frac{S^\mathrm{obs}_i(\lambda_j)-S^\mathrm{syn}_i(\lambda_j)}{\sigma_i} \right)^2.
\end{equation}
Although we have focused on wavelength-independent noise, the formalism allows to accommodate wavelength-dependent
noise by using $\sigma_i(\lambda_j)$ instead of $\sigma_i$ in the likelihood. Therefore, if the information is available,
it is possible to include other sources of uncertainty like reduction residuals, cross-talk, fringes, etc. 
The $\chi^2$ function that is typically used for the
inversion of Stokes profiles presents the weights $\{w_i,i=1\ldots4\}$ for each Stokes parameter. This 
differential weighting scheme is not applied here, but the method can accommodate it straightforwardly. The only
influence of this weighting is to change the width of the maximum likelihood regions (reducing or expanding the
confidence regions around the maximum). However, the location of the maximum is not changed.

\section{Markov Chain Monte Carlo}
In the Bayesian framework, the most plausible model is the one that maximizes the posterior distribution. Our
objective is then to sample the posterior distribution and to find the combination of parameters that
produce this maximum value. This will represent the most plausible model that matches the observed
Stokes profiles. For a small number of parameters $N_\mathrm{par}$, this brute force approach might be achievable. For
instance, assuming that ten values per parameter are desired, something like $10^{N_\mathrm{par}}$ evaluations
of the posterior distribution are needed. This implies that the forward model has to be evaluated a huge number of
times. When $N_\mathrm{par} < 5$, such a direct approach can be of applicability in case the computing
time per evaluation of the forward model is not very large. However, this brute force approach quickly becomes impractical
because the number of function evaluations increases exponentially with the number of free parameters. 
In order to overcome this difficulty, we have applied a Markov Chain Monte Carlo technique. Since this 
is the first time that such a method is applied to the inversion of Stokes profiles, we have decided to present in 
detail some important technical issues in Appendix A, although they are widely known in other research fields.
Briefly, our implementation of the Markov Chain Monte Carlo scheme is based on the Metropolis 
algorithm \citep{metropolis53,neal93}. The proposal density distribution is chosen to be a 
multi-variate gaussian with diagonal covariance matrix. Our code uses the convergence criterium of
\cite{dunkley05} although other criteria are discussed in the Appendix.

\section{Illustrative examples}
In order to show the capabilities of the newly developed code, several examples are shown. Some of them
deal with synthetic data where we can investigate the behavior of the method under controlled conditions.
After these synthetic tests, we apply the code to a realistic case obtained from spectro-polarimetric observations.

\subsection{Simple academic example}
\label{sec:academic_case}
The first example serves as an illustration of how the MCMC method is able to capture the presence of 
degeneracies. To this end, a very simplified example is presented, where we make use of a Zeeman
triplet line, namely the \ion{Fe}{i} line at 630.2 nm. The emergent Stokes profiles are 
calculated in a Milne-Eddington atmosphere. The value of the parameters are: $\Delta v_\mathrm{dopp}$=2.4 km~s$^{-1}$, 
$v_\mathrm{mac}=0$ km~s$^{-1}$, $\beta=9$, $\eta_0=9.8$, $a=0.3$, $B=100$ G, $\theta_B=45^\circ$ and $\phi_B=0^\circ$.
The main characteristic of such synthetic profiles is that the 
magnetic field strength is so weak ($B = 100$ G) that the Zeeman splitting is negligible compared to the Doppler
width of the line. As a consequence, the emergent Stokes profiles can be described in the weak field regime of the
Zeeman effect, in which the Stokes $V$ profile is proportional to the wavelength derivative of the
intensity profile \citep[e.g.,][]{landi92}. It is widely known that only the line-of-sight component of the magnetic field
vector (i.e., the product $B \cos \theta$, with $\theta$ the angle between the line-of-sight and the
magnetic field vector) can be obtained from the amplitude of the Stokes $V$ profile. 
On the contrary, when linear polarization is also present, the precise magnetic field vector can also be obtained.
Since linear polarization appears as a second order contribution to the emergent signal, they are difficult
to observe and a reduced noise level is fundamental. In this section we present an analysis with the aid of the
MCMC code on how the information
retrieved from the previously described synthetic Stokes profiles degrades with the presence of noise. The noise
is described by a Gaussian distribution parameterized by the value of $\sigma$, which is here given in units of the
continuum intensity, $I_\mathrm{c}$.

\begin{figure*}[!t]
\centering
\includegraphics[width=0.33\hsize,clip]{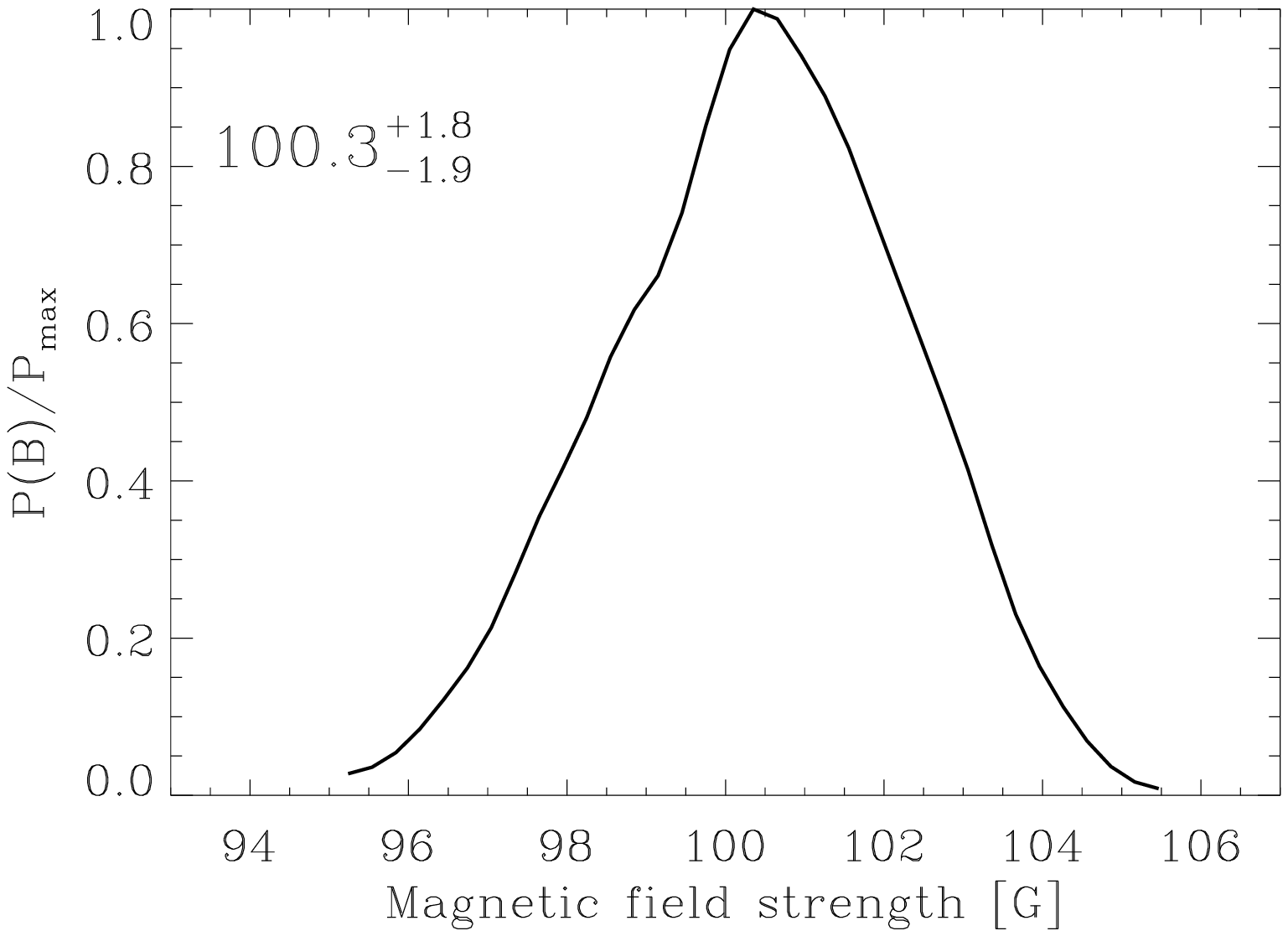}%
\includegraphics[width=0.33\hsize,clip]{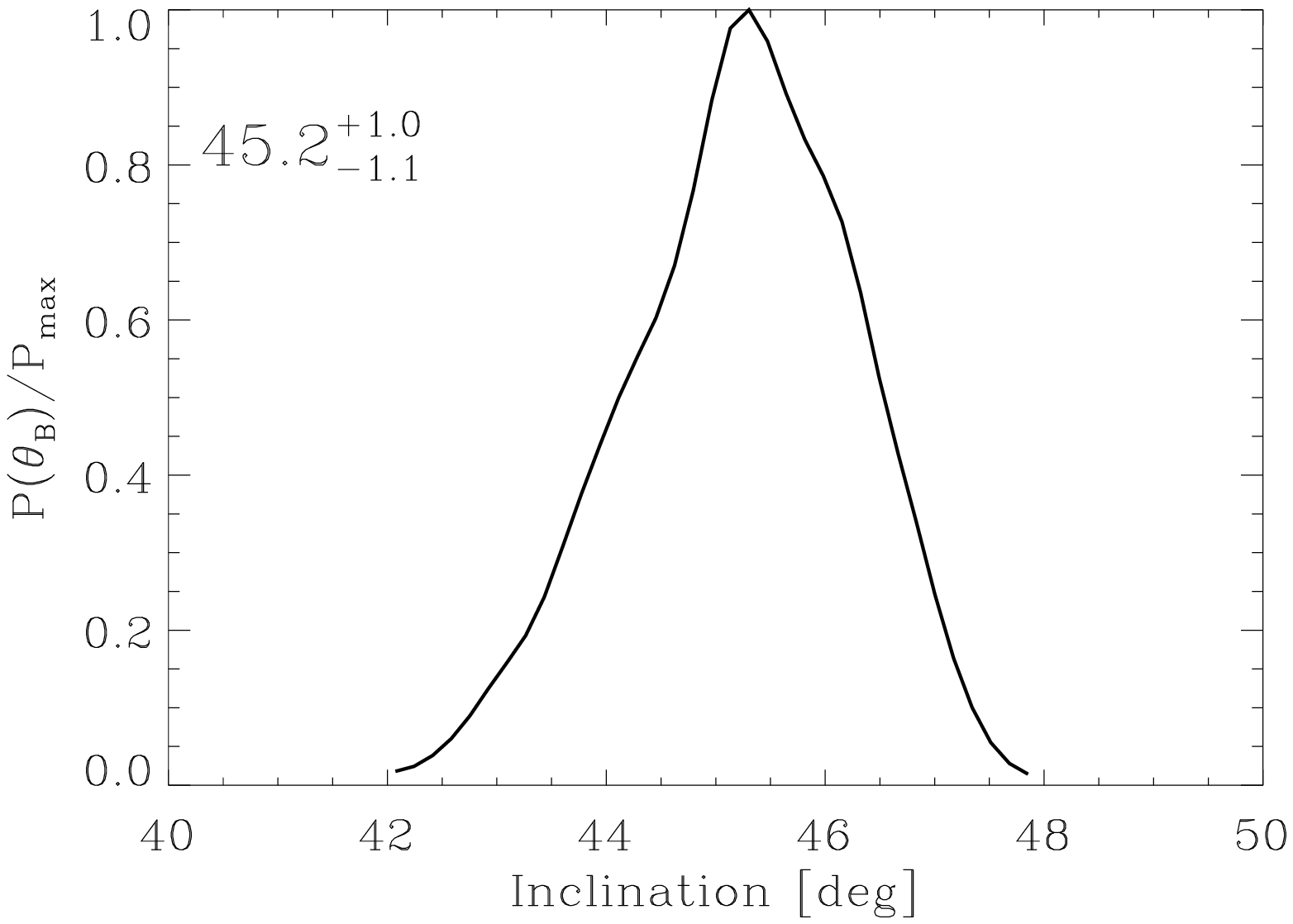}%
\includegraphics[width=0.33\hsize,clip]{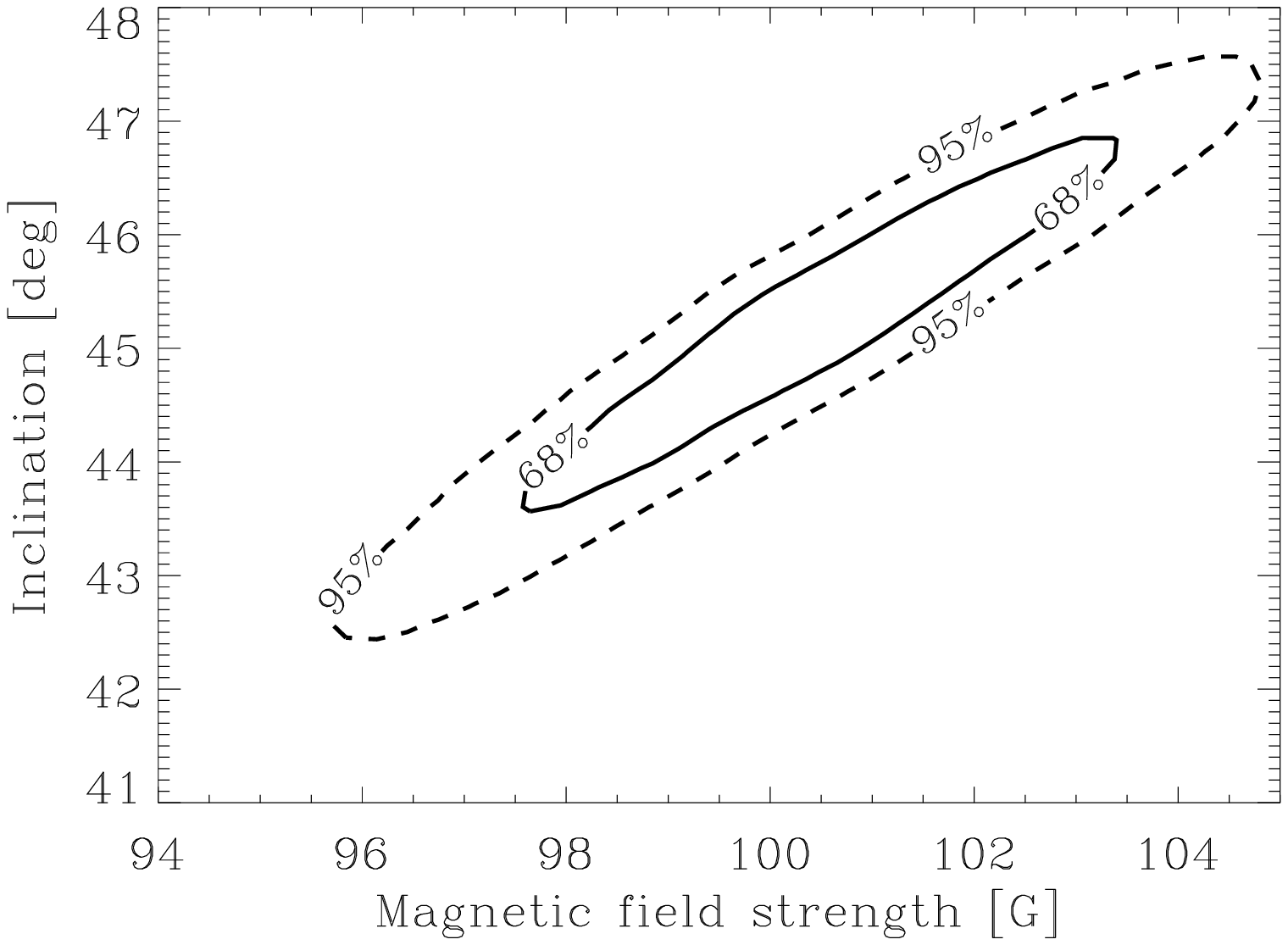}
\caption{Posterior probability distribution for the simple academic example with a noise $\sigma=10^{-5}$ in
units of the continuum intensity, $I_\mathrm{c}$. The full
Stokes vector ($I$,$Q$,$U$,$V$) is taken into account. The left panel 
shows the marginalized distributions for the magnetic field strength while the central
panel shows the distribution for the inclination. The right panel shows the two--dimensional
posterior distribution. The contours indicate the confidence levels at 68\% (solid line) and
95\% (dashed line).\label{fig:academic_noise1e-5}}
\end{figure*}

\begin{figure*}
\centering
\includegraphics[width=0.33\hsize,clip]{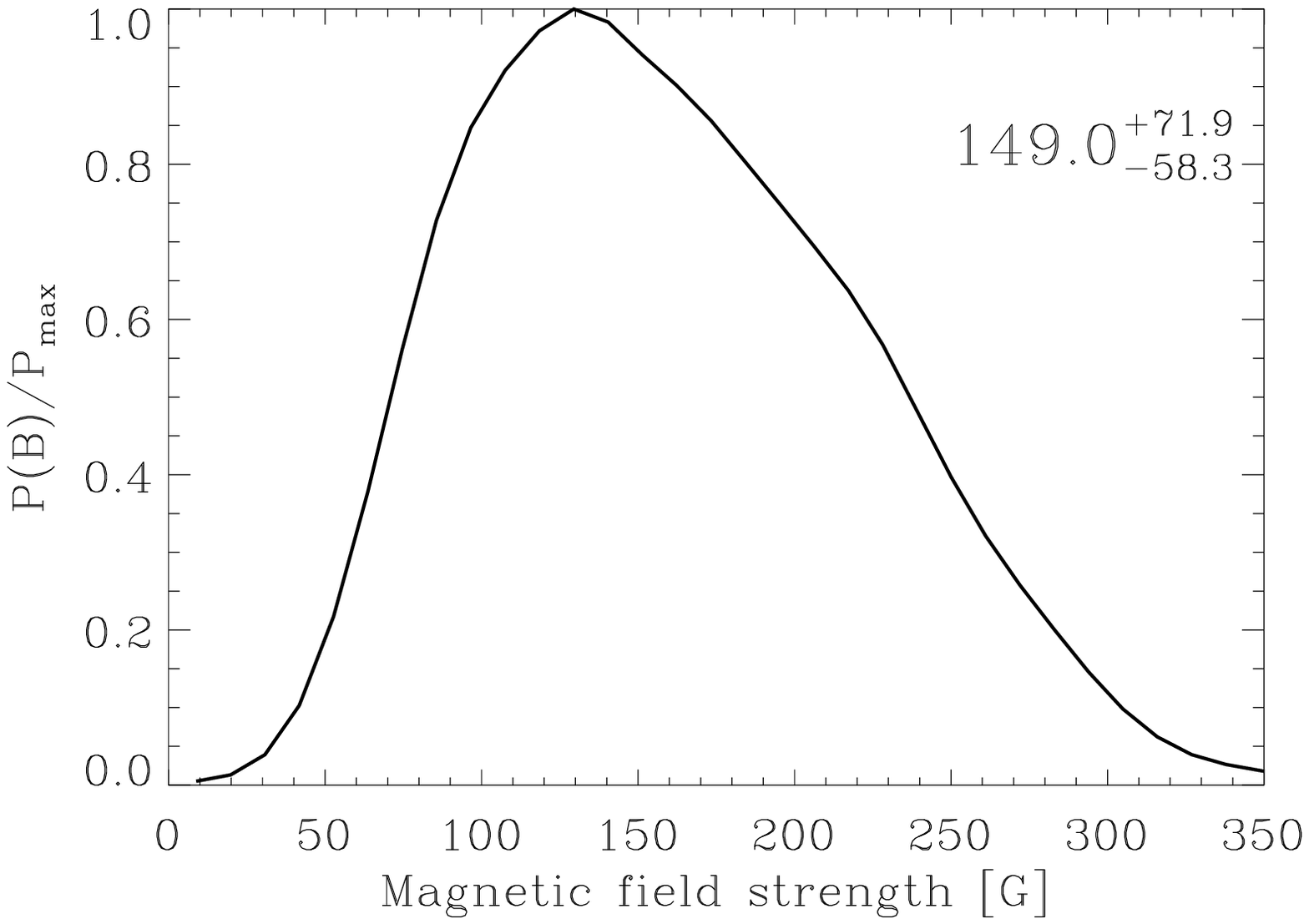}%
\includegraphics[width=0.33\hsize,clip]{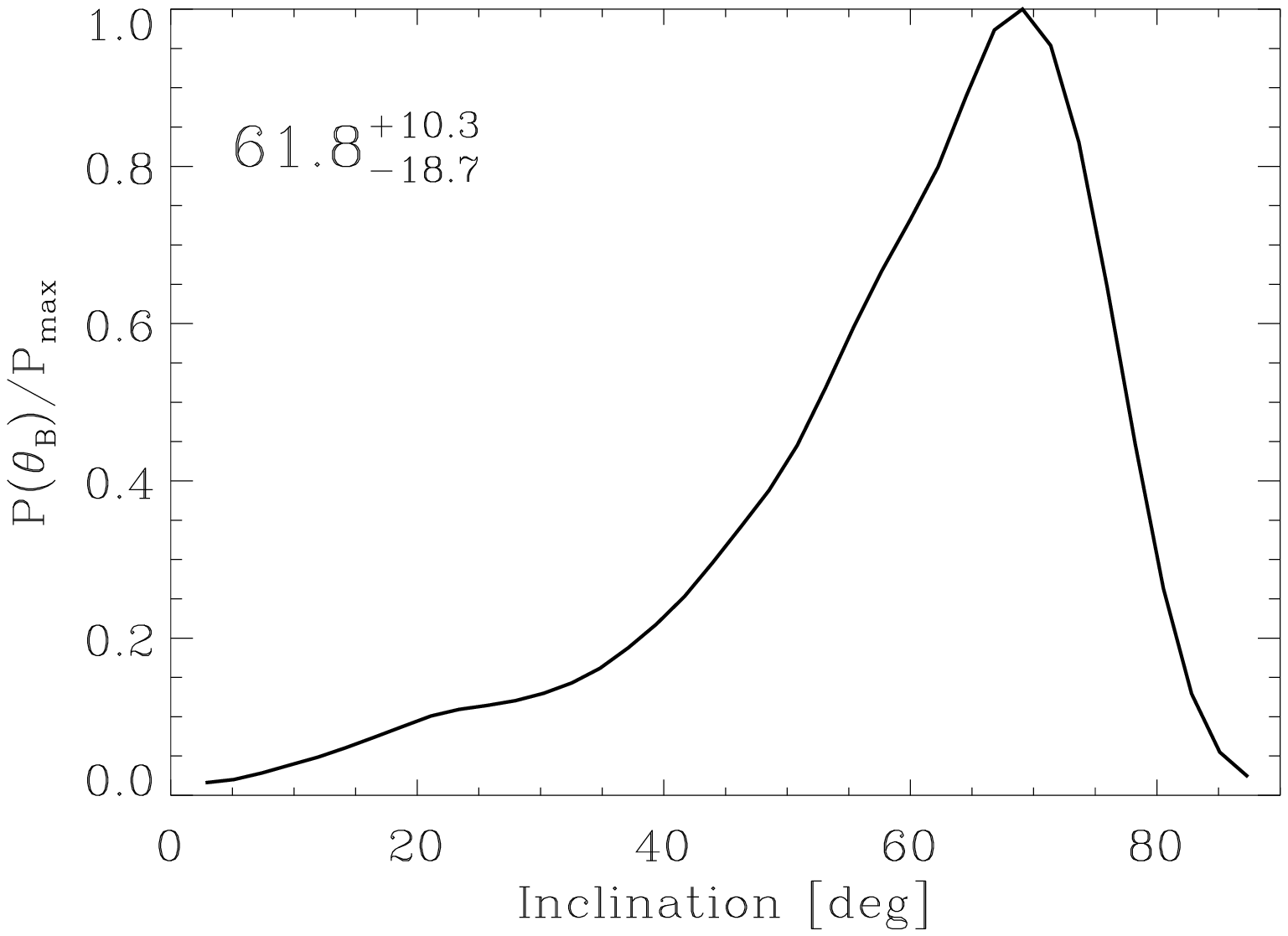}%
\includegraphics[width=0.33\hsize,clip]{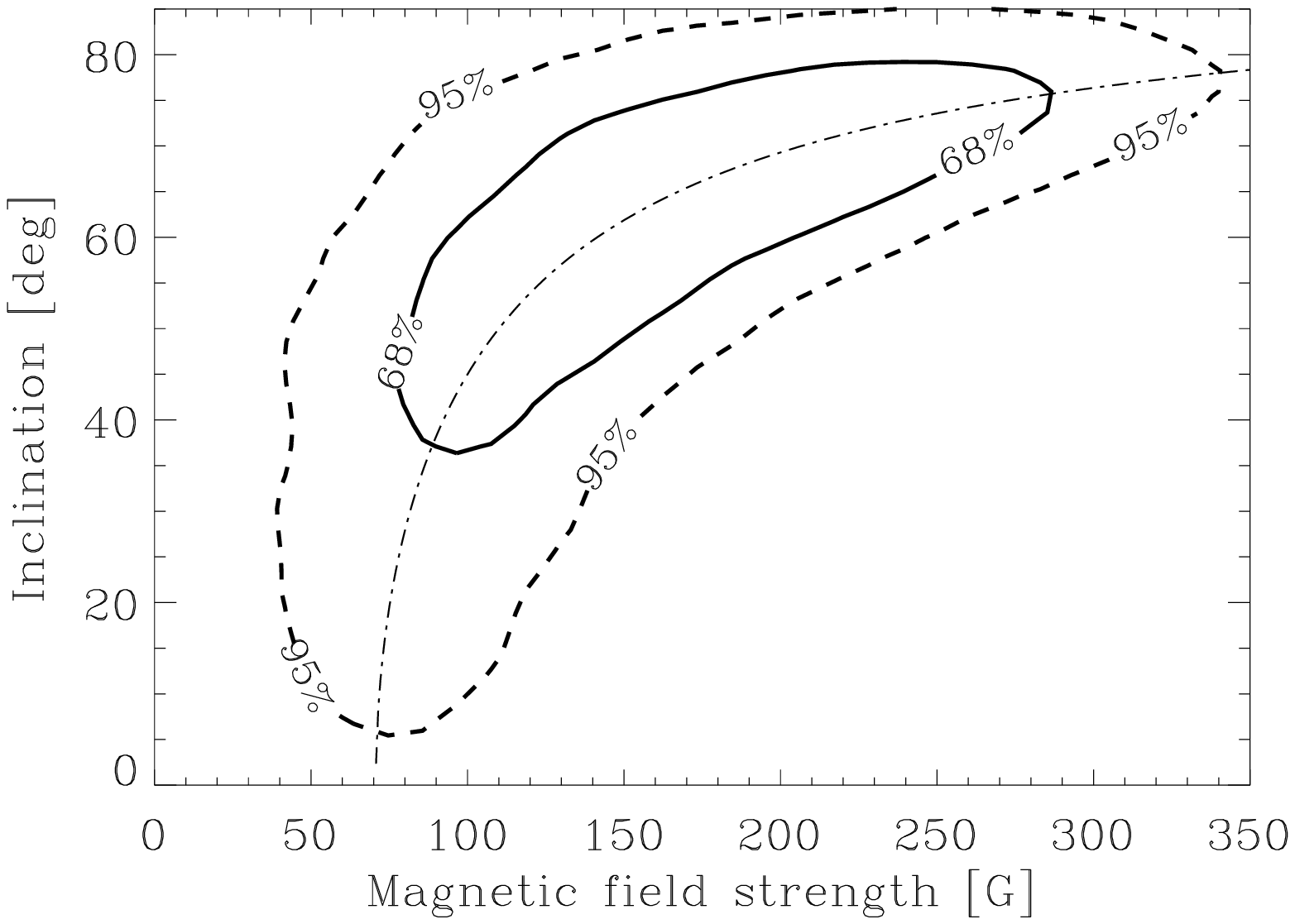}
\caption{Posterior probability distribution for the simple academic example with a noise $\sigma=10^{-3}$ in
units of the continuum intensity. The posterior clearly shows a degeneracy between the magnetic field strength
and the inclination. The dashed thin line in the right panel indicates the points where 
$B \cos \theta_B=B' \cos \theta_B'$, where the primed quantities are those belonging to the
synthetic profile, namely, $B'$=100 G and $\theta_B'$=45$^\circ$.
\label{fig:academic_noise1e-3}}
\end{figure*}

\begin{figure}
\centering
\includegraphics[width=0.9\columnwidth]{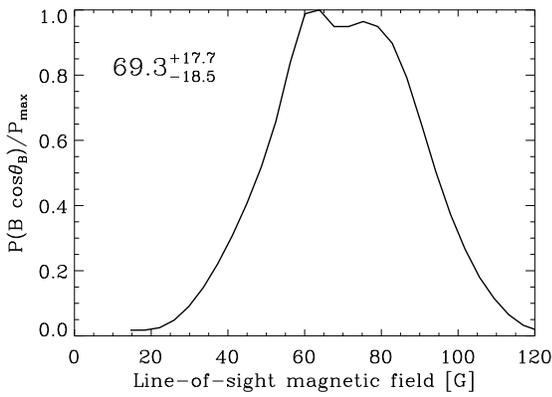}
\caption{Posterior probability distribution of the line of sight component of the magnetic field,
$B \cos \theta_B$ for the simple academic example with a noise $\sigma=10^{-3}$ in
units of the continuum intensity. The posterior clearly shows a peak compatible with the original value.
\label{fig:academic_noise1e-3_lineofsight}}
\end{figure}

We run the MCMC code on the synthetic Stokes profiles taking into account the full Stokes vector for the
calculation of the likelihood function given by Eq. (\ref{eq:likelihood}). All the 
thermodynamical parameters and the azimuth are assumed to be known and we only allow the magnetic field 
strength and the inclination of the
field to vary. We test that the obtained Markov chain is converged for the two parameters as indicated in 
Appendix A. For informative purposes, and although it depends on the 
complexity of the problem, our Markov chains require lengths of the order 50000 accepted samples to fulfill the convergence
criterion, with a total computational time of the order of 20 seconds on a standard computer. Finally, taking 
into account that the obtained Markov chain is sampling
from the posterior probability distribution, the posterior itself can be obtained simply by ``making
histograms''. Two different cases with different amounts of added noise have been considered. A case in which
the added noise level is $\sigma=10^{-5}$, whose results are shown in Fig. \ref{fig:academic_noise1e-5},
and a case with a much larger noise of $\sigma=10^{-3}$, whose results are shown in Fig. 
\ref{fig:academic_noise1e-3}. The two-dimensional histograms shown in the right panels of both figures
present a graphical representation of the posterior distribution of the magnetic field
strength and inclination, $p(B,\theta_B)$. We show two contours indicating confidence levels of
68\% and 95\%, respectively. 
The case $\sigma=10^{-5}$ shows a clearly peaked posterior
distribution, indicating that a very good estimation of the magnetic field strength and inclination is 
possible. On the contrary, the case $\sigma=10^{-3}$ presents a clear degeneration between both parameters,
manifested by the typical ``banana-shaped'' posterior distribution. The main reason for this extended
posterior distribution is that the Stokes $Q$ and $U$ signals are masked below the noise level. For 
such a high noise level, only the information encoded in the Stokes $V$ signal is available for retrieving
the magnetic field strength and inclination. Since the field is only 100 G, the line is in the weak-field
regime so that only the product $B \cos \theta_B$ can be estimated from Stokes $V$. In order to make sure
that this is indeed the case, we have overplotted the curve $B \cos \theta_B=B' \cos \theta_B'$ with
$B'=$100 G and $\theta_B'=$45$^\circ$, which closely follows the shape of the posterior.
Figure \ref{fig:academic_noise1e-3_lineofsight} shows the marginalized distribution of the
line-of-sight component of the magnetic field, $B \cos \theta_B$, showing that it can be recovered
with accuracy.
Marginalized posteriors\footnote{They are obtained by integrating the two-dimensional histogram with respect 
to one of the variables.} $P(B)$ and $P(\theta_B$) are also shown in the left and central panels of 
Figs. \ref{fig:academic_noise1e-5} and \ref{fig:academic_noise1e-3}. Sharp distributions are found for the
case with $\sigma=10^{-5}$ noise, while distributions with enhanced tails are found for the case 
$\sigma=10^{-3}$. Curiously, according to the marginalized distributions, a somewhat good estimation of the 
field strength is possible even for this highly noisy profiles, although there is a non-negligible tail for
larger field strengths. Concerning $P(\theta_B)$, it gives reduced information about the inclination, clearly
showing the $B \cos \theta_B$ degeneracy. For comparison purposes, we have also applied an inversion
code based on the Levenberg-Marquardt algorithm to estimate the parameters and their confidence intervals for
the case with $\sigma=10^{-3}$. The minimum of the $\chi^2$ function is correctly obtained for $B=98.5$ G 
and $\theta_B=44.1^\circ$. However, the symmetric confidence intervals that we obtain using the diagonal
elements of the covariance matrix \citep[e.g.,][]{numerical_recipes86} produce an estimation of
$98.5 \pm 180.9$ G for the magnetic field strength and $44.1 \pm 100.5$ degrees for the magnetic field
inclination. According to the estimated error, the field inclination is not constrained by the observations.
These results provide a poor estimation of the confidence intervals as compared to the
marginalized posterior distributions shown in Fig. \ref{fig:academic_noise1e-3}.

\begin{figure*}
\centering
\includegraphics[width=0.33\hsize,clip]{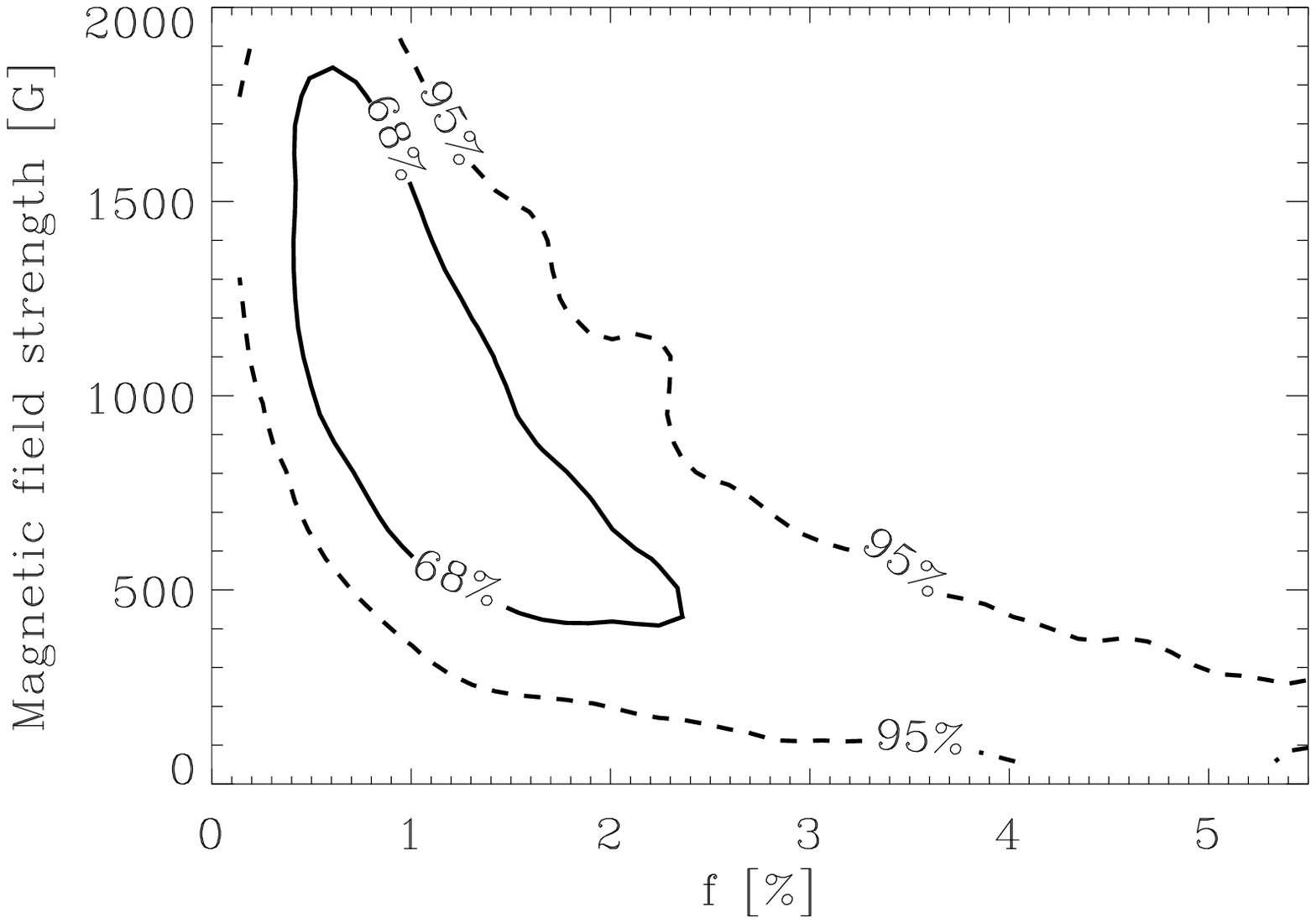}
\includegraphics[width=0.33\hsize,clip]{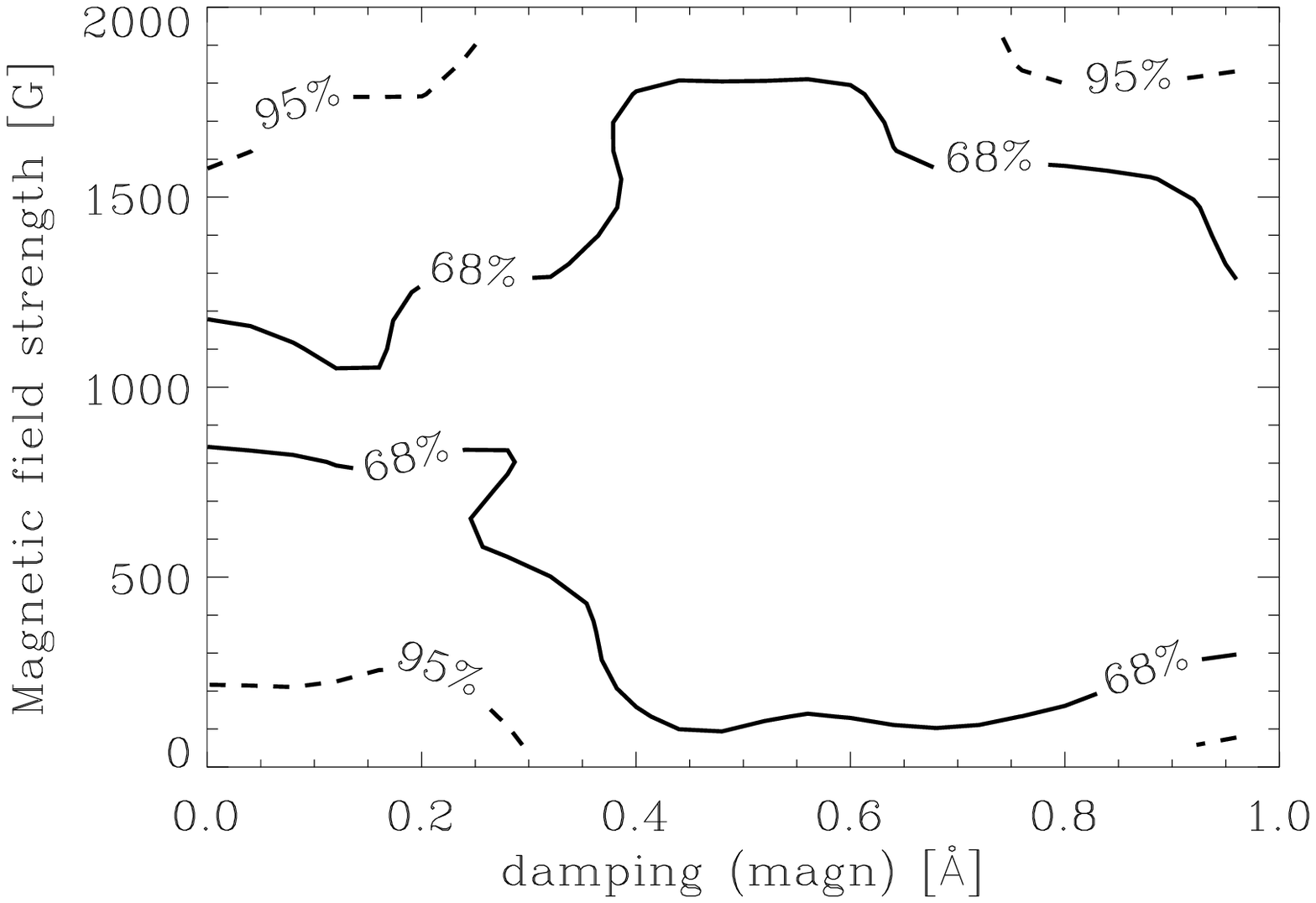}
\includegraphics[width=0.33\hsize,clip]{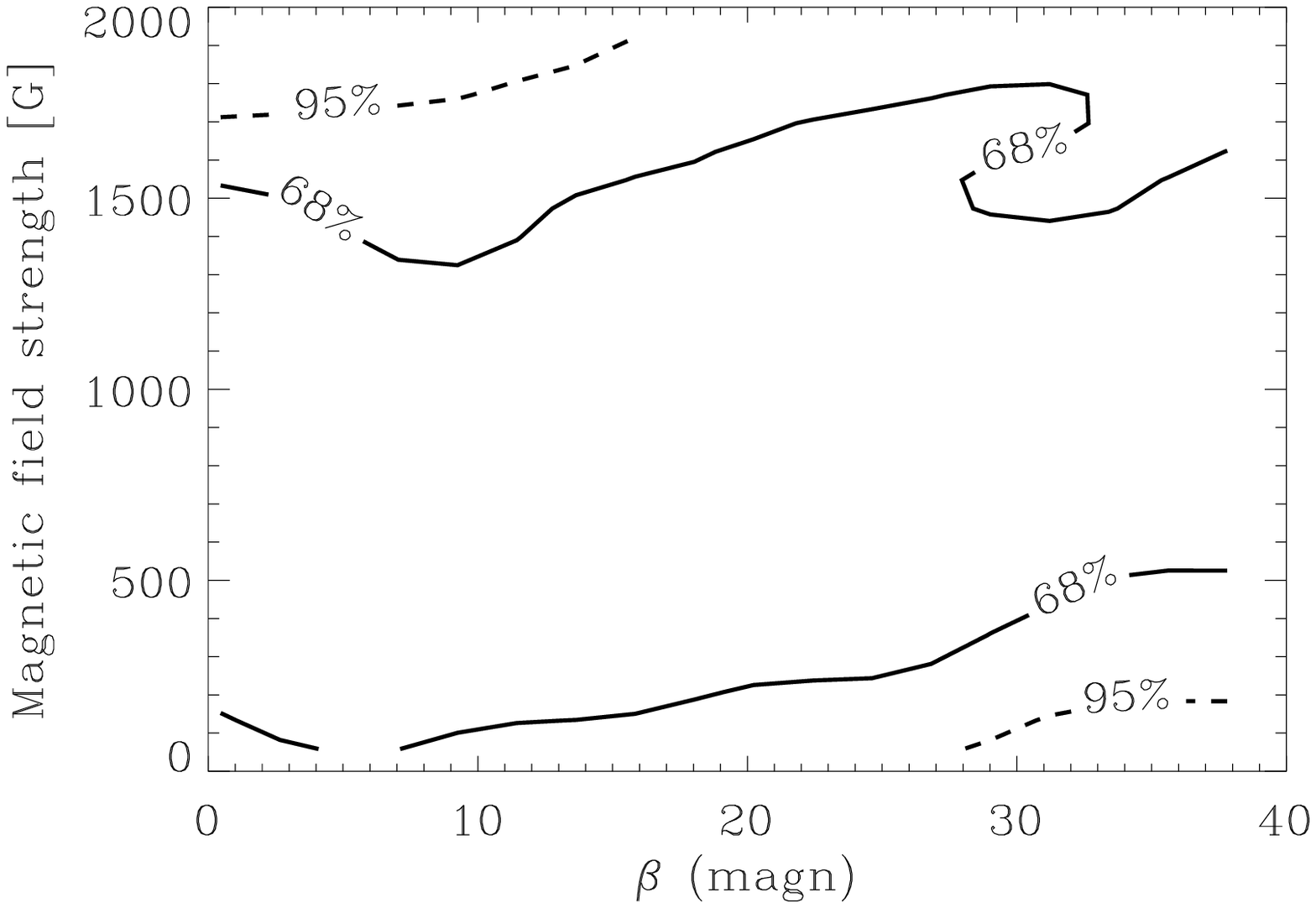}\\
\includegraphics[width=0.33\hsize,clip]{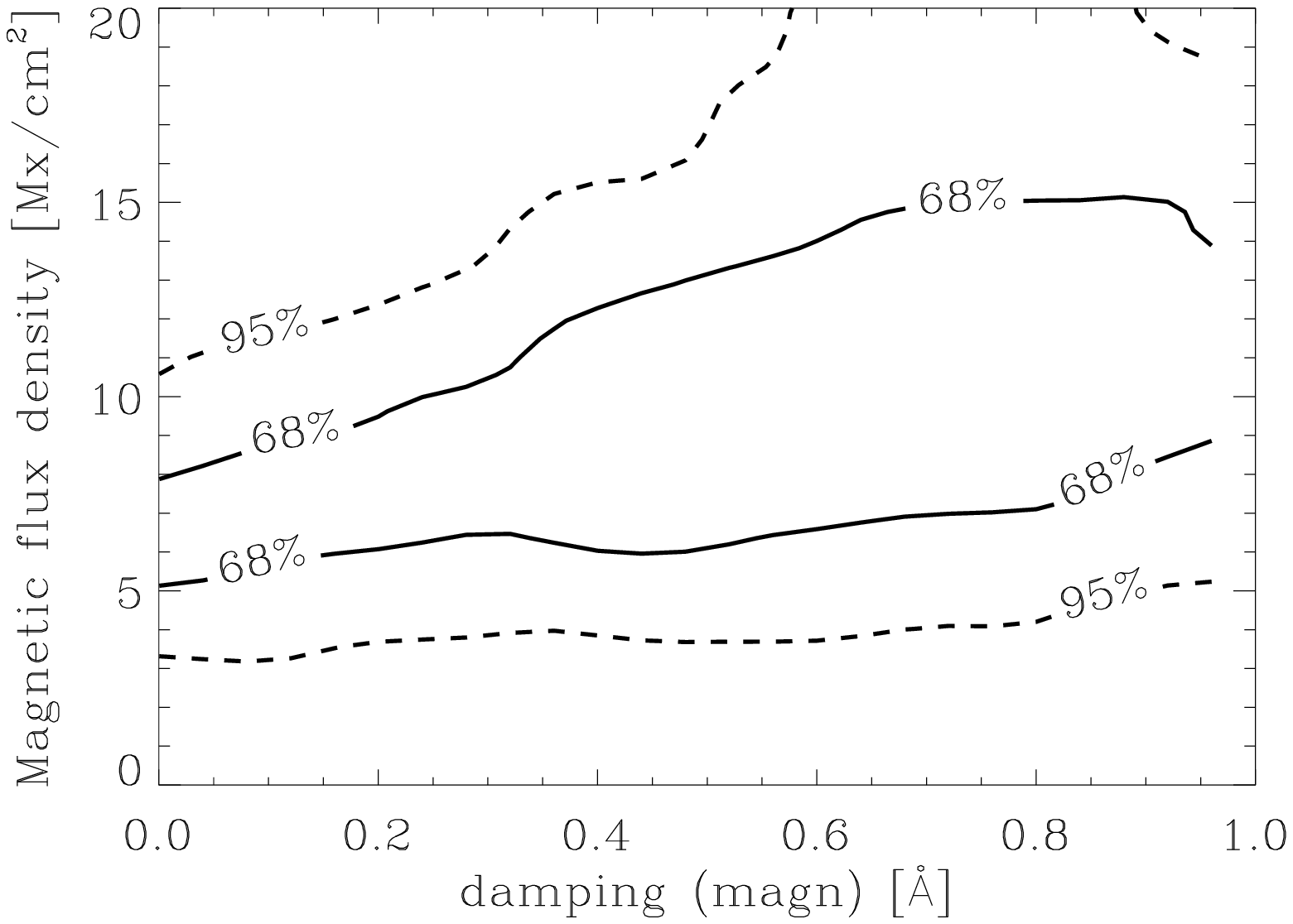}
\includegraphics[width=0.33\hsize,clip]{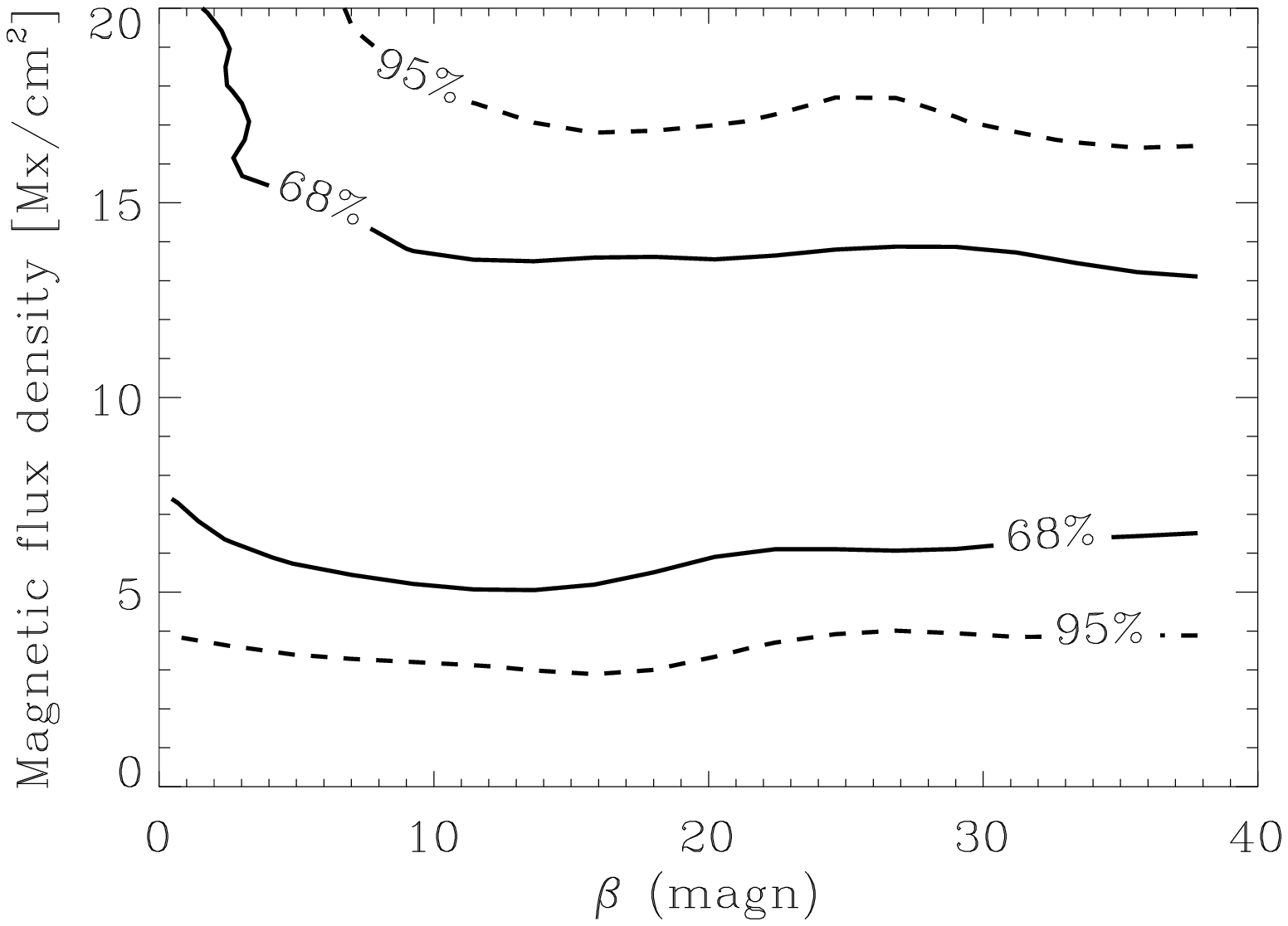}
\includegraphics[width=0.33\hsize,clip]{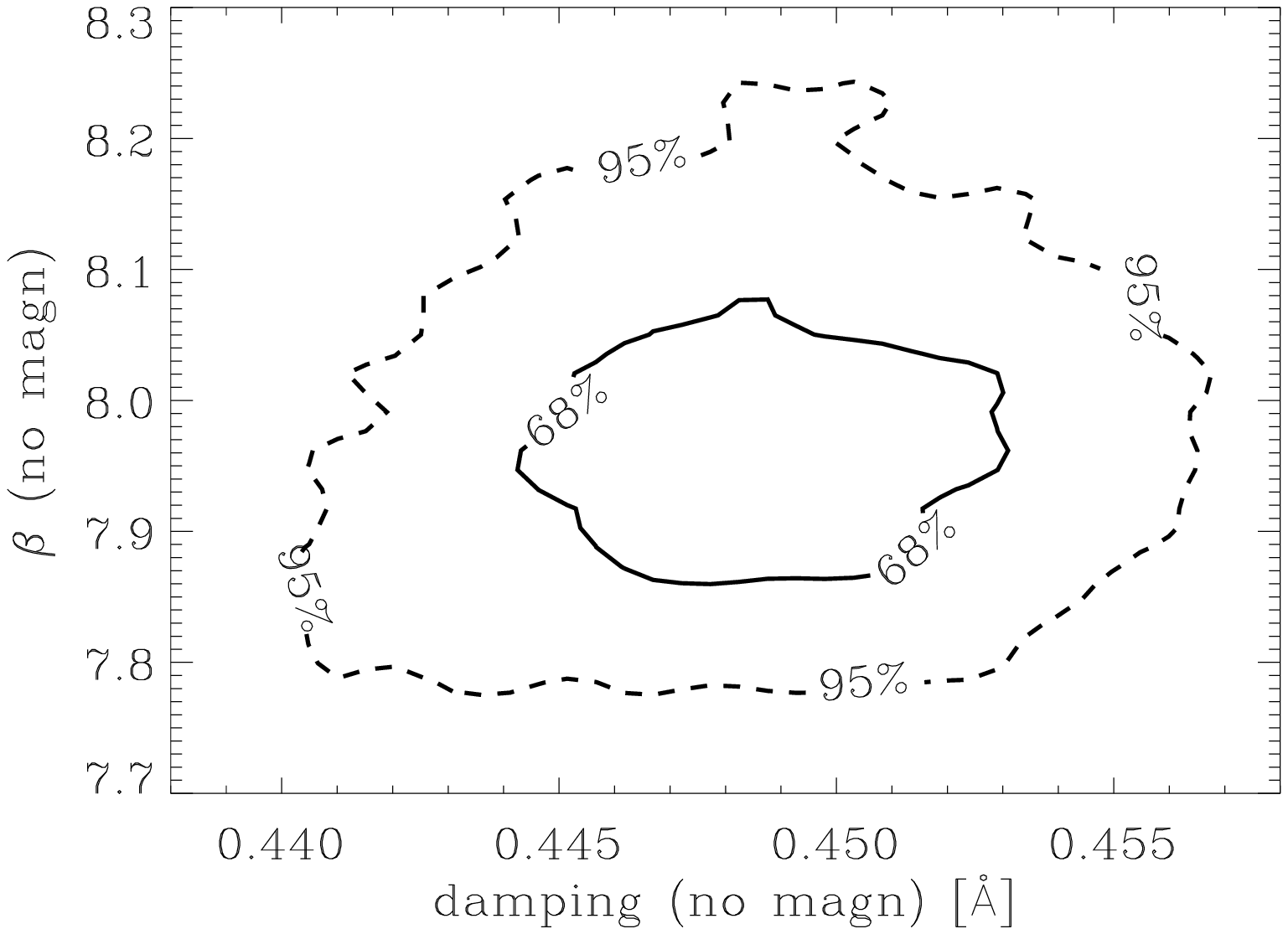}
\caption{Two-dimensional posterior distributions for several combinations of the parameters for the internetwork
synthetic example with a longitudinal magnetic flux density of $10$ Mx/cm$^2$. The contours indicate the regions
where 68\% and 95\% confidence levels are placed. Large degeneracies are present in almost all the parameters, 
except for the gradient of the source function and the damping in the non-magnetic component.}
\label{fig:qs1}
\end{figure*}

The
utility of this test is two-fold: on the one hand, we have demonstrated, with a simplified problem, the correct
operation of the MCMC inversion code; on the other hand, we point out the obvious importance of having 
accurate Stokes profiles in order to recover information about the magnetic field vector.

\subsection{The problem of the quiet Sun}
After the presentation of a simple instructive example, we focus now on a more realistic problem that
presents deep implications on the recovery of information about the magnetism of the quiet solar
photosphere.
The quiet Sun are those regions away from the most evident manifestations of magnetic 
activity. In the photosphere, it corresponds mainly to the network (magnetic flux concentrations 
in the supergranular boundaries) and the internetwork (filling up the interior of supergranular cells). 
At the present spatial resolution of ground-based spectropolarimetric observations ($0.5-1''$) the 
magnetic structures on the quiet Sun are thought to be not spatially resolved 
\citep[e.g.,][]{stenflo94,lin95,dominguez03,khomenko03,martinez_gonzalez06}. This has been
demonstrated by \cite{asensio_mn07}, presenting the first map of flux cancellation in the quiet Sun. The 
magnetism of the network is 
widely established as predominantly vertical kG structures filling approximately 
$10-20$\% of the resolution element. However, the problem turns out to be more 
complicated in the internetwork, where the polarization signals that we can measure 
by means of the Zeeman effect are unresolved, occupying only the $1-2$ \% of the resolution element. 
Typically the Stokes V profiles have an amplitude 
of $10^{-3}$ in units of the continuum intensity, I$_\mathrm{c}$. 
The noise that we can achieve in the observational data ($\sim 10^{-4}$ I$_\mathrm{c}$) 
is only one order of magnitude smaller than the polarimetric signals in the internetwork. 
As it has been 
shown in the previous section, it is important to have a reduced noise level in order to obtain
information about the magnetic field from the observed Stokes profiles.
In the internetwork, when the widely observed pair of \ion{Fe}{1} lines at $630$ nm are used, no linear
polarization signal above the noise level is found with the current instrumentation. However, even if there is 
a lack of signal in Stokes $Q$ and $U$, we could retrieve magnetic field strengths when the line is out from the 
weak field regime. In this particular pair of lines and for the typical photospheric physical conditions, the line
can be considered in the weak field regime for fields below $\sim 600$ G. This would mean that the kG magnetic
field strengths retrieved from this pair of lines would be reliable 
\citep[e.g.,][]{grossmann-doerth96,sigwarth99,dominguez03,sanchez_almeida03}.

However, \cite{martinez_gonzalez06} have shown that this results should be regarded with care. They 
show the most simple case in which the thermodynamics compensates the effect of a magnetic field. These authors 
used the SIR\footnote{Stokes Inversion based on Response functions.} code \citep{sir92} 
to synthesize the emergent Stokes profiles using the typical physical conditions of the 
internetwork. The inversion of such profile with random initializations showed that the resulting 
atmospheres depended on the initialization itself if a noise level of $5\times 10^{-5}$ I$_\mathrm{c}$ is 
assumed. In each case the change in the magnetic field was compensated by a small change in the magnetic 
temperature gradient (smaller than 300 K) and a slightly increase of the microturbulent velocity (below 1.5 km/s).
The change in the temperature gradient produces a modification on the Stokes V ratio of the two spectral lines 
while the increase in the microturbulent velocity leads to a broadening of the line profile. This procedure clearly
demonstrated the degeneracy of the inversion problem in this particular case.
Unfortunately, the Levenberg-Marquardt algorithm used in the SIR code for the inversion of Stokes profiles does not
produce a reliable and well-defined estimation of the errors in the parameters that describe the atmosphere. This 
is the reason why \cite{martinez_gonzalez06}
showed the degeneracy of the inversion problem by using repeated inversions with random initializations.
In this paper, we follow the study performed by \cite{martinez_gonzalez06} and we extend it to the cases in 
which we increase the filling factor (we improve the signal to noise ratio) or we add a particular inclination 
to the magnetic field vector (we generate linear polarization signal). However, this time the solution is based
on robust statistical techniques.

\begin{figure*}
\centering
\includegraphics[width=0.33\hsize]{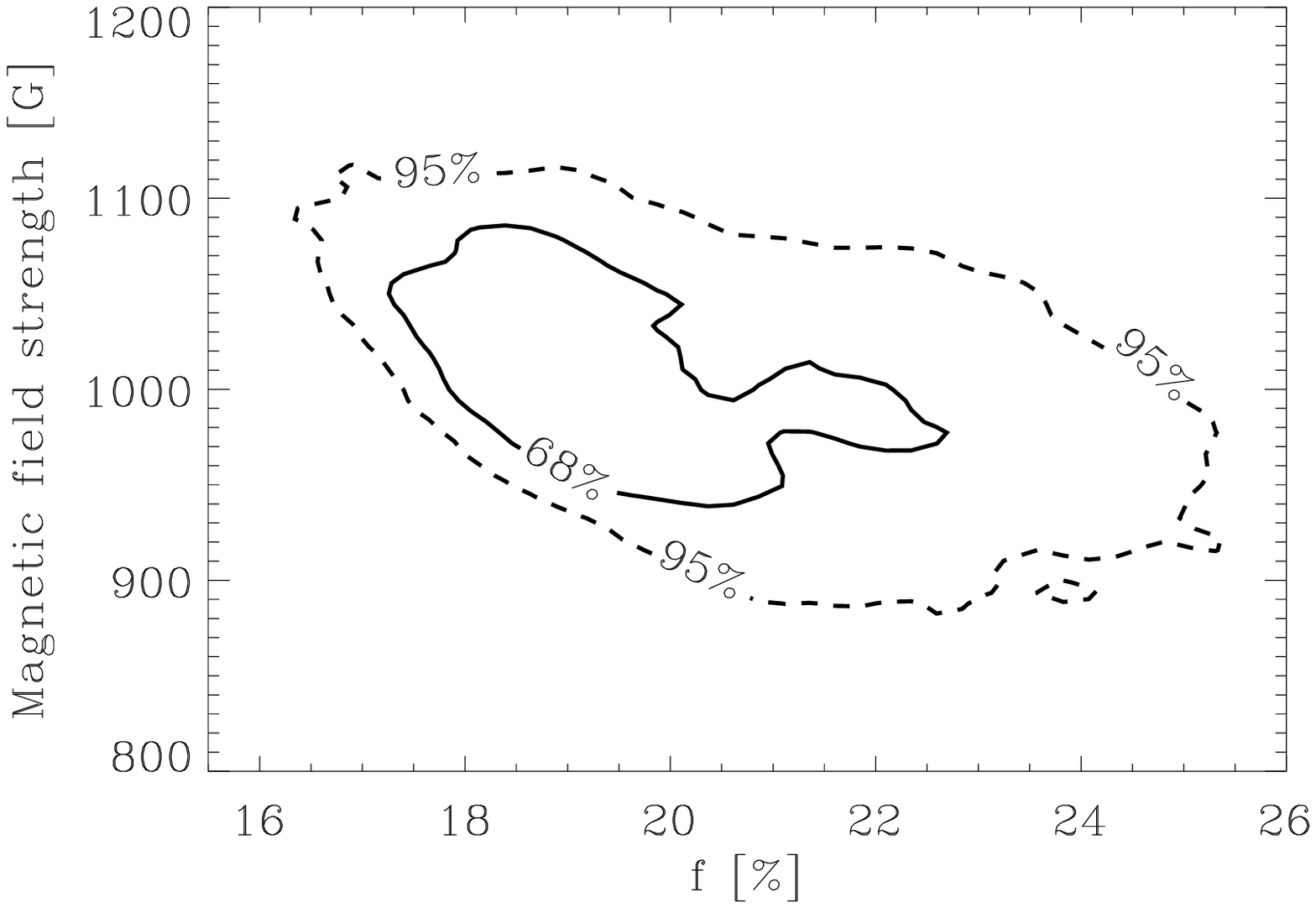}
\includegraphics[width=0.33\hsize]{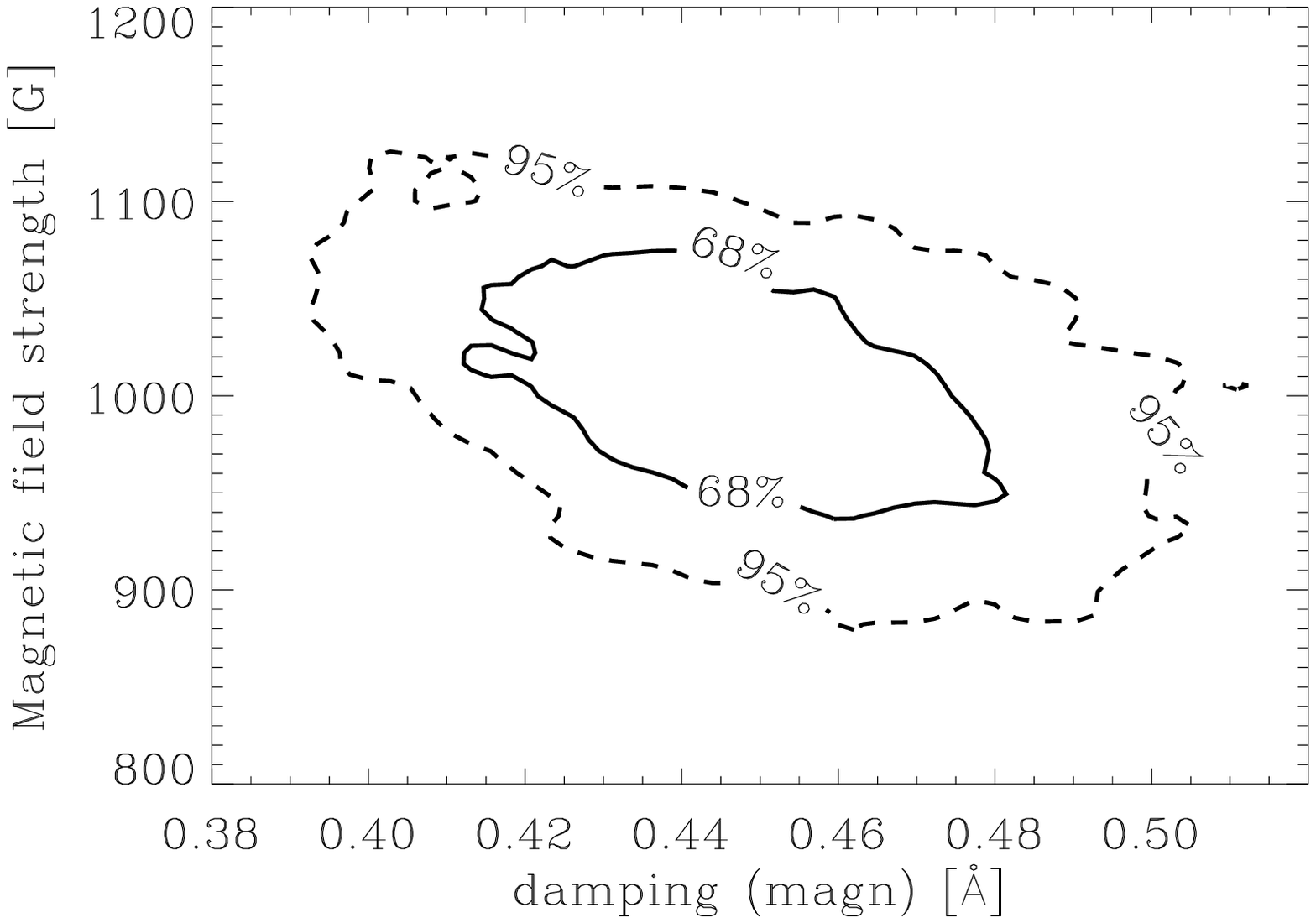}
\includegraphics[width=0.33\hsize]{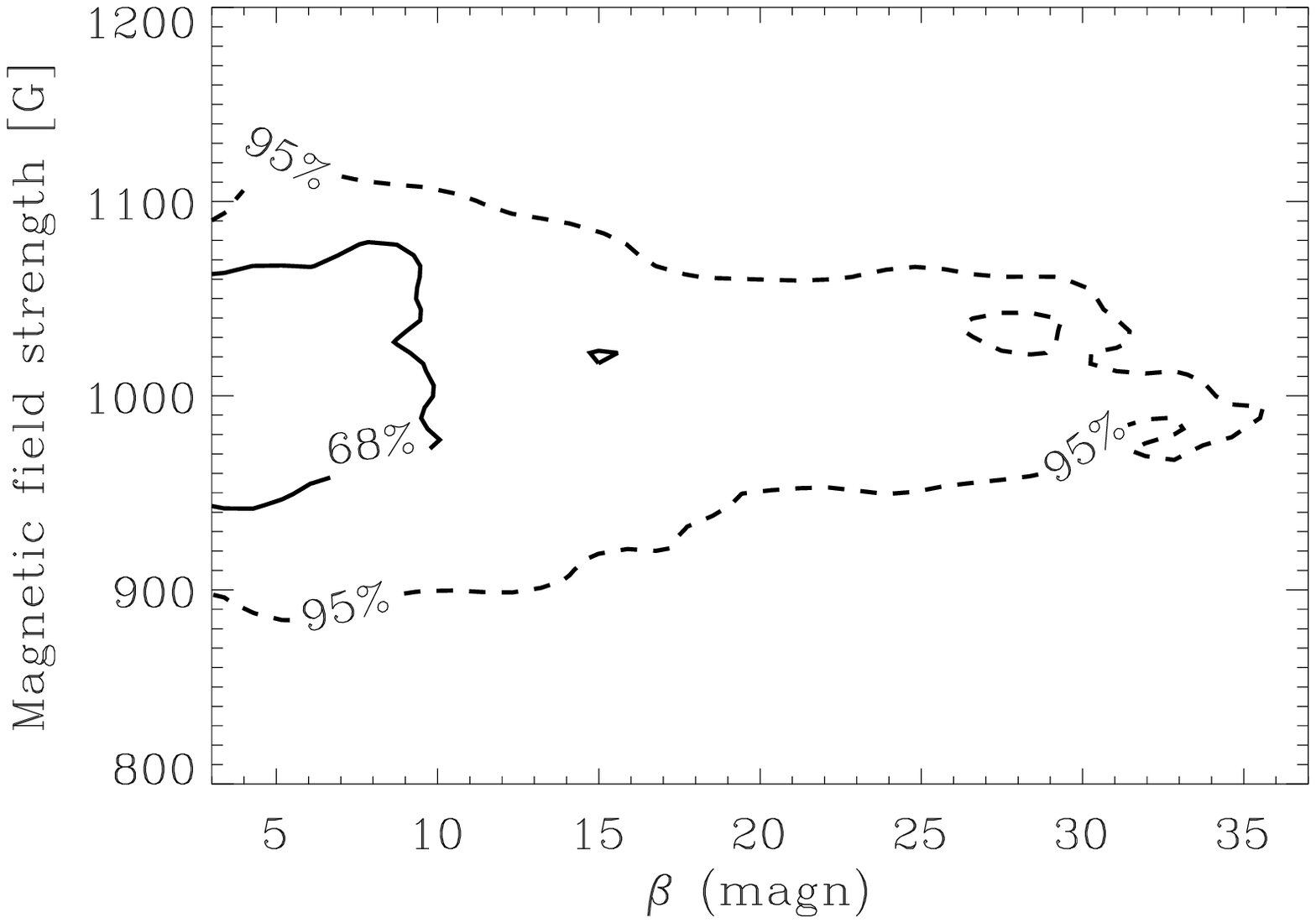}\\
\includegraphics[width=0.33\hsize]{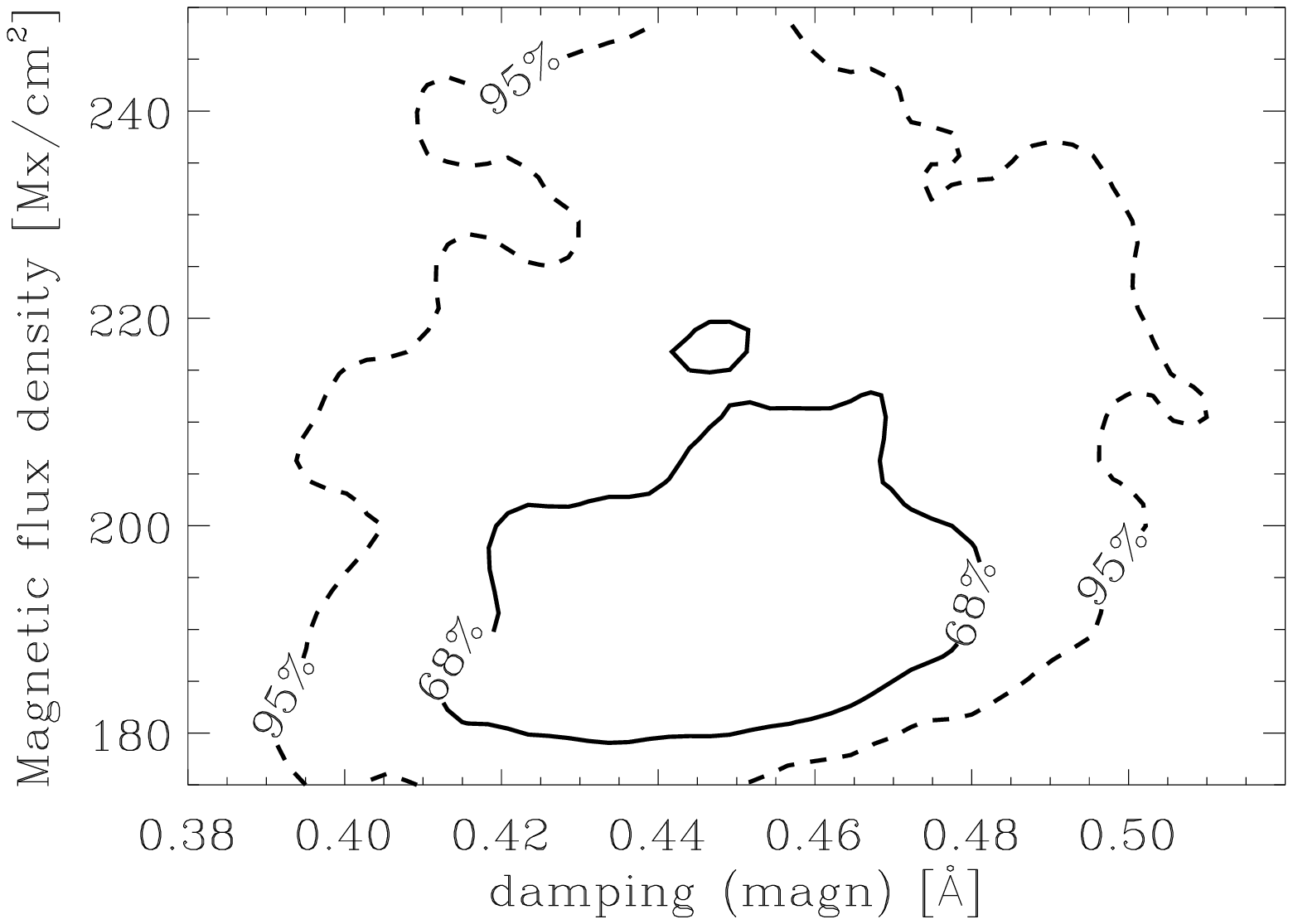}
\includegraphics[width=0.33\hsize]{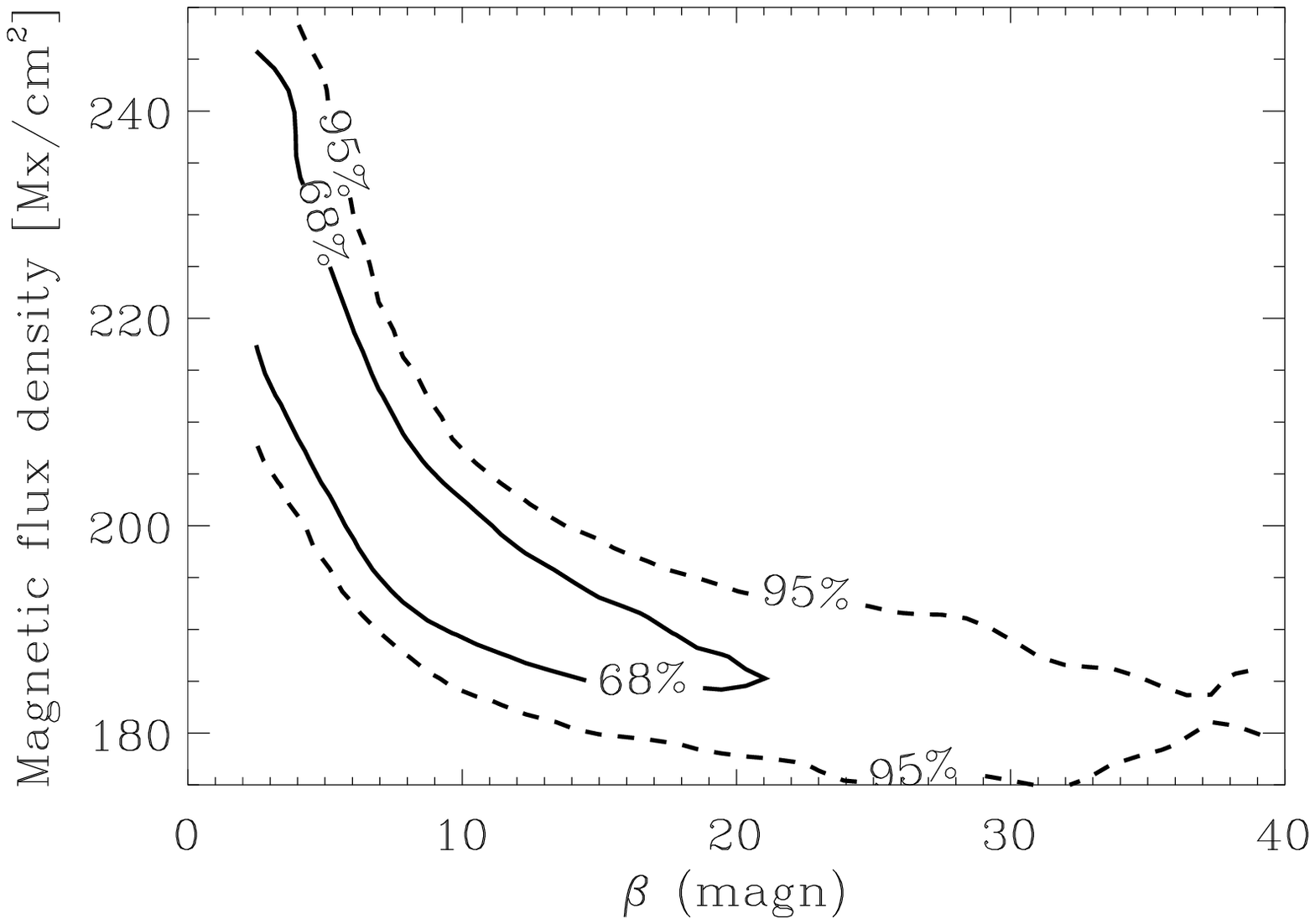}
\includegraphics[width=0.33\hsize]{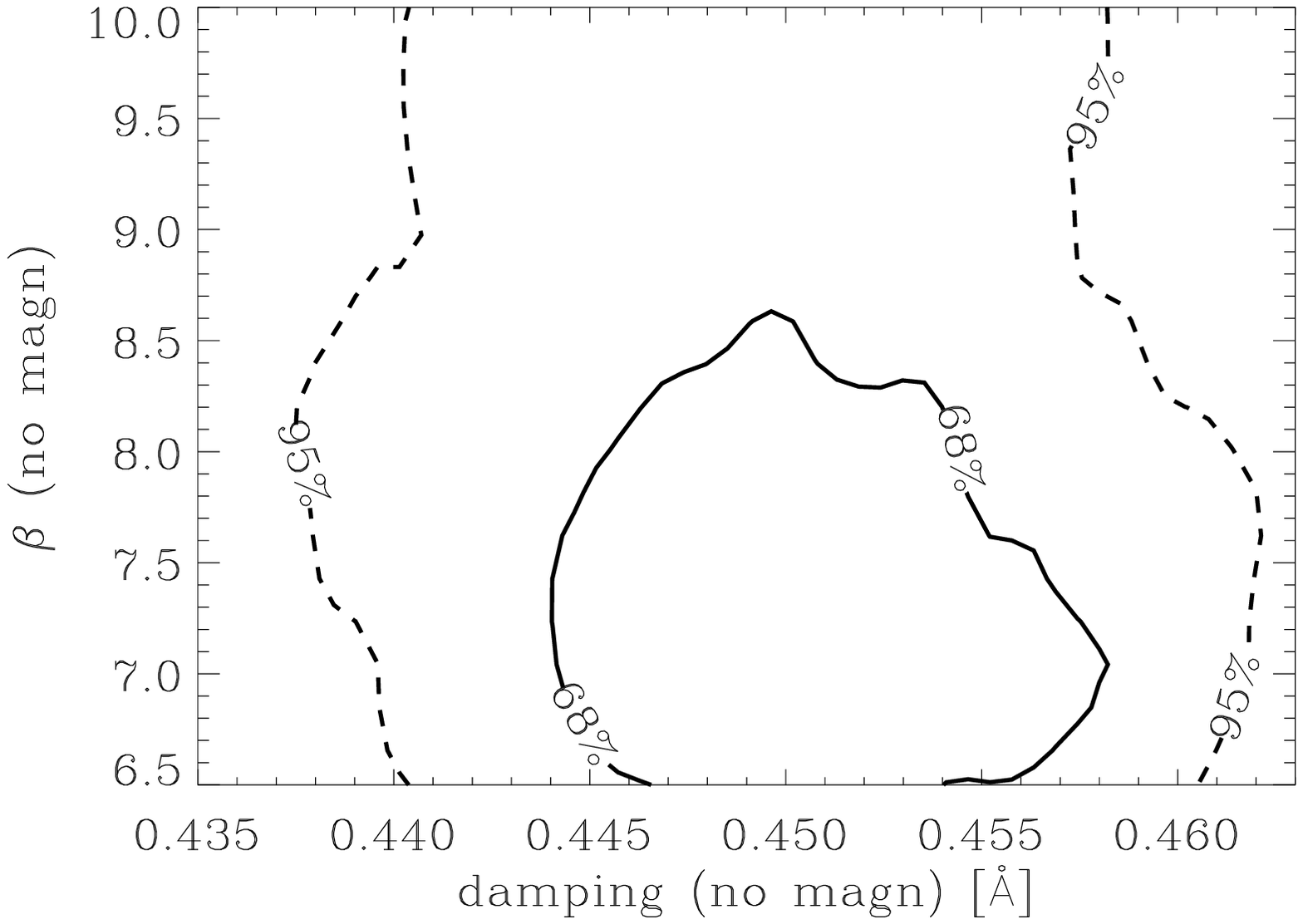}
\caption{Two-dimensional posterior distributions for several combinations of the parameters for the network
synthetic example with a longitudinal magnetic flux density of $200$ Mx/cm$^2$. The contours indicate the regions
where 68\% and 95\% confidence levels are placed. The magnetic field strength can be recovered better than
in the internetwork case shown in Fig. \ref{fig:qs1}. However, increased degeneracies are also seen in 
the parameters of the non-magnetic component due to the reduced surface covered by this component.}
\label{fig:qs2}
\end{figure*}

The interpretation of the weak field regime in the case that the magnetic feature is resolved is straightforward. 
In the weak field approximation, the radiative transfer equation has an analytical solution
\citep[see Chapter 9 of][for the conditions under which this approximation is valid]{landi_landolfi04}. The Stokes 
$V$ profiles can be written as:
\begin{equation}
V(\lambda)=-4.6686\times10^{-13} \bar{g} \lambda_0^2 B \cos\theta \frac{\partial I(\lambda)}{\partial \lambda},
\label{eq:weak_field}
\end{equation}
where $\bar{g}$ is the effective Land\'e factor of the line, $\lambda_0$ is the central wavelength of the 
spectral line given in \AA\ and $B$ is the magnetic field strength given in G. The simultaneous observation 
of the Stokes $I$ and $V$ profiles allows us to 
compute the product $B\cos\theta$. The situation in the quiet Sun is not so straightforward since 
the magnetic structures occupy a very small portion of the resolution element. Then, the modelization of these 
areas requires at least two components: a magnetic component that gives rise to the polarization signals and a 
non-magnetic one that accounts for the rest of the pixel that is field-free\footnote{The term field-free might
result confusing since this component can indeed present a magnetic field that, due to its special structure,
presents a zero Zeeman signal (e.g., microturbulent distribution, isotropic distribution, etc.).}. Two complicated
problems arise due to this particularity. First, the right-hand side of Eq. (\ref{eq:weak_field}) has to be 
multiplied by the filling factor 
and the value that we will recover would be the longitudinal magnetic flux density $\alpha B \cos\theta$. Second, 
the intensity profile that applies in Eq. (\ref{eq:weak_field}) is the one coming from the magnetic component. Then, 
the product $\alpha B \cos\theta$ cannot be computed from the ratio of the Stokes $V$ profile and the wavelength
derivative of Stokes $I$, since the observed Stokes $I$ is coming mainly from the non-magnetic component. 
If one still wants to use the previous approach, the only 
way to recover the product $\alpha B \cos\theta$ would be by computing a calibration curve. This means that 
we have to assume a model atmosphere and 
compute the Stokes $V$ profile for different values of the longitudinal magnetic flux density. 
As a result, the inferred magnetic field is model dependent. In other words, the Stokes $V$ profiles 
depend on the magnetic and thermodynamic properties so, if one wishes to infer the magnetic properties of the 
plasma, it is fundamental to fix the thermodynamical properties first. The only technique available to
overcome this difficulty is to apply inversion techniques. However, one has to have in mind that the 
information encoded in the Stokes $I$ profile (that has $\sim$99 \% contribution from the non-magnetic component)
and in the Stokes $V$ profile is not enough to constrain the problem and to recover in a trustable way all 
the atmospheric parameters in a two component model.

\begin{figure*}
\centering
\includegraphics[width=0.33\hsize]{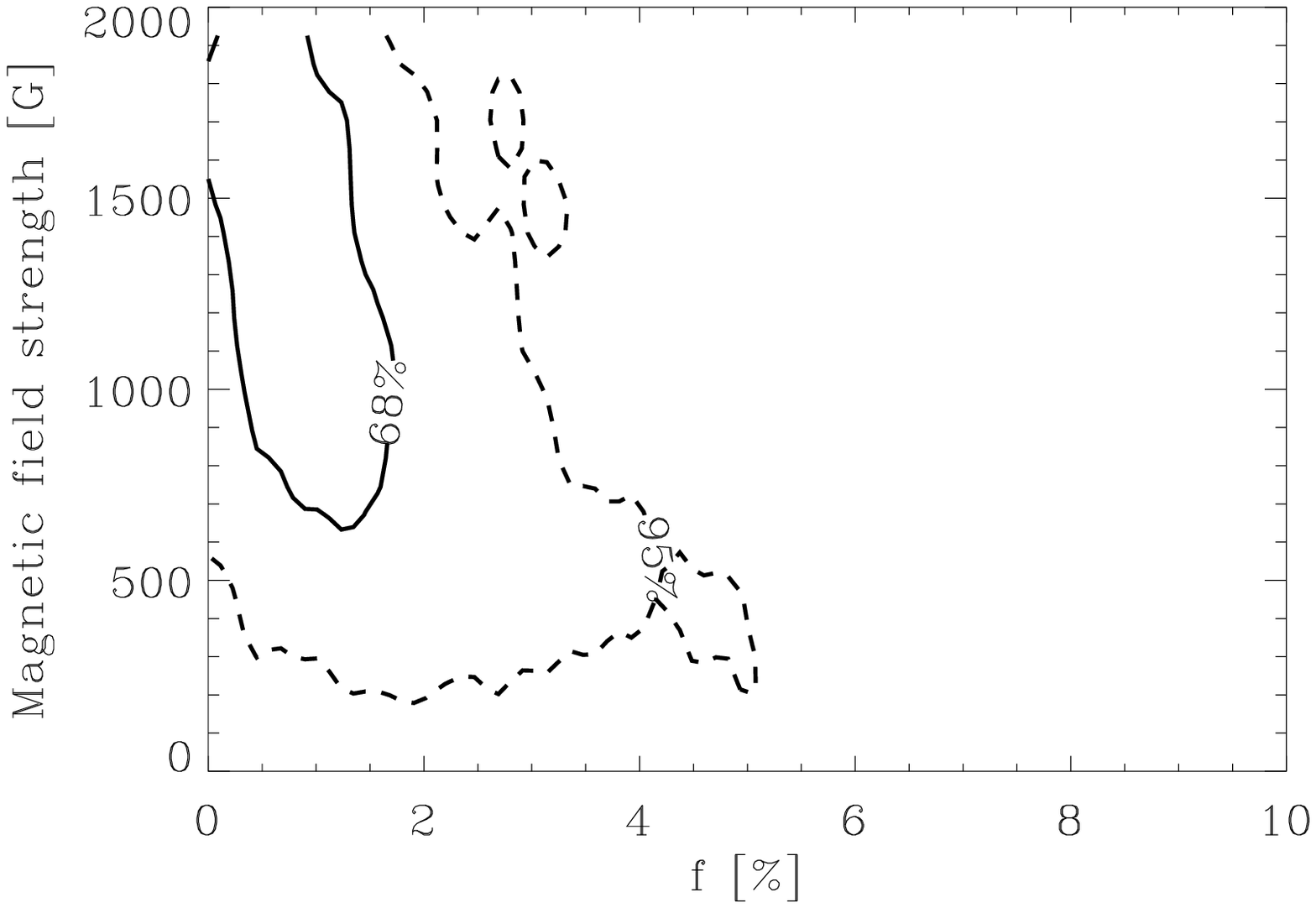}
\includegraphics[width=0.33\hsize]{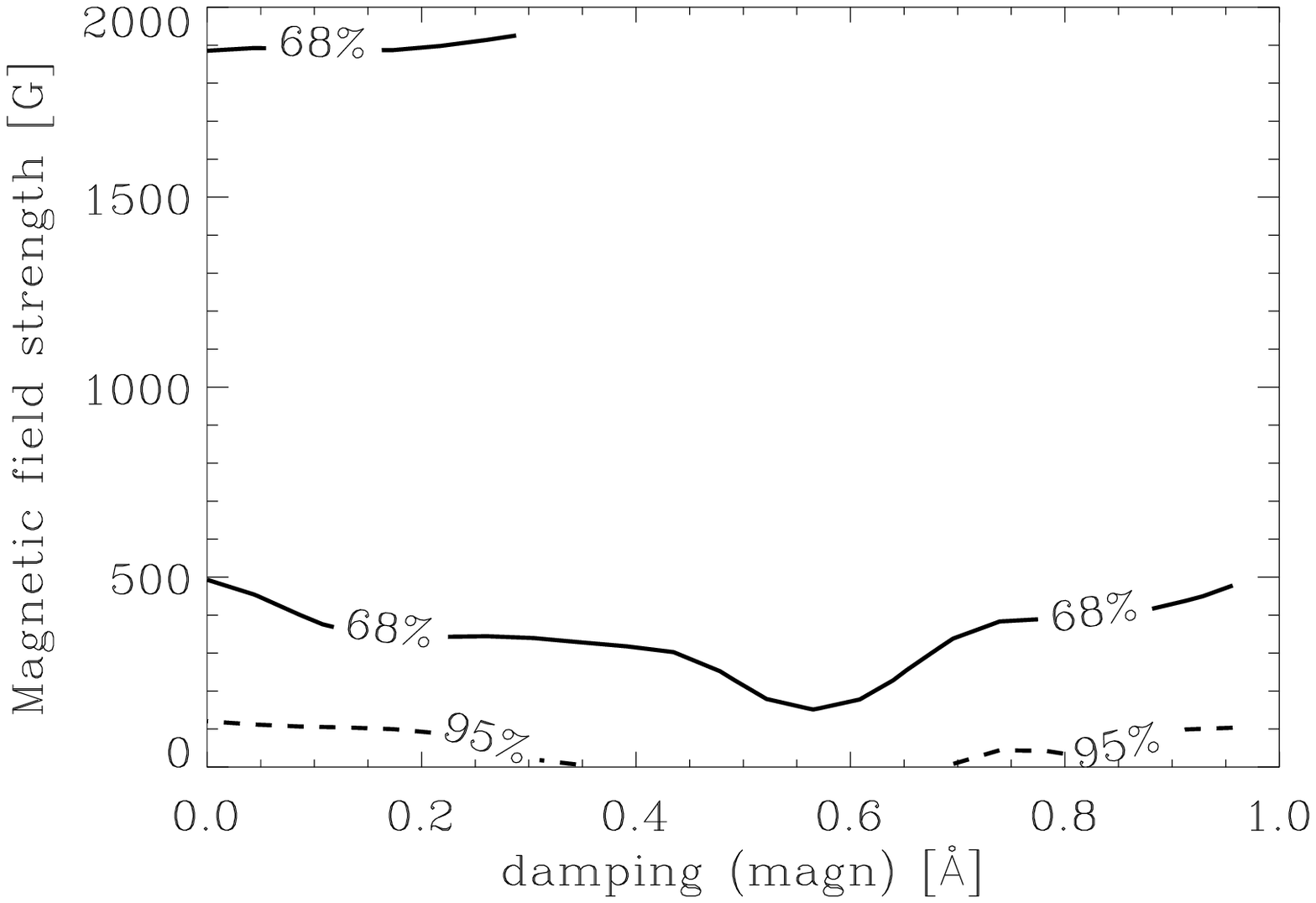}
\includegraphics[width=0.33\hsize]{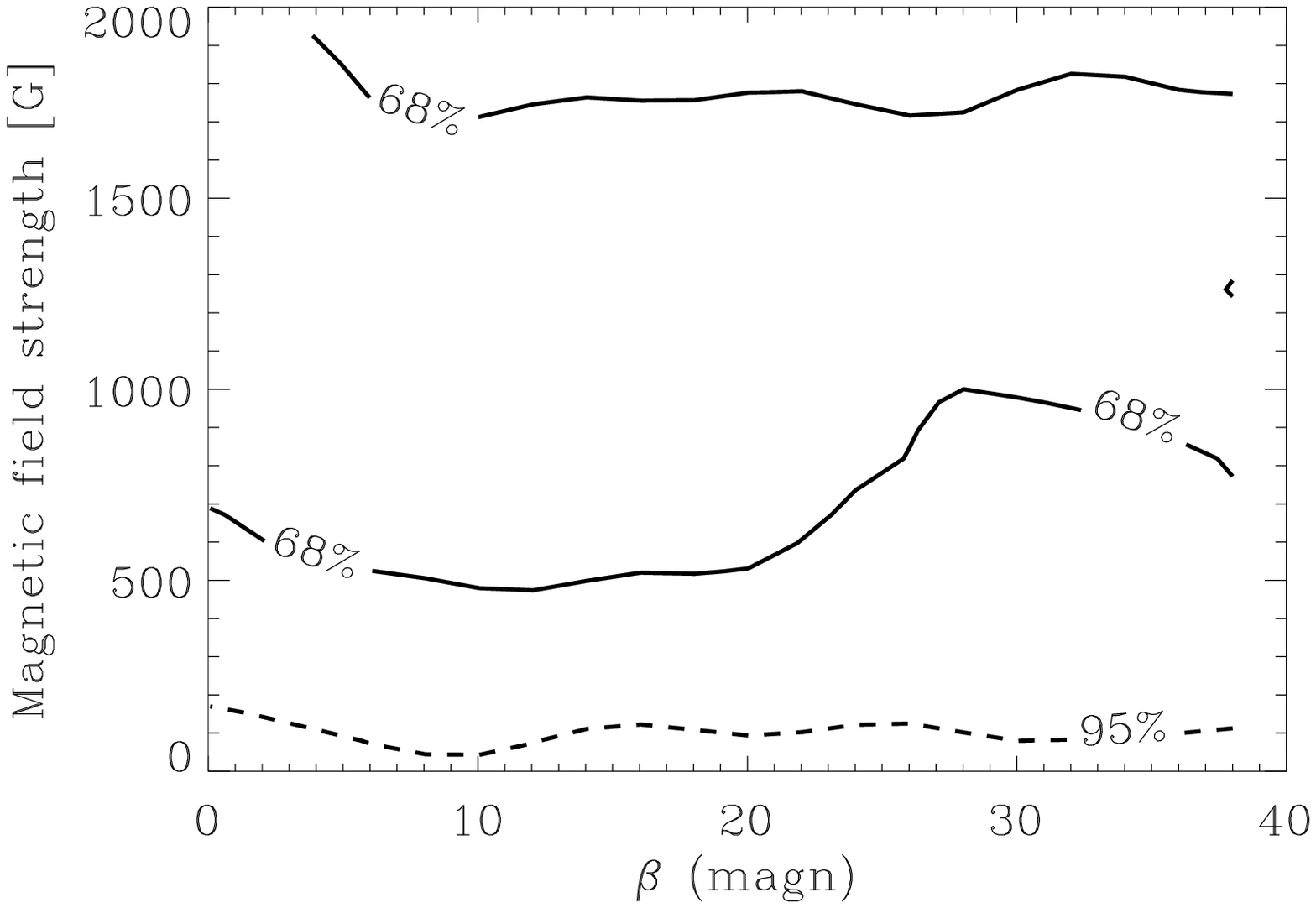}\\
\includegraphics[width=0.33\hsize]{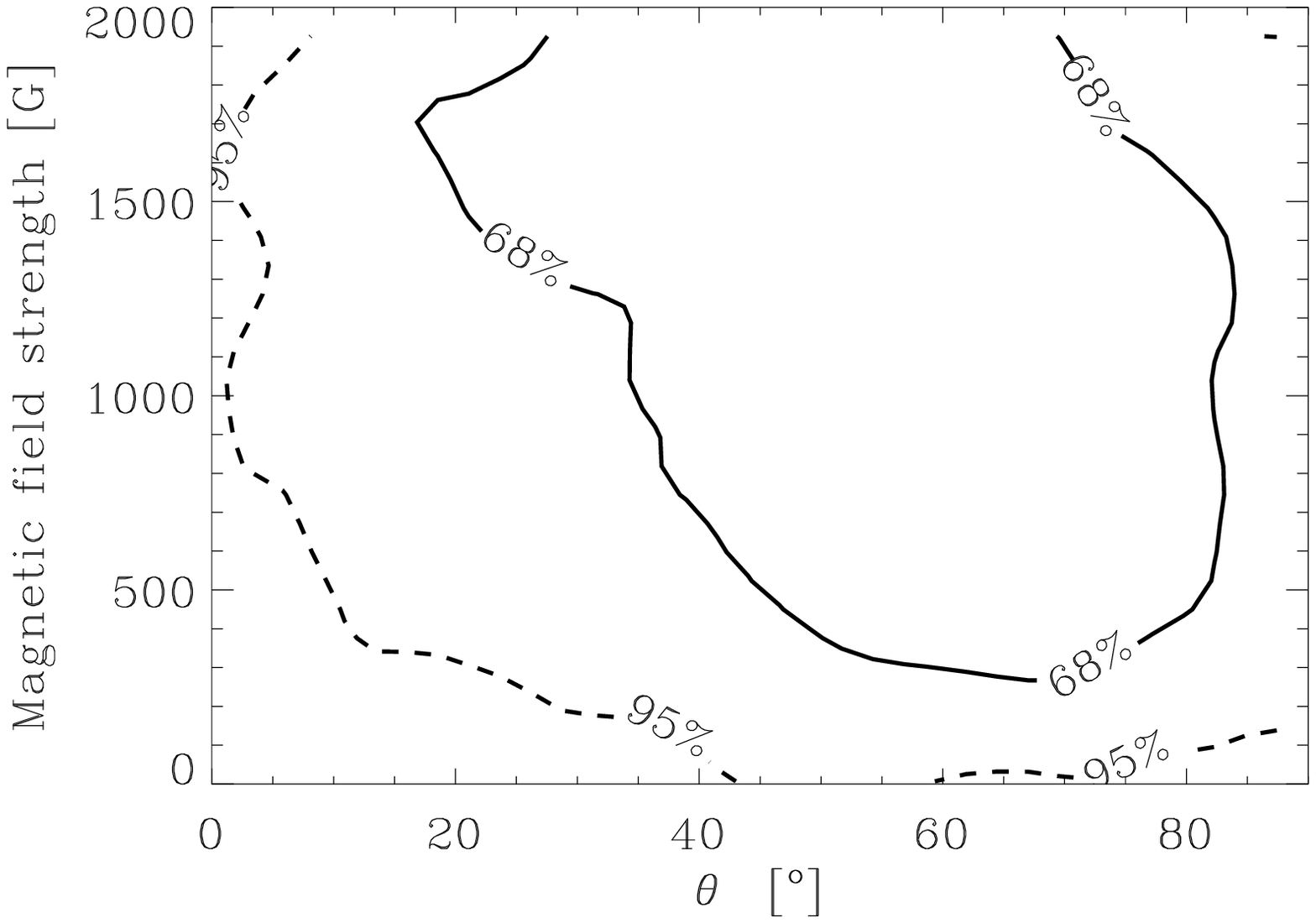}
\includegraphics[width=0.33\hsize]{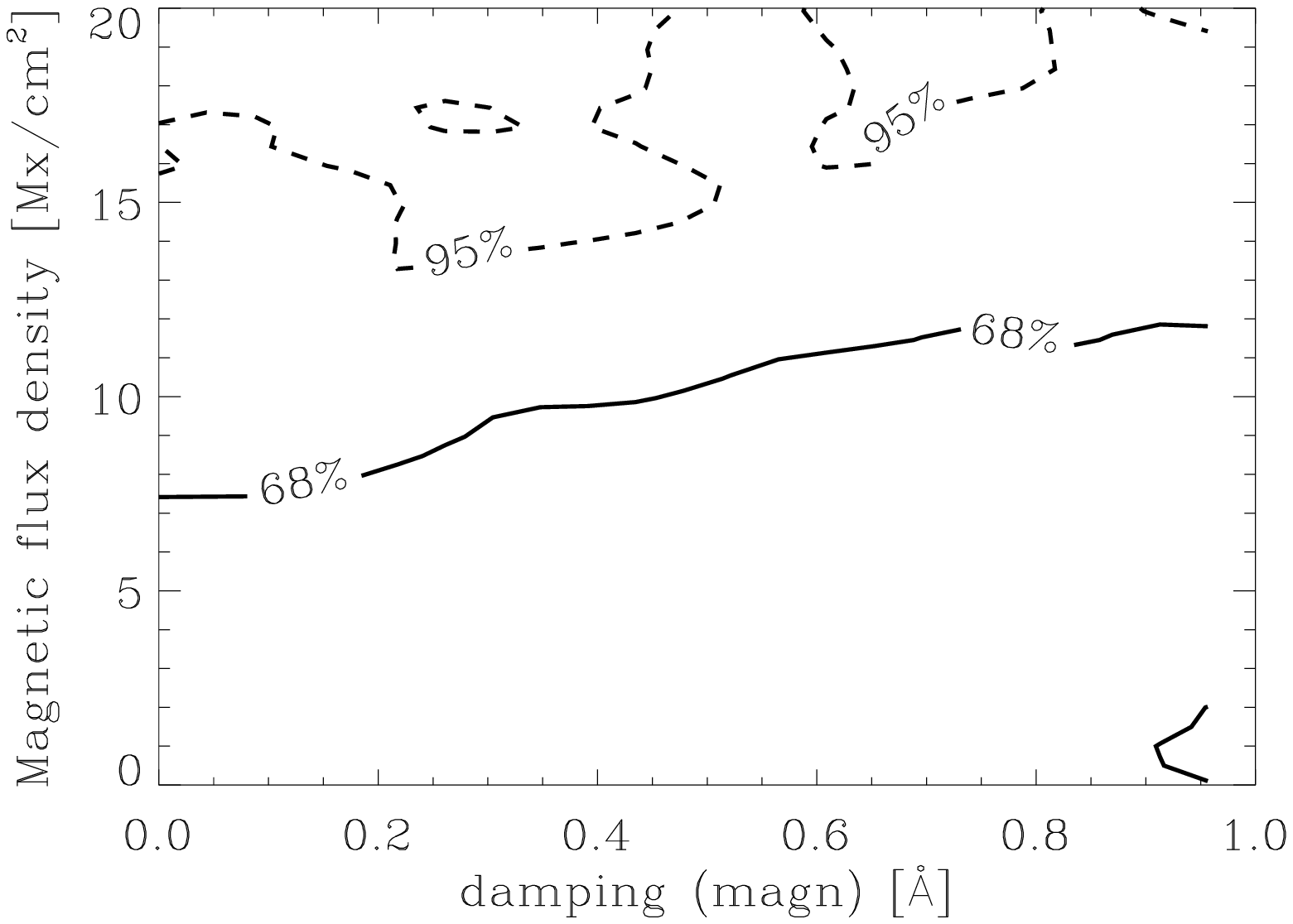}
\includegraphics[width=0.33\hsize]{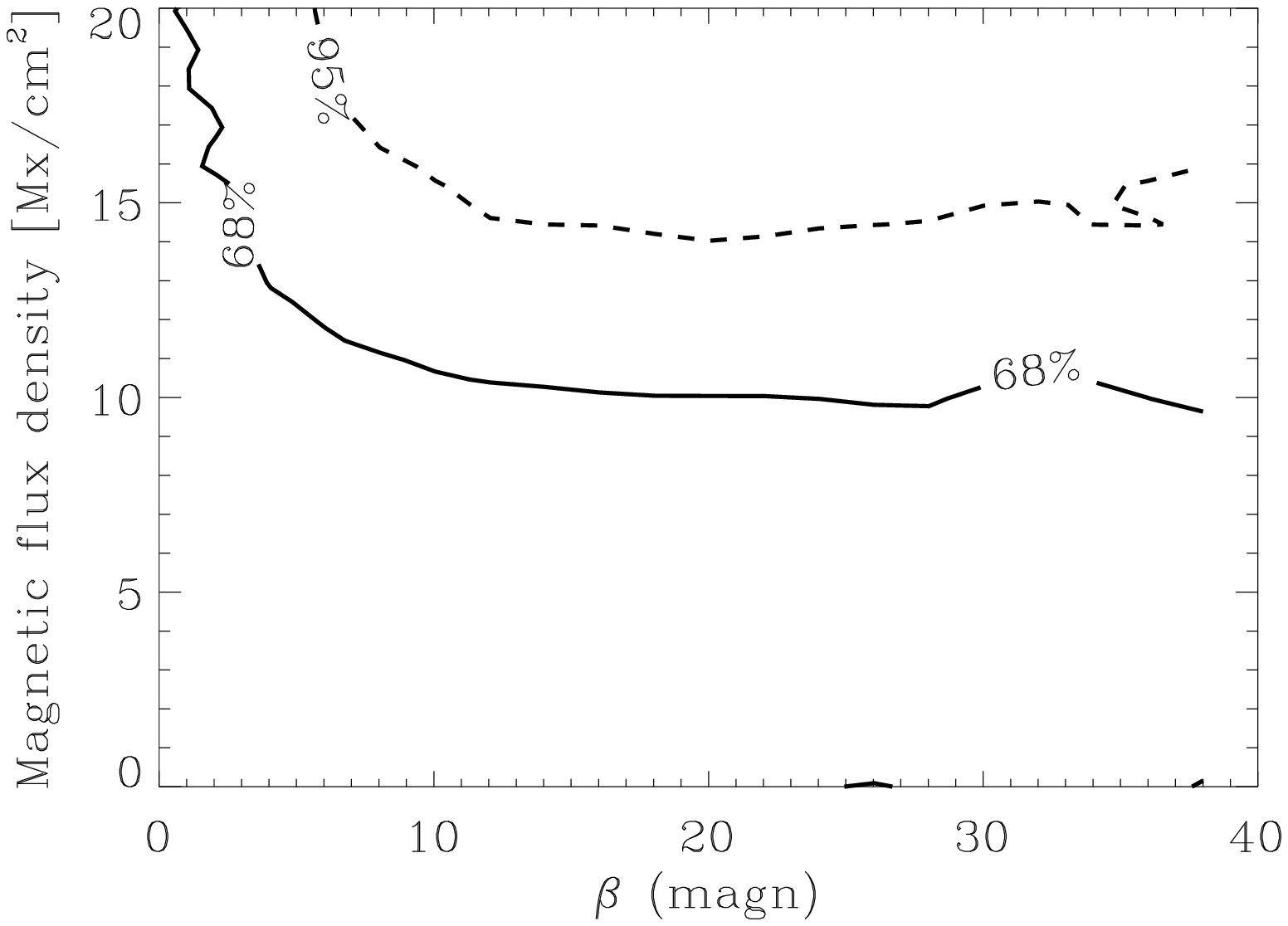}
\caption{Two-dimensional posterior distributions for several combinations of the parameters for the internetwork
synthetic example when the magnetic field vector has an inclination of 45$^\circ$. The contours indicate the regions
where 68\% and 95\% confidence levels are placed. In spite of the presence of more information encoded in the
linear polarization Stokes profiles, the marginalized posterior distributions show shapes very similar to those
found in Fig. \ref{fig:qs1}.}
\label{fig:qs3}
\end{figure*}

\subsubsection{Recovering the magnetic field strength in the internetwork}
In order to investigate in detail this last point, we deal with synthetic profiles that can be representative 
of the quiet Sun to see how well we can recover the magnetic field separately from the rest of the 
parameters. We synthesize the Fe\,{\sc i} lines at $630.1$ and $630.2$ nm using a two component model. 
Both atmospheres have the same values of the ME parameters except for the magnetic 
field strength and the filling factor. For informative purposes, the value of the parameters 
are: $\Delta v_\mathrm{dopp}$=0.05 \AA\ , $v_\mathrm{mac}=0$ km~s$^{-1}$, $\eta_0^{630.1}=5.0$, 
$\eta_0^{630.2}=4.5$, $a=0.45$ \AA\ , and $\beta=8$. The magnetic flux density is fixed to $10$ Mx/cm$^2$,
representative of the typical value in the internetwork. Since the magnetic field strength is set to 1000 G and it has 
been assumed to be vertical, the filling factor of the magnetic component is set to 1\%. A certain amount
of noise, characterized by a normal distribution with a standard deviation of $10^{-4}$ I$_\mathrm{c}$, is 
added to the profiles.

It is important to point out that, under the framework of a Milne-Eddington atmosphere, none of the parameters is
strictly equivalent to the temperature or the microturbulent velocity that are present in the LTE
approximation used by SIR. Accordingly, we select the gradient of the
source function, $\beta$, and the damping coefficient, $a$, in both components together with the magnetic field 
strength and the filling factor as the free parameters in our test. Figure \ref{fig:qs1} summarizes the
results of the Bayesian inversion. All the upper panels show the enormous degree of degeneracy between
the magnetic field strength and the rest of parameters. The upper left panel indicates that magnetic fields 
with all values below 2000 G can reproduce the profile with an accuracy better than two times the noise level. 
Furthermore, magnetic field strengths between 400 and 1800 G fit the profile with a confidence level smaller than 
the noise level (see an example of two possible fits with different field strengths in Appendix B).
An interesting behavior is shown in the central and right upper panels.
The posterior distribution presents almost no variation along these directions (damping parameter and gradient of the
source function) and they are only limited by the ranges that we have assumed for them. Therefore, this means that 
the data has provided no new information for constraining these parameters (flat likelihood) and we are only 
recovering information about the priors. This is the typical example in which, due to the lack of 
information, the result depends critically on the prior information and one should be very cautious with the 
conclusions inferred from the calculations.
Finally, the lower left and central panels show the marginalized posterior distribution for the longitudinal
magnetic flux density and the damping $a$ and the gradient of the source function $\beta$, respectively. 
Again we see that the line profiles carry reduced information about these parameters. Even more striking is the
fact that the longitudinal magnetic flux density is recovered with $\sim 50$\% error at a 68\% confidence level. 
On the contrary, the parameters of the non-magnetic component are recovered with precision, as stated in the 
lower right panel of Fig. \ref{fig:qs1}, with differences with respect to the input values that are 
well below $2.5$\% in both cases. 

\begin{figure}[t]
\centering
\includegraphics[width=0.9\columnwidth]{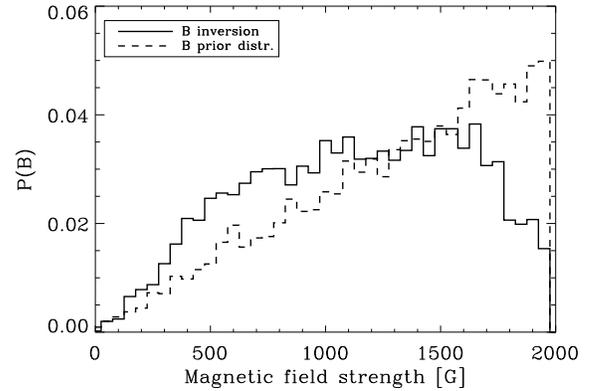}
\caption{Marginalized distribution of the inferred magnetic field strength (solid line) together with
the prior one (dashed line). Both distributions are very similar, making us consider that the information contained in 
the emergent line profiles is very small.}
\label{fig:prior}
\end{figure}

The previous analysis demonstrates that it is difficult to obtain reliable information from Stokes profiles 
representative of internetwork regions. However, what happens in strongly magnetized areas like the network, where 
the magnetic fluxes are 10-20 times higher than in the internetwork? To investigate this issue, we use the very
same ME parameters but we assume an enhanced magnetic flux density of $200$ Mx/cm$^2$, where we have increased the 
filling factor of the magnetic component to 20\%. The marginalized posterior distributions are shown in
Fig. \ref{fig:qs2}. In this case, both the magnetic field strength and the longitudinal magnetic flux density 
are well recovered, together with the damping, $a$. However, it is interesting to point out that, as a consequence
of the reduced filling factor of the non-magnetic component with respect to the internetwork case, the gradient
of the source function of the non-magnetic component presents a more extended posterior distribution. This 
behavior is easy to understand because there is less information about the non-magnetic component
encoded in the Stokes profiles produced by the large filling factor of the magnetic component.

\subsubsection{Inclined fields}
Coming back to the internetwork, it is of interest to investigate the shape of the marginalized posterior distributions
when the magnetic field vector is inclined with respect to the line of sight. In this case, the information provided
by the linear polarization profiles can lead to better constraints. We use the same synthetic profile with a 
magnetic flux density of $10$ Mx/cm$^2$ and we assume inclinations of 20$^\mathrm{\circ}$, $45^\mathrm{\circ}$ and 
70$^\mathrm{\circ}$. In the first case, the Stokes $Q$ signal is below the noise level and it is not surprising that the 
results are comparable to the ones in which the magnetic field vector was assumed to be vertical. The 
case of an inclination of $70^\mathrm{\circ}$ shows the same behavior since, in this case, the 
Stokes $V$ signal is below the noise 
level. Figure \ref{fig:qs3} shows the results of the inversion for the intermediate case of 
$\theta=45^\mathrm{\circ}$. In this case, strong signatures of degeneracy are detected. The upper left panel
shows that the magnetic field strength is concentrated in high values. However, contrary to what one could
think, this does not mean that the value of the field strength is better recovered. First, we can see the 
large degeneracy with the other parameters. Second, Fig. \ref{fig:prior} shows that the marginalized
posterior distribution for the magnetic field strength strongly resembles that of the prior. It seems that 
even the inclination angle cannot be constrained with the available information. One of the reasons is that, 
although the Stokes $Q$ and $U$ signal might become larger than the noise, the Stokes $V$ signal decreases 
and gets closer to the noise. Therefore, some information available in Stokes $V$ is hidden by the presence of noise.

\begin{figure*}
\includegraphics[width=0.33\hsize]{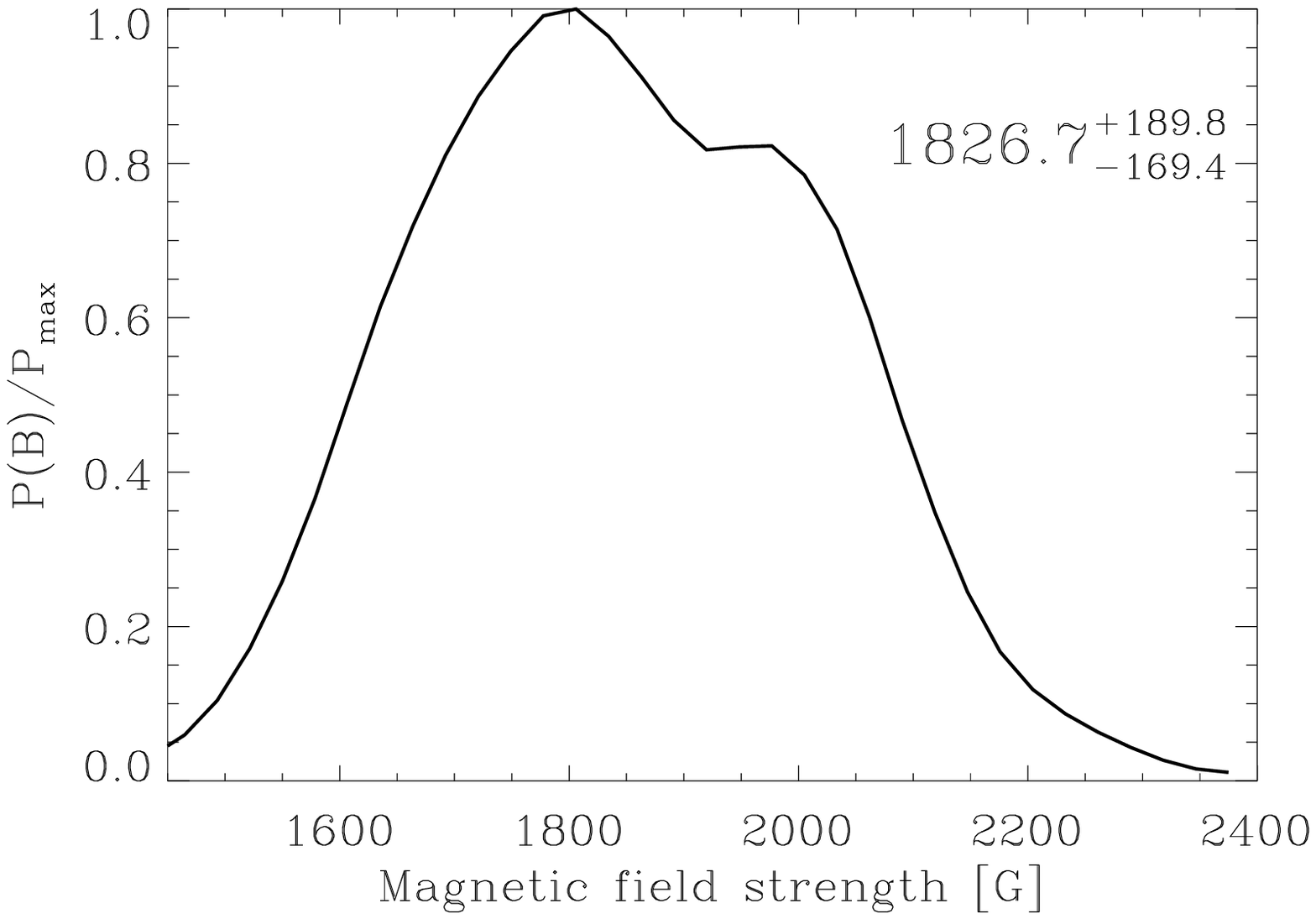}%
\includegraphics[width=0.33\hsize]{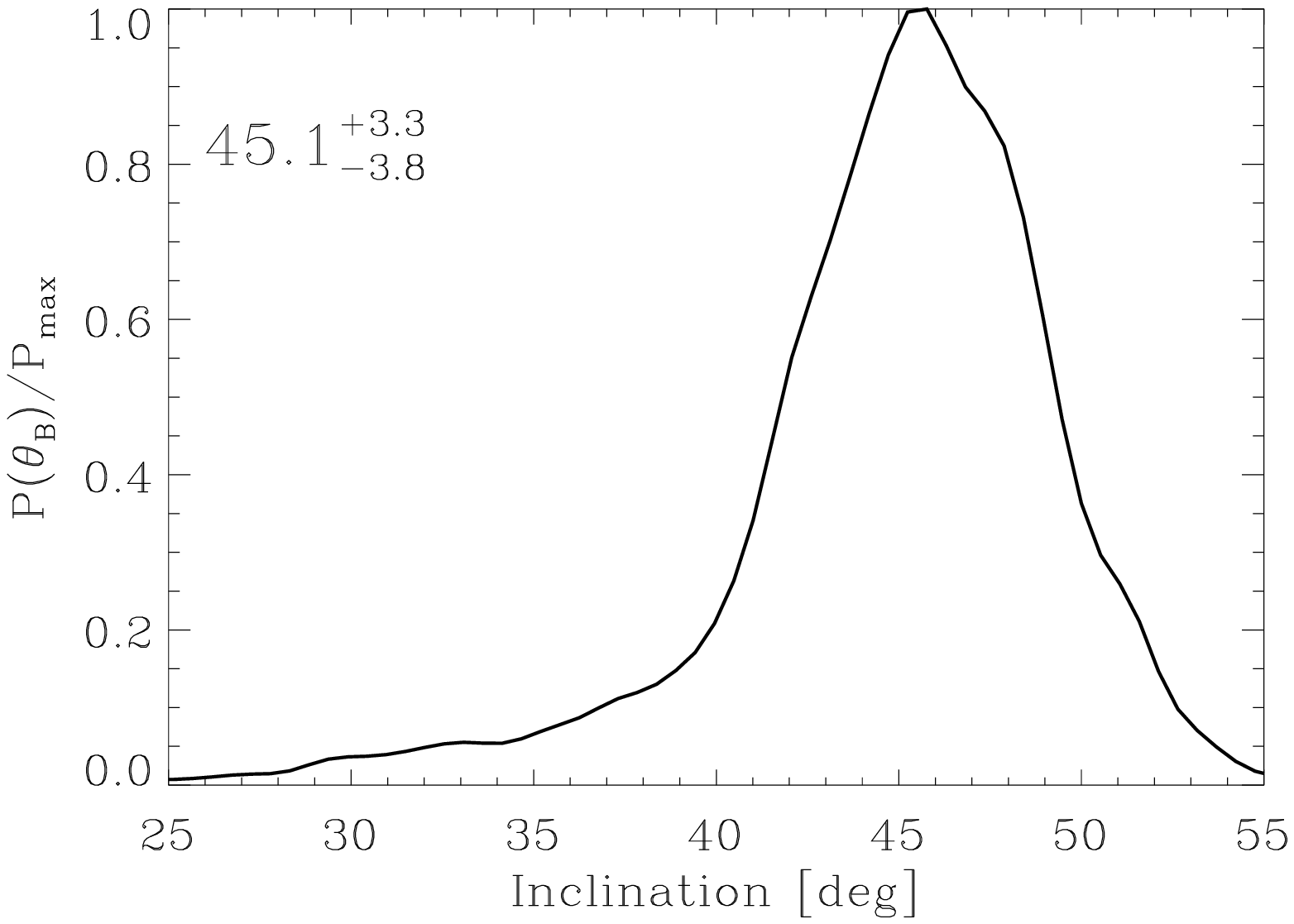}%
\includegraphics[width=0.33\hsize]{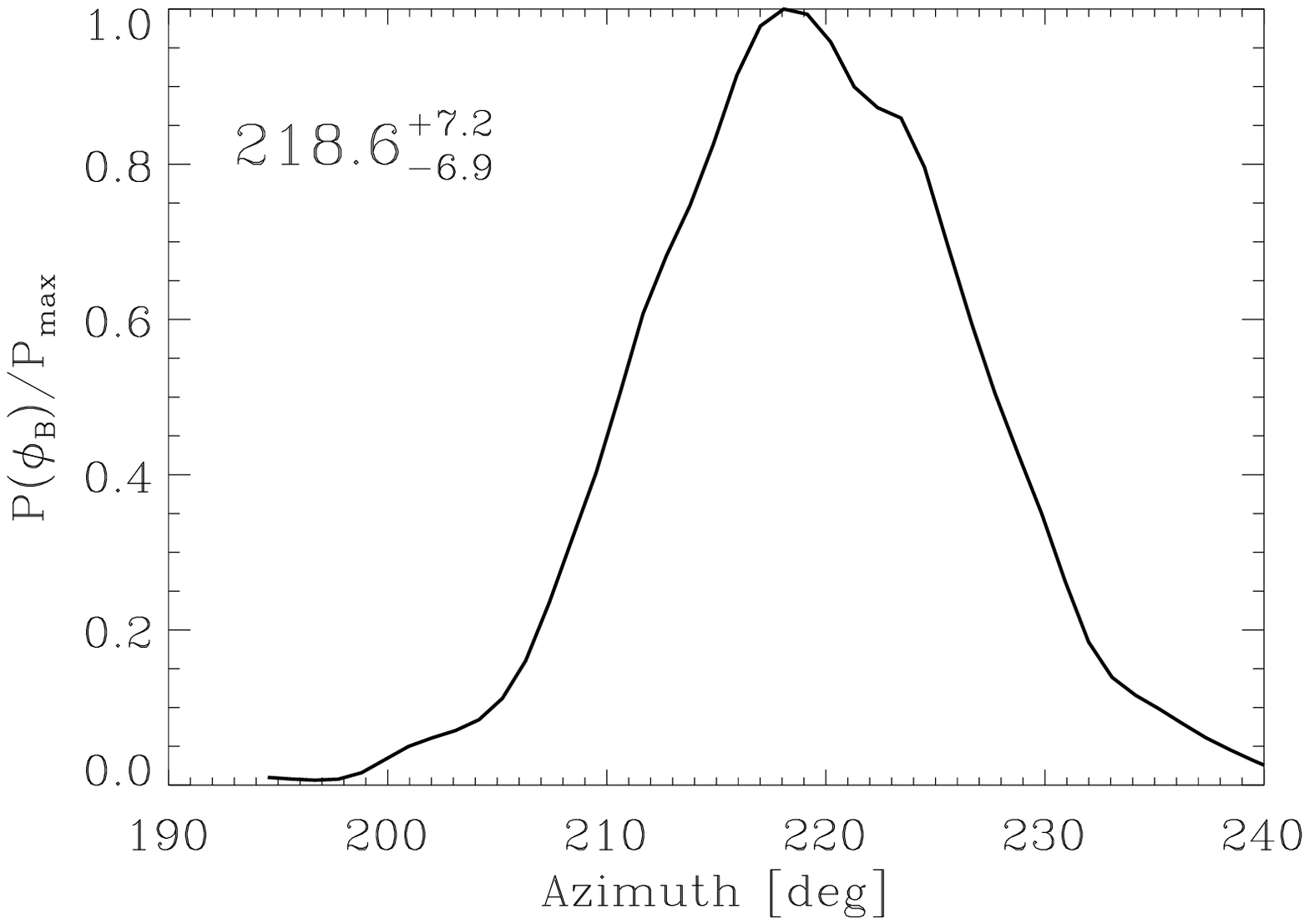}\\
\includegraphics[width=0.33\hsize]{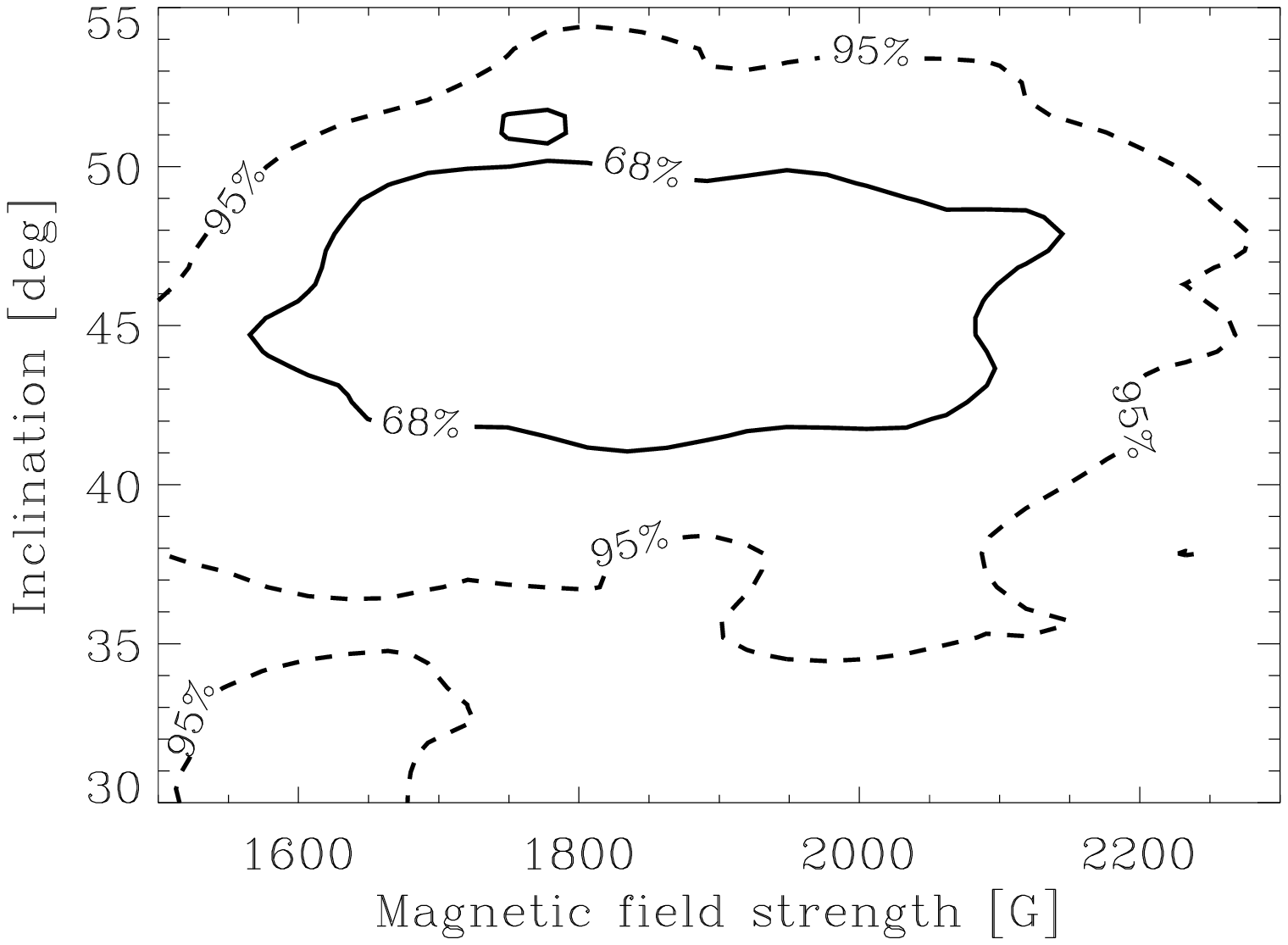}%
\includegraphics[width=0.33\hsize]{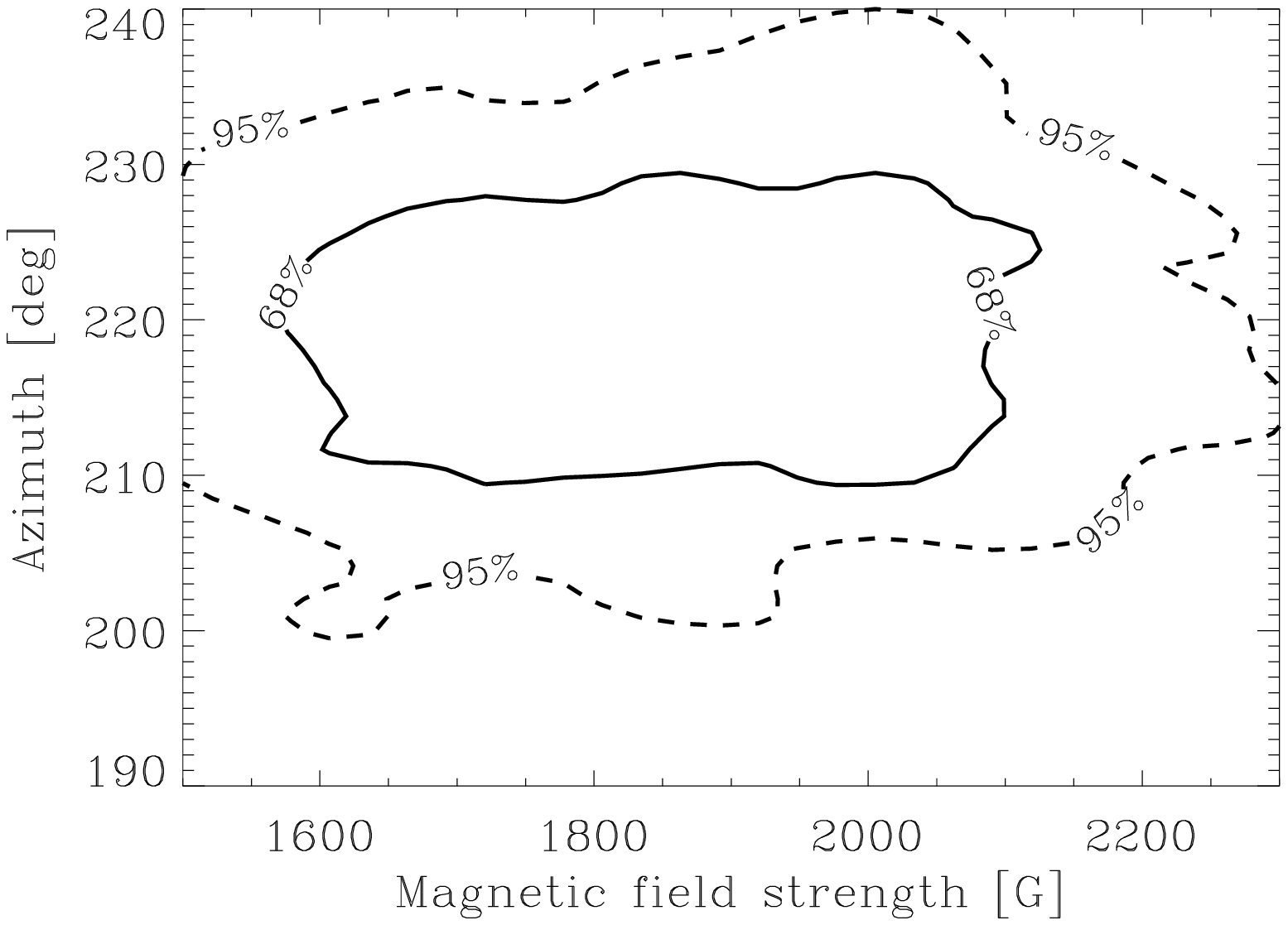}%
\includegraphics[width=0.33\hsize]{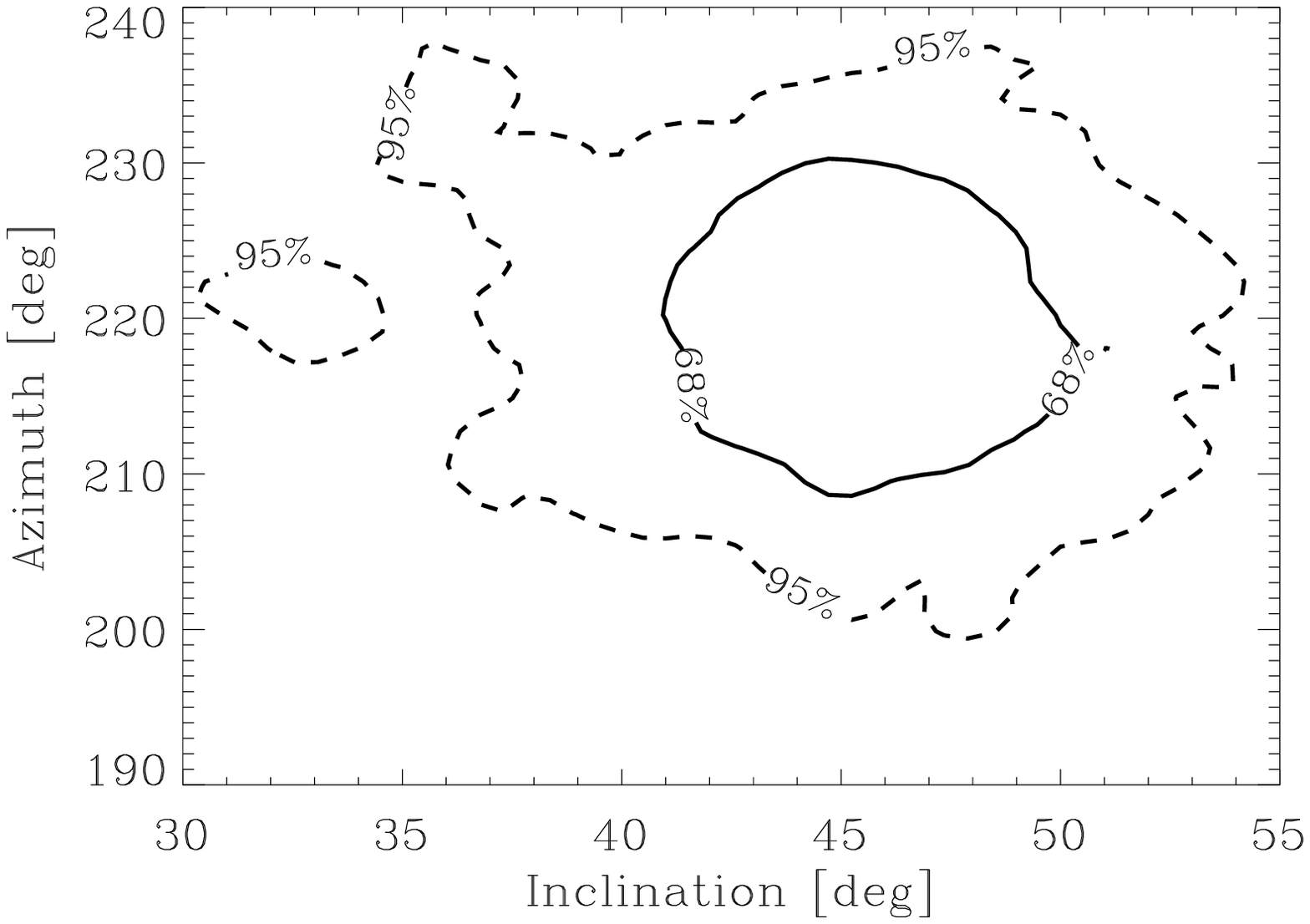}
\caption{One--dimensional and two--dimensional marginalized posterior distributions for the parameters that
define the magnetic field vector for the inversion of the \ion{Fe}{i} line at 630.2 nm. The distributions 
clearly indicate the presence of a peak, belonging to the
value of the parameters that produce the best fit. We have indicated in the one--dimensional posteriors the 
most probable value of each parameter (large arrow), together with the 68\% confidence interval (small arrows).
Note the presence of asymmetric confidence intervals. The two--dimensional posteriors have been represented
with contours indicating the confidence levels at 68\% (solid line) and 95\% (dashed line).}
\label{fig:sunspot6302}
\end{figure*}

\subsection{Realistic examples}
\begin{figure*}
\centering
\includegraphics[width=0.4\textwidth,clip]{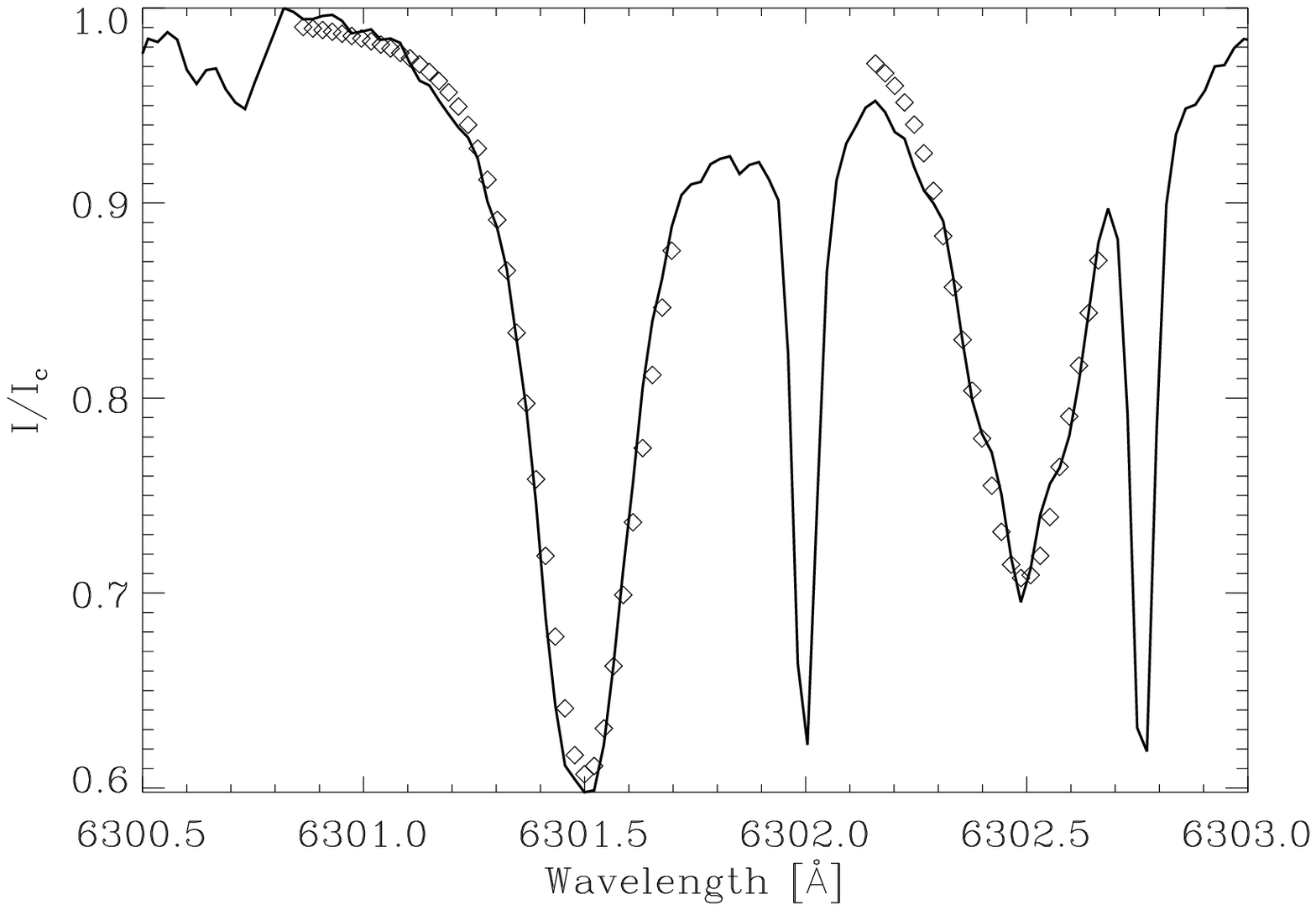}%
\includegraphics[width=0.4\textwidth,clip]{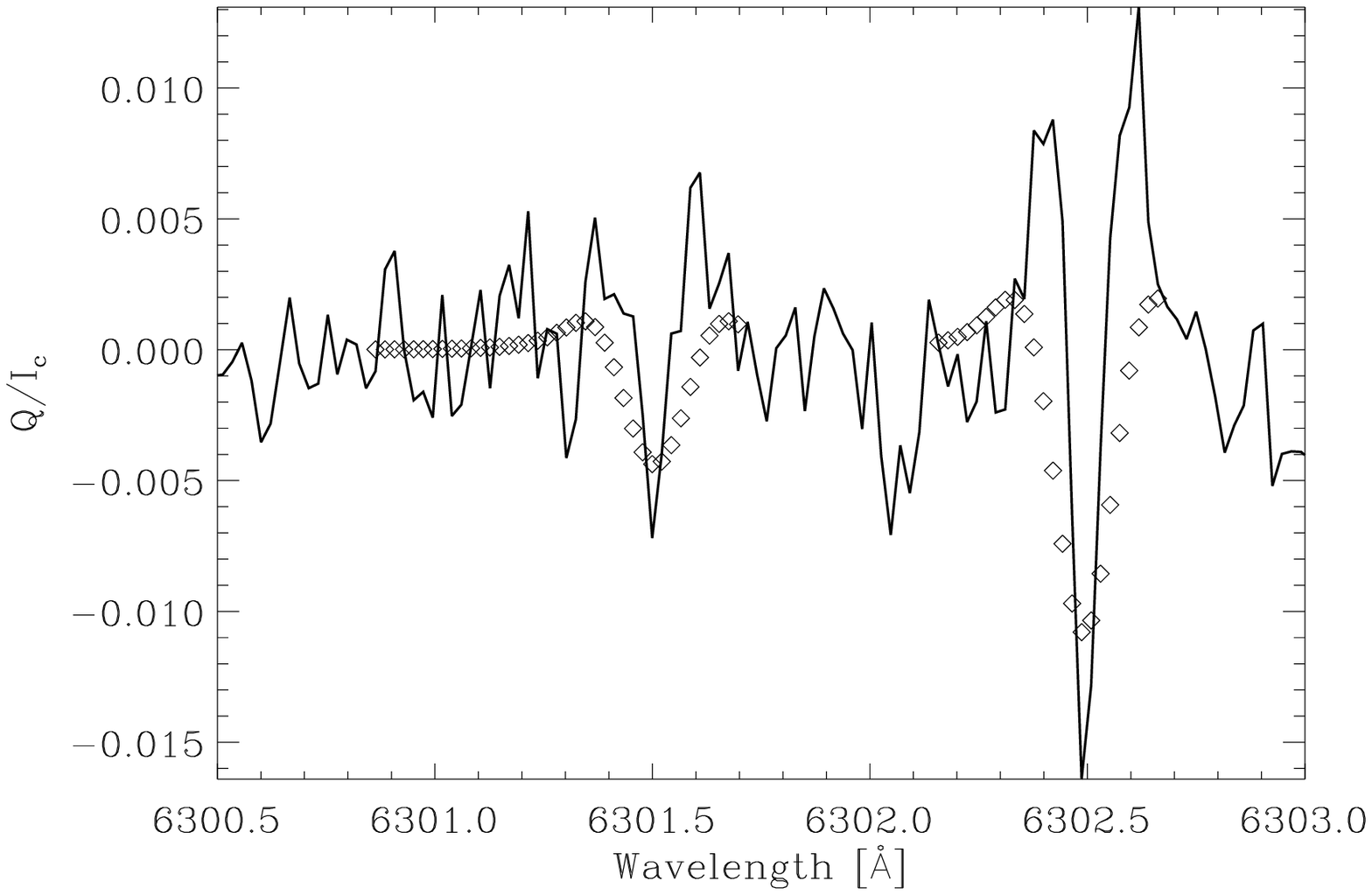}\\
\includegraphics[width=0.4\textwidth,clip]{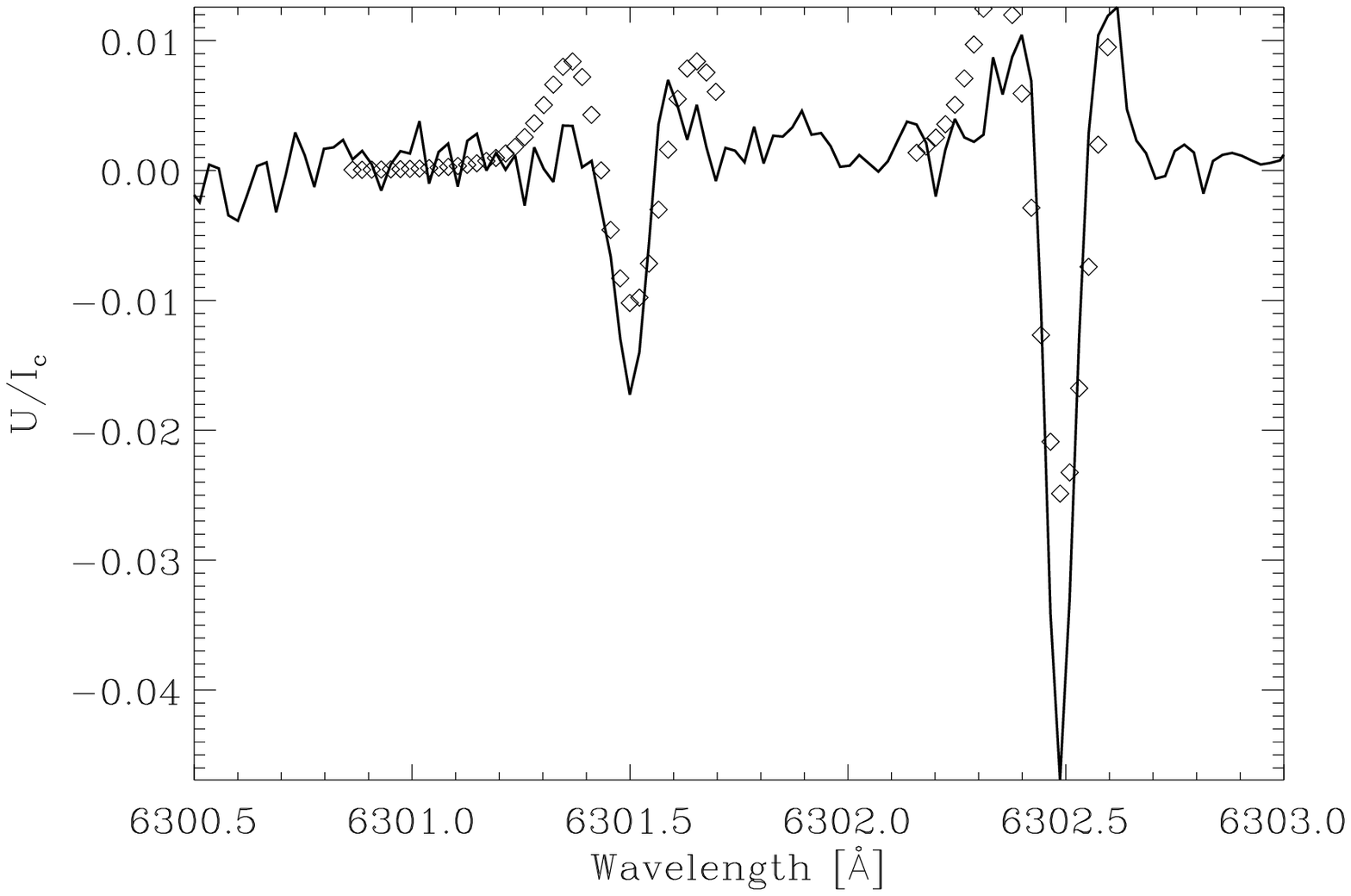}%
\includegraphics[width=0.4\textwidth,clip]{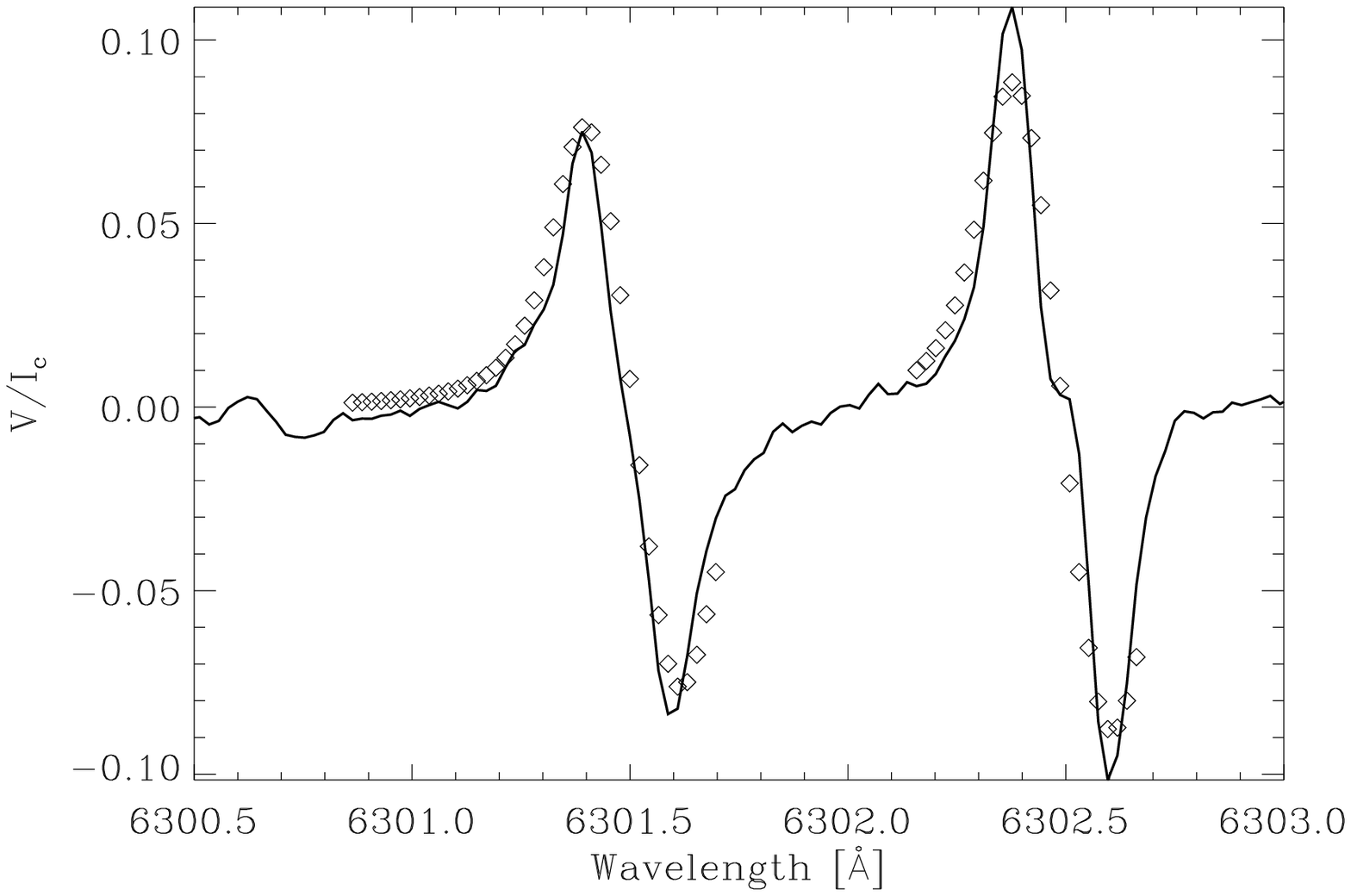}
\caption{Stokes profiles observed in the umbra of a sunspot (solid lines). The well known 630.2 nm 
\ion{Fe}{i} line presents a larger magnetic sensitivity that translates into an enhanced Zeeman splitting. The
noise in the observations is of the order of 1.5$\times$10$^{-3}$ in units of the continuum intensity. The 
symbols show the fit obtained when taking the most probable value of the parameters inferred from the 
Markov Chain.}
\label{fig:sunspot_stokes}
\end{figure*}

In order to demonstrate the capabilities of the MCMC code, we show an application to realistic Stokes profiles.
They correspond to a position on an umbra of a sunspot observed during August 17, 2004 \citep{alberto_arturo07}. 
The observation was carried out with the TH\'EMIS telescope at the Observatorio del Teide (Spain). The telescope
was operated in the MTR mode, so that the polarization analysis was performed for each wavelength at each
pixel. Although the observation consisted on a scan over a sunspot, for the purpose of demonstrating the
capabilities of the MCMC code, we only focus here on the information obtained in one pixel of the whole scan.
The observed spectral region contains the previously mentioned 630 nm pair of \ion{Fe}{i} lines. 
Figure \ref{fig:sunspot_stokes} presents,
in solid line, the observed Stokes profiles. The noise level estimated from the continuum where no polarization
signal is detected is of the order of 1.6$\times$10$^{-3}$ in units of the continuum intensity. This spectral
region consists on two \ion{Fe}{i} lines at 630.1 and 630.2 nm, together with two telluric contributions. 
The wavelength calibration has been carried out with the aid of the two
telluric lines. The 630.2 nm line presents a higher magnetic sensitivity and this translates into an
enhanced Zeeman splitting that can be also detected in the Stokes I profile.

\begin{figure*}
\includegraphics[width=0.33\hsize]{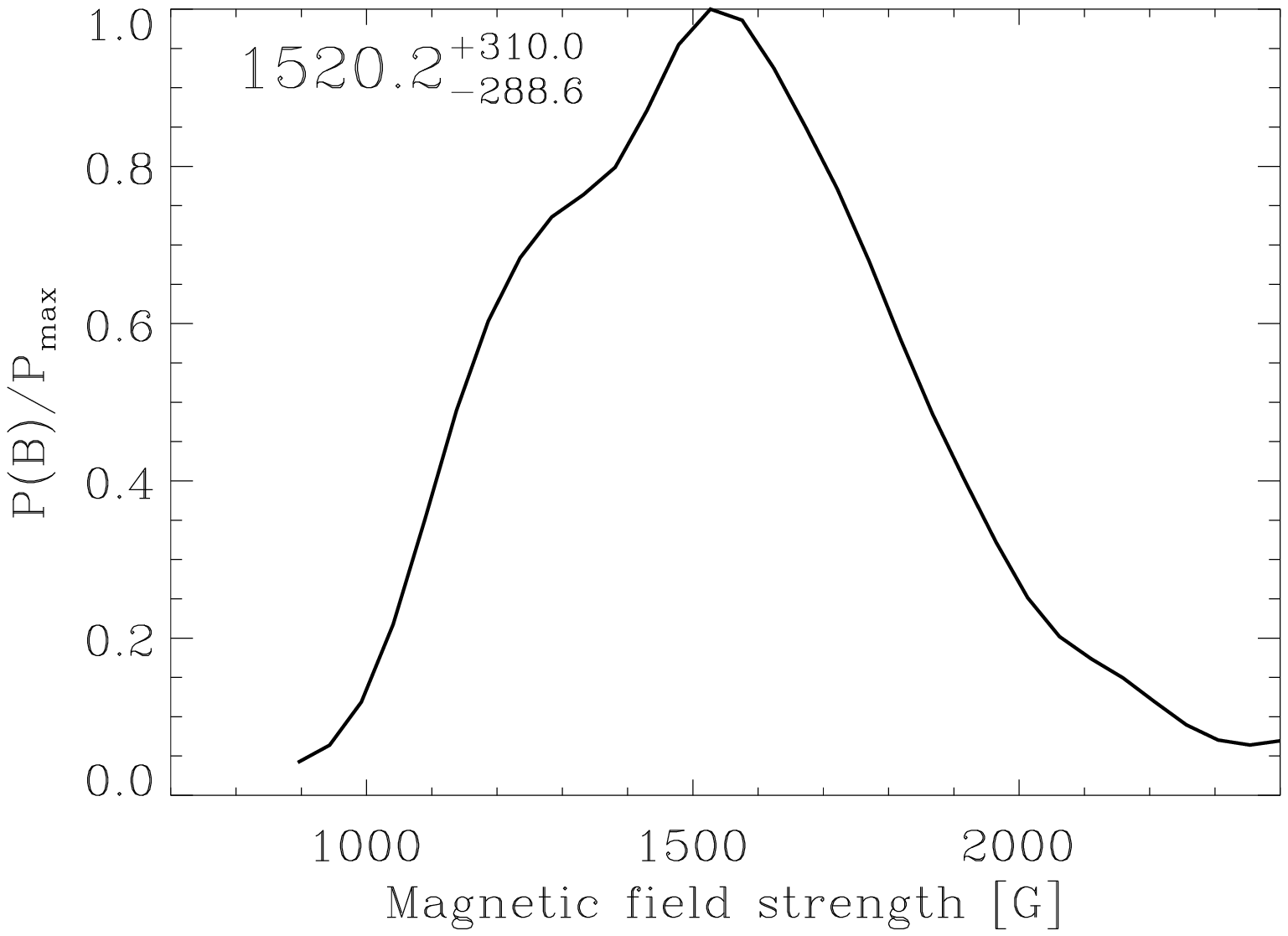}%
\includegraphics[width=0.33\hsize]{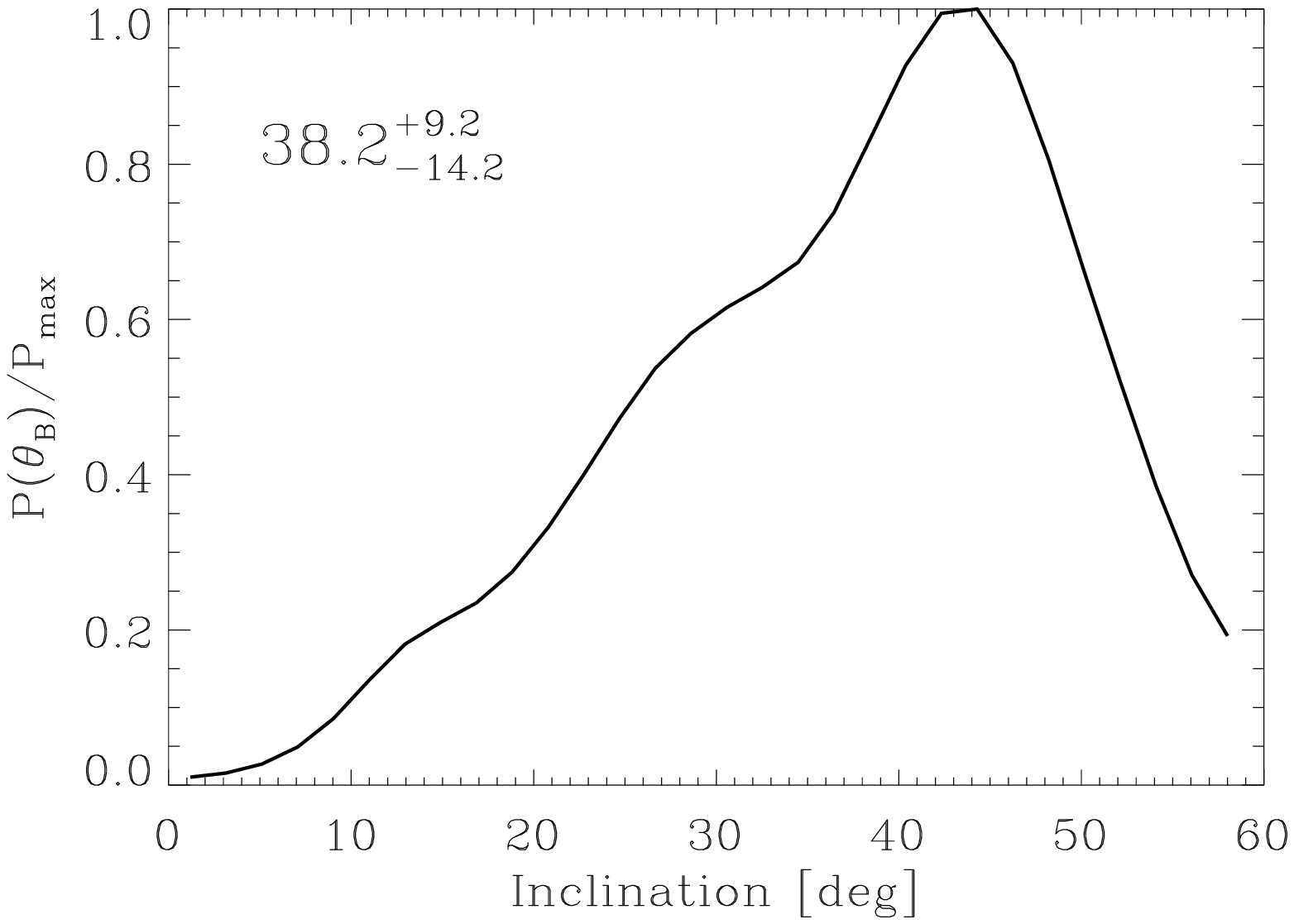}%
\includegraphics[width=0.33\hsize]{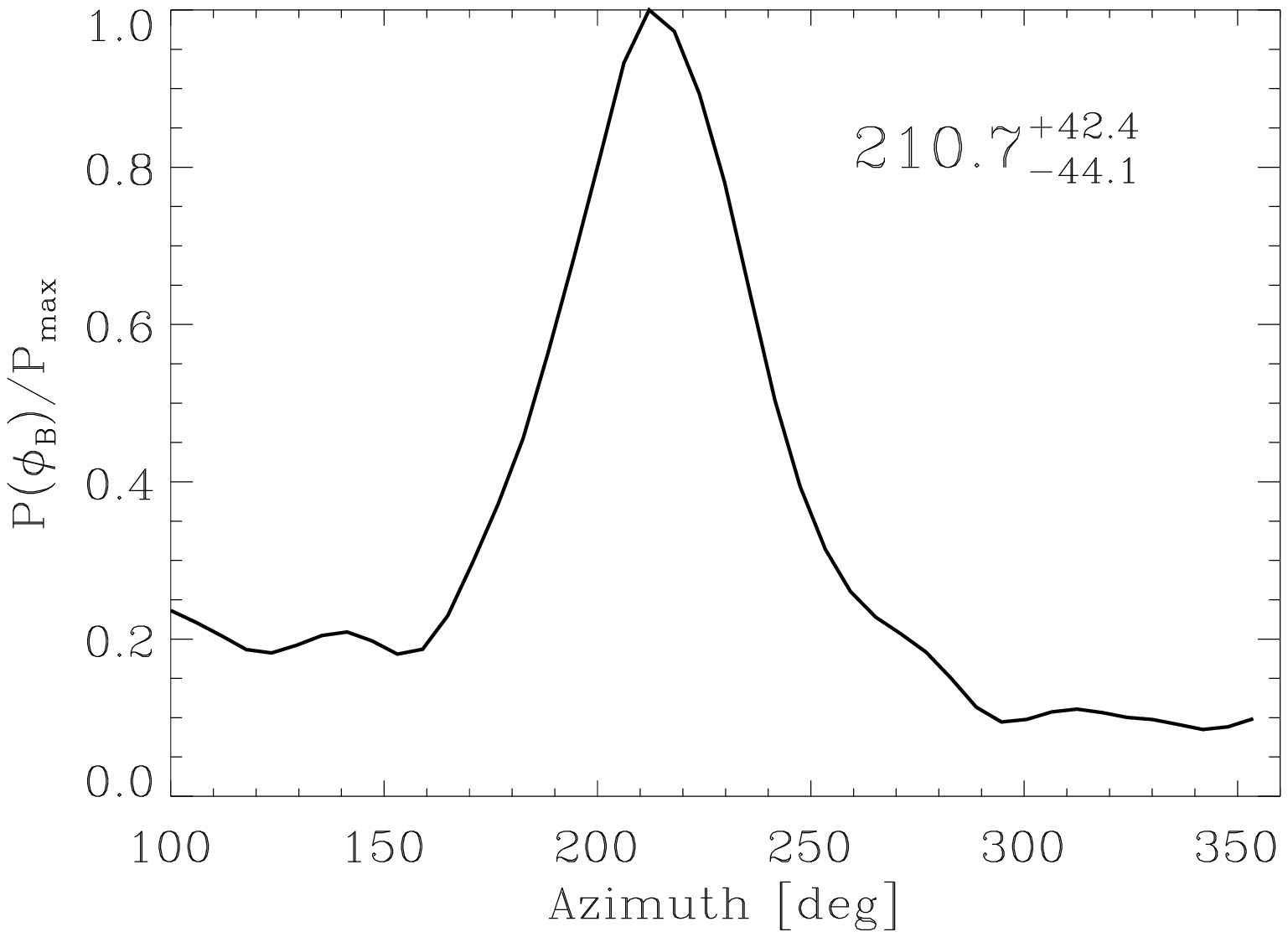}\\
\includegraphics[width=0.33\hsize]{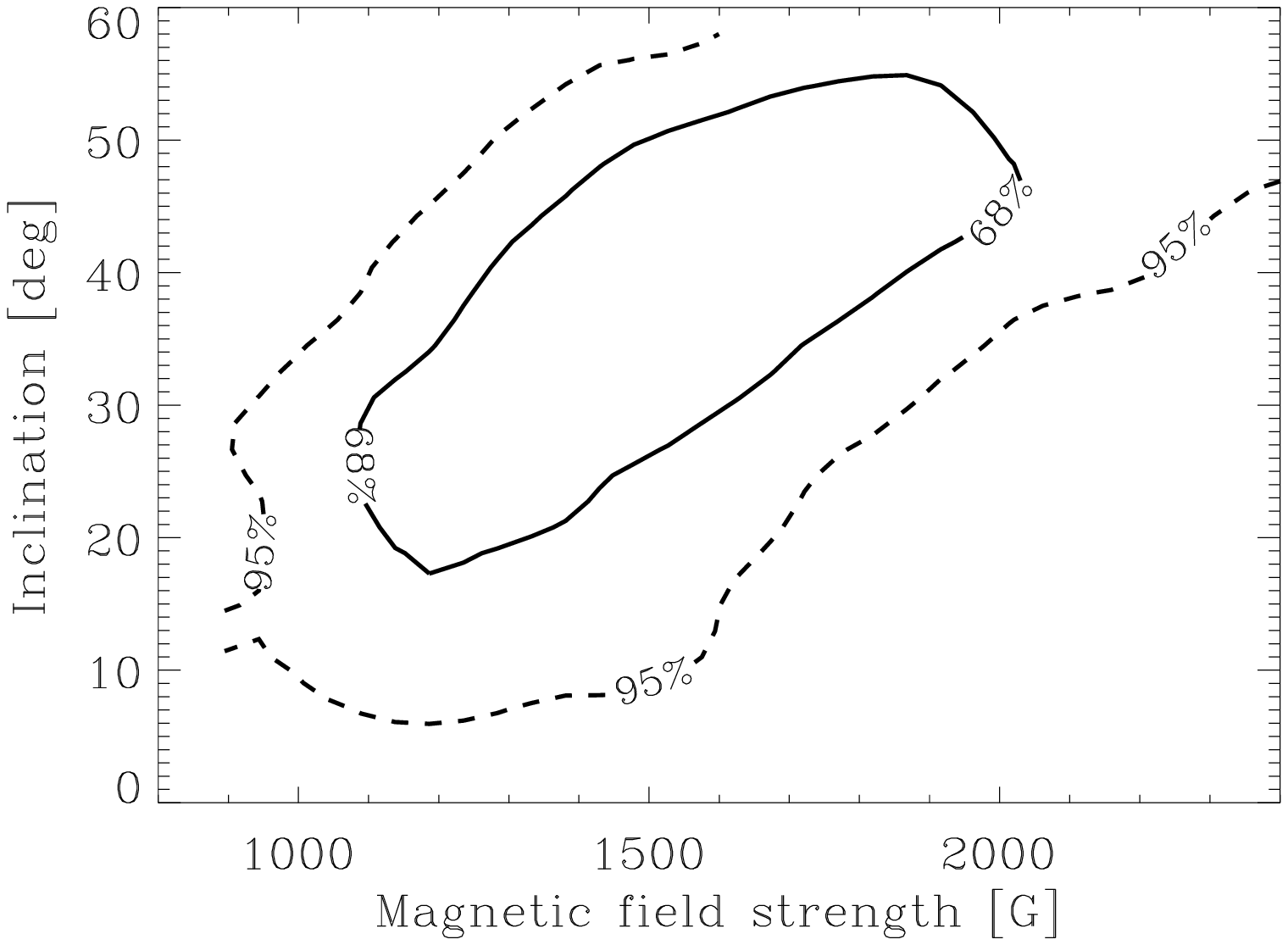}%
\includegraphics[width=0.33\hsize]{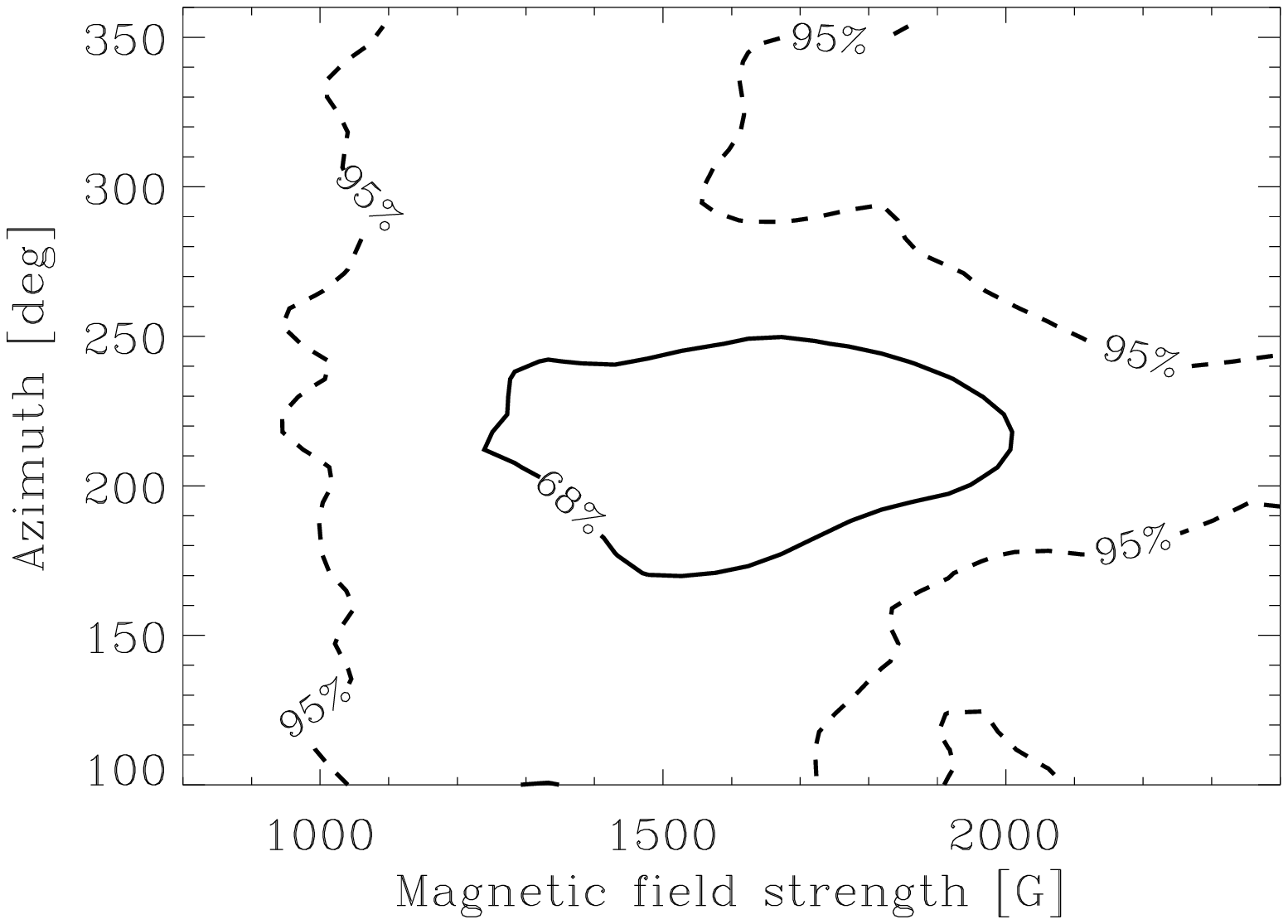}%
\includegraphics[width=0.33\hsize]{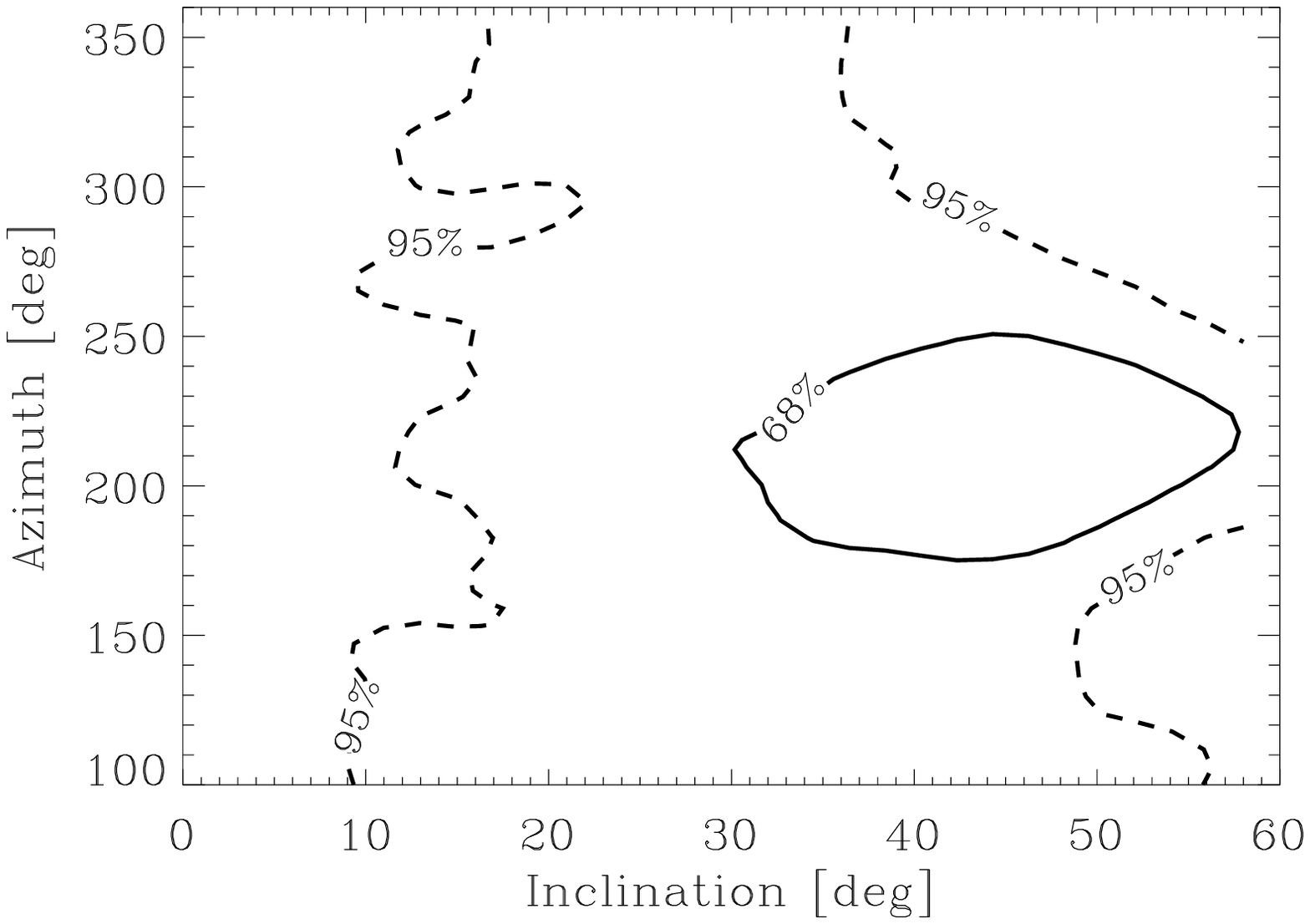}
\caption{Same as Figure \ref{fig:sunspot6302} but for the \ion{Fe}{i} line at 630.1 nm. The two--dimensional
representations of the posterior distribution clearly indicate the presence of a degeneracy between the
inclination and the magnetic field strength. This translates into very broad one--dimensional marginalized
distributions.}
\label{fig:sunspot6301}
\end{figure*}

The MCMC code has been applied to both spectral lines separately. We leave all the Milne-Eddington
parameters free but we only focus on the results concerning the magnetic field vector. Stray-light
contamination from the surrounding quiet Sun is also taken into account. The results indicate a filling
factor of the umbral component in the range 91-94\%, with a confidence interval of the order of $\pm$3\%.
Figure \ref{fig:sunspot6302} shows the results obtained from the inversion of the 630.2 nm 
\ion{Fe}{i} line. We show posterior probability
distributions marginalized over all parameters except for one and except
for two. The results shown in Fig. \ref{fig:sunspot6302} indicate that the information
encoded in the observed data is enough to constrain the characteristics of the magnetic field vector. Except for
the case of the azimuth of the field, the marginalized one--dimensional probability distribution functions 
present an asymmetric non-gaussian 
shape, with extended wings. The parameters are nicely constrained by the observations and the inferred values are given
in each plot, together with the 68\% confidence interval. Concerning the two--dimensional distributions, 
we show them as contour plots, where the 68 and 95\% confidence levels are indicated.

Concerning the \ion{Fe}{i} line at 630.1 nm, the results are definitively worse. The posterior distributions
are much broader than for the 630.2 nm line and they present strong degeneracies. Several 
points deserve a more profound discussion. First, 
the elongated shape of the $p(B,\theta_B)$ posterior indicates a certain degree of degeneracy between
both parameters. The reason for this behavior is that the 630.1 nm line is still in the transition from
the Zeeman weak-field regime to a saturation regime. As a consequence, the $B$-$\theta_B$ degeneracy that
we have discussed in \S\ref{sec:academic_case} introduces problems in the unique determination of the field
strength and inclination. 
Second, it is important to point out the fact that these results have been obtained assuming
$\sigma=1.5 \times 10^{-3}$ in units of the continuum intensity. This is a relatively large value which
poses a relaxed tolerance in the quality of the fit, thus resulting in increased tolerance in the inferred
parameters. Clearly, due to the different magnetic sensitivity, noise affects differently to both spectral lines.
Since the Zeeman splitting in the 630.2 nm line is clearly visible, the information about the 
magnetic field strength is readily available from the peak separation in the Stokes $V$ profile. This separation
is much less affected by noise. Once the field strength is fixed, the inclination and 
azimuth of the field are easily obtained. Contrarily, since the 630.1 nm line is partially in the weak-field
regime, the magnetic field strength has to be obtained from the amplitude of the Stokes $V$ profile, together
with the rest of Stokes parameters. The estimated value of the field strength crucially depends on the value of the
tolerance $\sigma$. As a proof of this, we have verified that the shape of the $p(B,\theta_B)$ surface shown in Fig. \ref{fig:sunspot6301}
approximately follows $\cos \theta_B \propto 1/B$ and that the width is related to the tolerance $\sigma$.

\begin{figure*}[!t]
\centering
\includegraphics[width=0.33\hsize,clip]{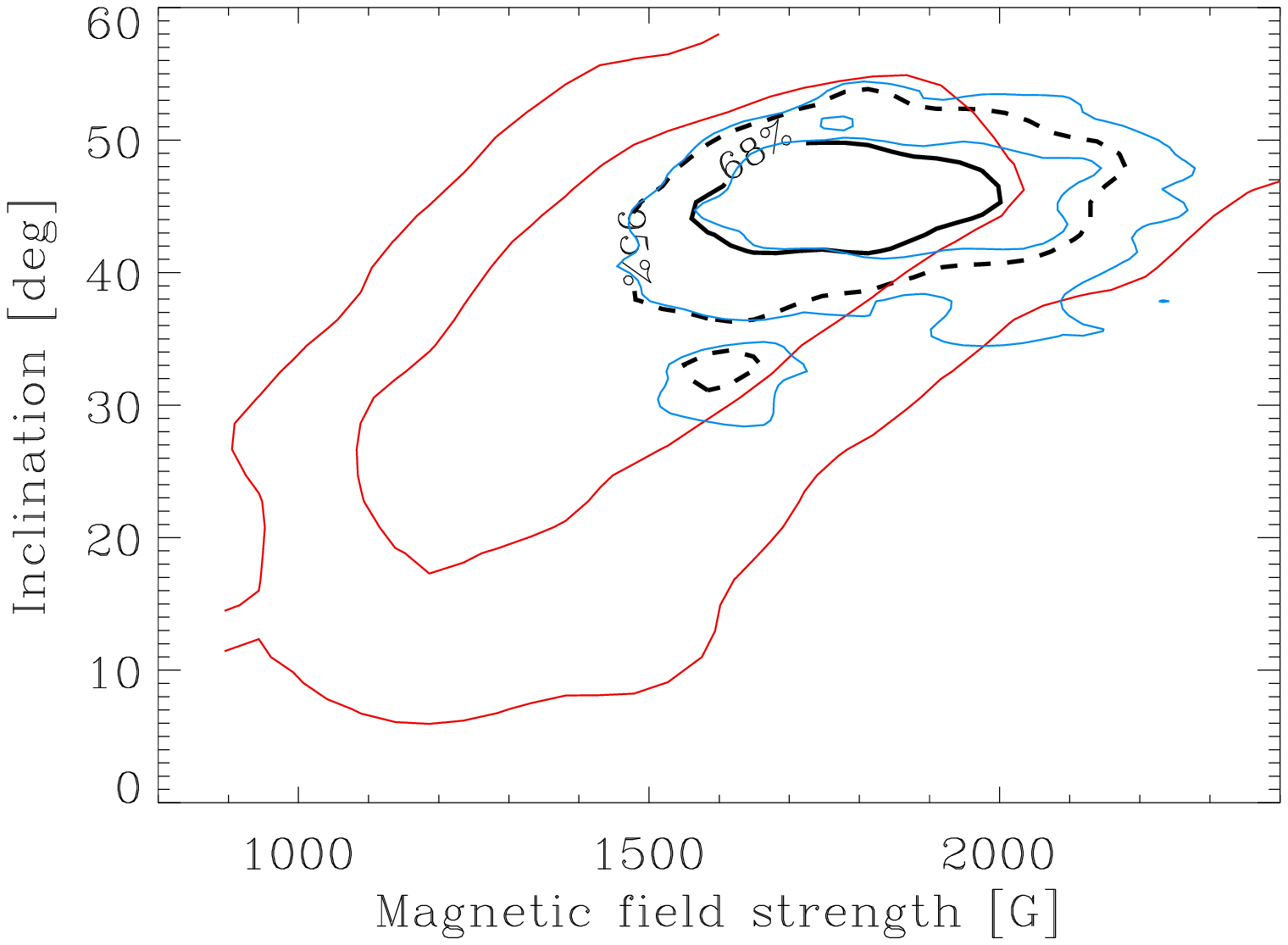}%
\includegraphics[width=0.33\hsize,clip]{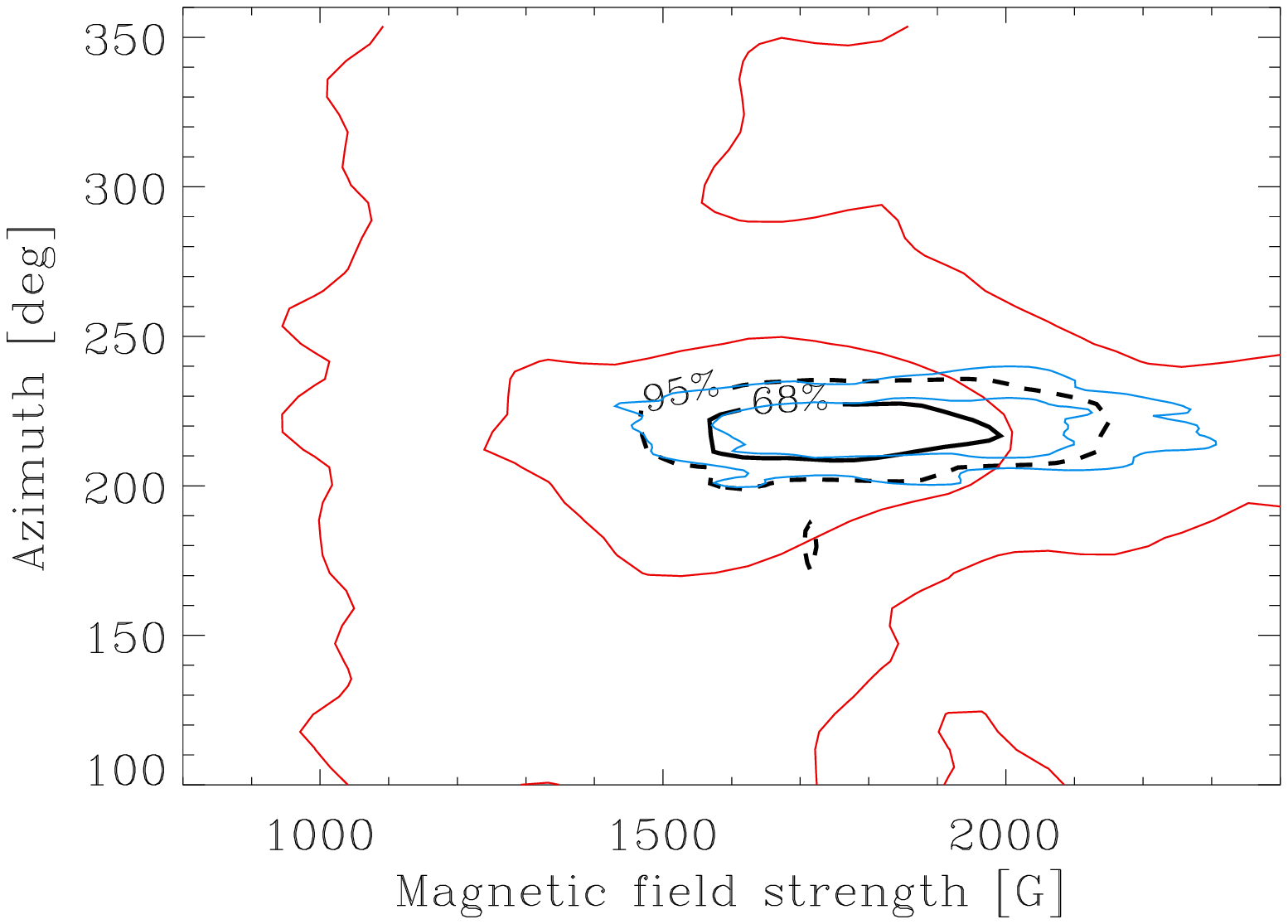}%
\includegraphics[width=0.33\hsize,clip]{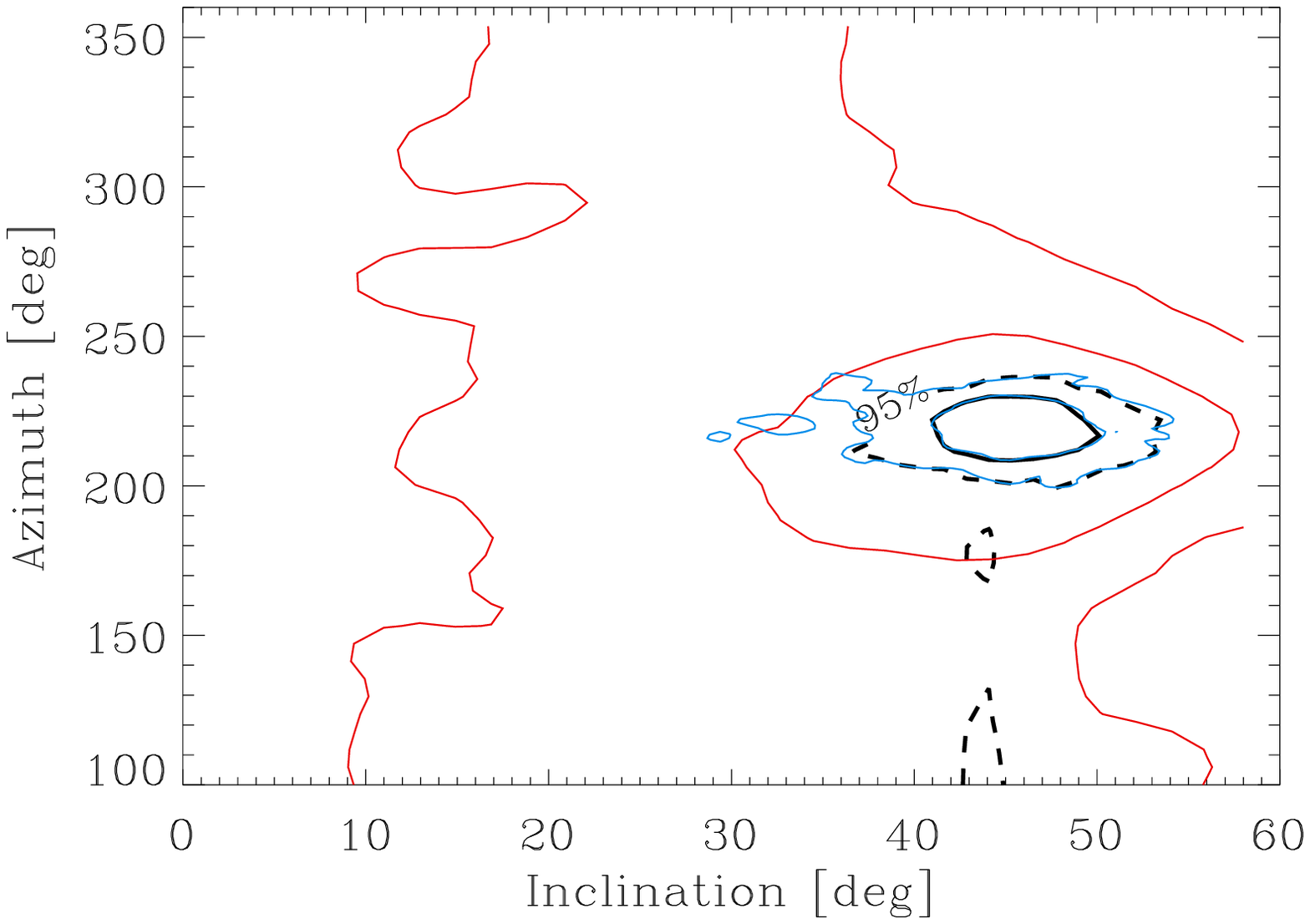}
\caption{Combined posterior probability distribution for the 630.1 and 630.2 nm lines. Note that 
there is a region of compatibility between both lines, that gives a magnetic field 
of $\sim$1700 G, an inclination of $\sim 45^\circ$ and an azimuth of $\sim 220^\circ$.\label{fig:composition}}
\end{figure*}

The results of both inversions should also be regarded in conjunction, as shown in Fig. \ref{fig:composition}. 
If the line formation region of
both lines would have been exactly the same, one would expect to find equivalent results from both lines. Since this
is not the case \citep[e.g.,][]{shchukina_trujillo01}, some differences might exist. In spite of this, 
the posterior distributions clearly overlap in a region of the space of parameters that describe the
magnetic field vector. The results clearly demonstrate that the combination of the two lines
produces a slight improvement on the restriction of the parameters. However, the result is very
similar to what we find using only the line at 630.2 nm. The field azimuth
is compatible with a value of $217.4^{+7.0}_{-7.8}$ degrees. It is important to point out that, for the purpose of
showing a less crowded plot, we have restricted the range of variation of the azimuth arbitrarily 
to $[\pi,2\pi]$, thus avoiding the presence of ambiguities. However, we have verified that the 
code is able to correctly capture the intrinsic azimuth ambiguity when the range of variation is set to
$[0,2\pi]$. The field inclination given by both lines is consistent with $45.4^{+3.0}_{-3.2}$ degrees, while the
magnetic field strength is consistent with $1768.6^{+165.9}_{-133.8}$ G. 
Figure \ref{fig:composition}
shows what we consider one of the most appealing properties of the Bayesian method for the inversion of Stokes profiles
that we are presenting in this paper. It is possible to assess the amount of information given by
one spectral line individually and combine many lines in order to investigate whether the
added information helps in better constraining the model parameters.

Our results tend to indicate that the information obtained from the 630.1 nm line alone is very
reduced and that it can hardly be used to restrict the magnetic field vector for the noise level that
we have in the observations. As another exercise, we have
inverted both lines simultaneously following the very same scheme as that presented above. We do not show
a graphical representation of the results because they are very similar to those inferred from the
630.2 nm line which can be found in Fig. \ref{fig:sunspot6302}, as also suggested
by Fig. \ref{fig:composition}. At the light of the results presented here, it is desirable
to accumulate information from many spectral lines, with the hope that the combined effect helps
us to better constrain the physical parameters \citep{semel81,socas_navarro_multiline04,asensio_dimension07}.

\section{Concluding remarks}
The framework that we have presented here is of very general nature and allows its application 
to any existing Stokes inversion code. Once a model that can be used to calculate the emergent Stokes 
profiles is available, the MCMC method can be used to 
efficiently explore the posterior probability distribution function.
Presently, we are witnessing an enormous input of Stokes profiles observations from existing ground-based 
instrumentation like THEMIS \citep{lopez_ariste00}, TIP \citep{martinez_pillet99} and 
POLIS \citep{beck_polis05} and with the space-based instrumentation like the 
recent mission HINODE. The pressure will be even larger once the new generation of big solar telescopes
like GREGOR and ATST arrives. Therefore, a huge effort is been put into developing fast inversion codes that can
cope with such an amount of observations. Inversion codes based on PCA \citep{rees_PCA00} and artificial neural networks 
\citep{socas_navarro_aann05} are good candidates for such a demanding work. 

Our approach here has a completely different point of view. We understand that Bayesian inversions cannot
compete in speed with these fast algorithms (they cannot even compete with standard inversion codes based on 
Levenberg-Marquardt optimization). However, the Bayesian approach is the only one that can be used to investigate
in detail the accuracy of inversions, the sensitivity of the parameters to the noise and give confidence
intervals to all the inferred parameters. Furthermore, it can be used to rule out a given model for its lack of
ability to fit a given observed Stokes profile. It is also important to point out that our approach can make use of 
the well-developed machinery behind the Bayesian formalism \citep[e.g.,][]{marshall06,liddle07}. For instance, 
model selection techniques based on 
the calculation of the evidence can be introduced. Similarly to the results presented by \cite{asensio_ramos06},
the simplest model that better fits the observations is preferred with respect to more complicated models (even
if they produce a slightly better fitting).

In spite of the intrinsic high computational load of the MCMC method, one of its advantages is that it is
easily parallelizable. Many Markov chains can be run simultaneously in different isolated threads with
no communication between them. Once the chains are finished, they can be combined into a large chain. Since each
Markov chain (after the burn-in period) is sampling from the posterior distribution, we end up with a very large
chain that also samples from the posterior distribution. Except for the presence of a burn-in period in each
chain, the gain in computational time is roughly proportional to the number of threads. A more refined
way of parallelization is to start a chain and, after the burn-in period, subdivide it into different threads. At the
end, all the threads are combined and we end up with a long chain. In this case, the gain in computational time
is slightly larger than in the previous case. 

The inversion code for the Milne-Eddington case (Bayes-ME) is made available after contact with any of 
the authors. The present version of the code is
extremely versatile and it presents a very good convergence rate. However, we plan to introduce different 
refinements in the future. The most straightforward is the modification of the proposal density so that
non-diagonal elements of the covariance matrix can be taken into account. Although the convergence rate
assuming a diagonal covariance matrix is acceptable, this refinement can lead to a reduction
in the length of the chains because more underlying structure of the posterior distribution is captured in
the proposal density.

\begin{acknowledgements}
We thank R. Manso Sainz and A. L\'opez Ariste for illuminating discussions. We also acknowledge the help of
A. Sainz Dalda for gently providing us with the umbra profiles shown in this paper. This research has been partly 
funded by the Ministerio de Educaci\'on y Ciencia through project AYA2004-05792.
\end{acknowledgements}



\begin{appendix}
\section{Markov Chain Monte Carlo}
\subsection{Metropolis algorithm}
The idea of this
approach is to directly sample the posterior distribution using a Markov Chain. The elements of the
chain are the vector of parameters $\thetabold$ that are used to describe every model. The Markov Chain is a
stochastic process $\{ \thetabold_0,\thetabold_1,\ldots,\thetabold_n \}$ in which every element 
$\thetabold_i$ only depends on the previous one $\thetabold_{i-1}$. The key idea of the MCMC method is to
choose the next point in the chain depending on the previous point such that the distribution of
the chain asymptotically tends to be equal to the posterior distribution, i.e.:
\begin{equation}
\lim_{n \to \infty} p(\{ \thetabold_0,\thetabold_1,\ldots,\thetabold_n \}) = p(\thetabold|D).
\end{equation}
Several methods are available although we will focus on the Metropolis
algorithm \citep{metropolis53,neal93} that, in spite of its simplicity, gives extremely good results. The algorithm
can be defined as follows:
\begin{description}
\item 1. Choose a starting vector of parameters $\thetabold_0$. If some information is available about the
value of some of the parameters, it is advantageous to start close to the solution. However, this
condition is not mandatory for the convergence of the Markov chain.
\item 2. Calculate the posterior probability given the data $p(\thetabold_0|D)$. This includes the calculation
of the priors and the likelihood (including the calculation of the forward modeling problem).
\item 3. Obtain a new vector of parameters $\thetabold_i$ sampling from a \emph{proposal density distribution}
$q(\thetabold_i|\thetabold_{i-1})$. We will explain this step more in detail afterwards.
\item 4. Evaluate the posterior probability $p(\thetabold_i|D)$.
\item 5. Evaluate the ratio 
\begin{equation}
r=\frac{p(\thetabold_i|D) q(\thetabold_i|\thetabold_{i-1})}{p(\thetabold_{i-1}|D) q(\thetabold_{i-1}|\thetabold_{i})}.
\end{equation}
Admit $\thetabold_i$ in the Markov Chain with probability
\begin{equation}
\beta = \min [1,r].
\end{equation}
If a point is rejected, include $\thetabold_{i-1}$ in the chain.
\item 6. Go back to step to 3.
\end{description}
It has been shown that the previous numerical scheme leads to a Markov Chain whose
probability distribution converges towards the posterior distribution \citep[e.g.,][]{metropolis53}. The 
advantage with respect
to the brute force approach is that the number of evaluations of the posterior distribution
is no more exponentially increasing with the number of parameters, but linearly. As a consequence,
we can treat much more complicated problems with a reduced computational effort. The reason for this
behavior is that since the chain is sampling the underlying posterior distribution, the regions of 
larger probability are evaluated more times. 
It is important to point out that the proposal density distribution is usually chosen to
be symmetric, thus $q(\thetabold_i|\thetabold_{i-1})=q(\thetabold_{i-1}|\thetabold_{i})$. As a consequence, the ratio
that have to be evaluated at step 5 simplifies to $r=p(\thetabold_i|D) / p(\thetabold_{i-1}|D)$.

\subsection{The proposal density}
The key ingredient of the Metropolis MCMC algorithm is the proposal density. In the ideal case, one should
choose $q(\thetabold_i|\thetabold_{i-1})$ as close to the posterior distribution as possible. In the 
limiting case that the proposal distribution is exactly matching the posterior one, one is carrying out
a perfect sampling: more samples are performed in the regions of larger probability. Consequently, all 
the proposed steps will be included into the Markov Chain. This case is obviously unrealistic because
it assumes that our aim (i.e., the evaluation of the posterior distribution) has been already achieved. 

The power of the MCMC scheme lies in the fact that, even na\"ively chosen proposal densities
lead to an algorithm that efficiently samples from the posterior distribution. However, it is also
true that a smart election of the proposal density greatly improves the convergence rate of the
algorithm. Common proposal densities include gaussian, normal or uniform distributions centered
at the current value of the parameters to propose a new value of the parameters. In our case,
we have chosen a combination of gaussian and uniform distributions. Both cases lead to a
symmetric proposal density. For the initial $N_\mathrm{unif}$ steps of the
chain, we have decided to propose parameters following a uniform distribution in each parameter. The 
limits of the uniform distribution are free parameters chosen to be equal to their range of variation.
The minimum values for all the parameters are put into the vector $\thetabold^\mathrm{min}$ while the maximum values
are included into the vector $\thetabold^\mathrm{max}$. Then:
\begin{equation}
q(\thetabold_i|\thetabold_{i-1}) \sim U(\thetabold^\mathrm{min},\thetabold^\mathrm{max}).
\end{equation}

After the first $N_\mathrm{unif}$ steps, some information about the posterior probability is known. 
Therefore, statistical properties like the covariance matrix $\mathbf{C}$ can be estimated. At this point, we
change to a gaussian proposal density centered at the current value of the parameters. Ideally,
one should propose with the following distribution:
\begin{equation}
q(\thetabold_i|\thetabold_{i-1}) \sim \exp \left[ -\frac{\alpha}{2} \mathbf{u}^\dag \mathbf{C}^{-1} \mathbf{u} \right],
\end{equation}
where $\mathbf{u}=\thetabold_i-\thetabold_{i-1}$, $\mathbf{u}^\dag$ stands for the transpose of the $\mathbf{u}$
vector and $\alpha$ is a constant whose meaning will be discussed later. Sampling from such a proposal 
density would require the diagonalization of the covariance matrix due to the matrix inversion \citep[e.g.,][]{dunkley05}.
This proposal density is very useful for problems in which strong degeneracies are present in the
problem, so that the posterior distribution shows very elongated maxima. However, in the first version of our
inversion code, we neglect the non-diagonal elements of the covariance matrix. We have verified that this 
approximation gives extremely good results in our case (in spite of the degeneracies present in the problem).
The inclusion of non-diagonal terms in the covariance matrix is left for future revisions of the code.

When we only take into account the diagonal elements of the covariance matrix, the proposal of each parameter can 
be done independently of the rest of parameters. Random numbers following a normal distribution with unit
variance are picked and the proposed value for each parameter is obtained by multiplying them with their 
corresponding variances. The variances are updated after a fixed number of iterations of the Markov Chain. It is not
necessary to use the whole Markov Chain to estimate the variances, because the following updating rule can
be applied to update the variance of parameter $i$ at step $n$:
\begin{eqnarray}
\sigma^2_i(n) &=& \frac{n-1}{n} \sigma^2_i(n-1) \\
&+& \frac{\left[ \theta_{i}(n)-\overline{\theta_i}(n-1) \right]^2}{(n+1)^2}  + 
\frac{\left[ \theta_{i}(n)-\overline{\theta_{i}}(n) \right]^2}{n},
\end{eqnarray}
where $\theta_i(n)$ is the value of the proposed parameter, $\overline{\theta_i}(n-1)$ stands for the average of 
the parameter $i$ taking into account the first $n-1$
elements of the chain, while $\overline{\theta_i}(n)$ takes also into account element $n$ in calculating
the average. The average
can also be updated following the rule:
\begin{equation}
\overline{\theta_i}(n) = \overline{\theta_i}(n-1) + \frac{\theta_i(n)-\overline{\theta_i}(n-1)}{n+1}.
\end{equation}

Concerning the constant $\alpha$, it is used to tune the convergence process. It has been demonstrated 
\citep{gelman96,dunkley05} that, in
order to efficiently sample from a posterior distribution, the acceptance rate of models should be of the order of
25\%. We use $\alpha$ to shrink or broaden the proposal density so that such an acceptance rate is assured. We have 
verified with an extensive test phase that this technique behaves nicely and the chain rapidly
samples the posterior distribution.

\subsection{Convergence}
\label{sec:convergence}
The convergence of the Markov Chain is a critical issue \citep[e.g.,][]{gelman_rubin92,lewis02}. A chain is 
said to be converged when the
statistical properties of its elements reflect with ``enough accuracy'' the statistical properties of the 
underlying distribution that is being sampled. A problem arises for what ``enough accuracy'' means.
Great efforts have been put into the development of powerful convergence tests \citep[e.g.,][]{gelman_rubin92}. 
The key ingredient
in dictating the convergence rate is the proposal density distribution. One of the most widely applied
methods of convergence test is the one proposed by \cite{gelman_rubin92}. The main drawback is that
it works by generating several Markov chains with random initial points. A posterior analysis of their 
statistical properties helps us to distinguish when a chain is sampling from the posterior
distribution. At this point, the elements of the chain can be used to obtain information about the
statistical properties of the posterior distribution that we are sampling. 
Our code uses the alternative of \cite{dunkley05} for testing for convergence. It is based on the idea 
that the Fourier power spectrum of a the Markov chain would be flat and equal to the variance of the underlying 
distribution in case complete convergence is obtained. However, the chain can be considered as converged
in much less restrictive conditions \citep[see][for details]{dunkley05}.

At the beginning of the MCMC algorithm, the chain typically proposes large jumps through the parameter space 
until the regions of high posterior probability distribution are located. This is specially true when the
initial point of the chain is very far away from the regions of large posterior density. The chain, once migrated to these
regions, proposes smaller jumps. The initial steps of the chain are not representative of the underlying 
posterior $p(\thetabold|D)$. They are usually known as the ``burn-in'' of the chain and these elements
are typically thrown away. Following \cite{dunkley05}, one easy way to locate the number of elements of the
``burn-in'' is to locate the maximum value of the posterior $p_\mathrm{max}$ and discard the first elements of the chain
until $p(\thetabold|D) / p_\mathrm{max} > f$, with $f \sim 0.1-0.2$. When the initial point of the
chain is close to the high probability region, this scheme leads to a ``burn-in'' of a few (or even zero) elements.

\end{appendix}

\begin{appendix}
\section{Profiles}
\begin{figure*}
\centering
\includegraphics[width=0.5\hsize,clip]{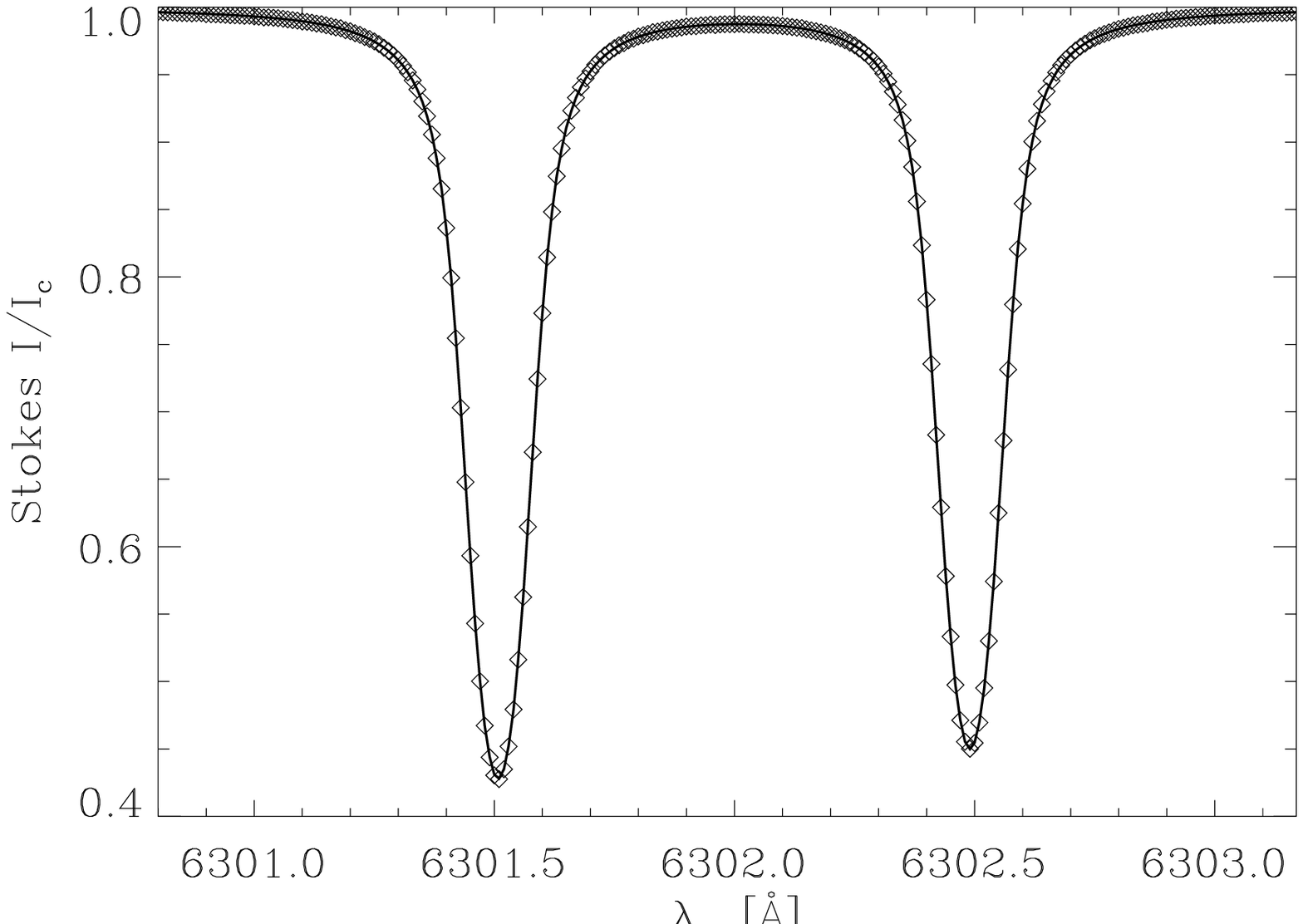}%
\includegraphics[width=0.5\hsize,clip]{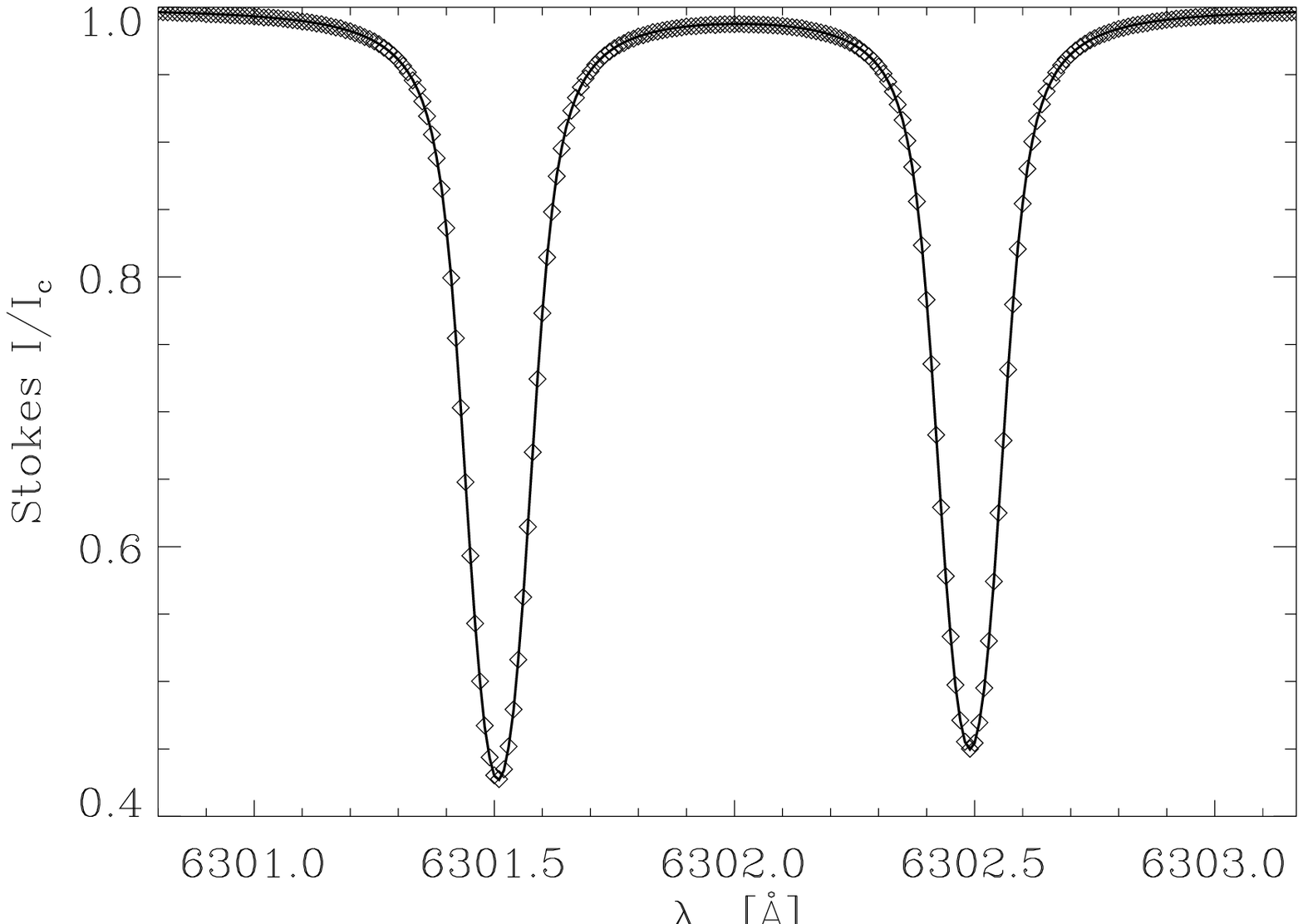}
\includegraphics[width=0.5\hsize,clip]{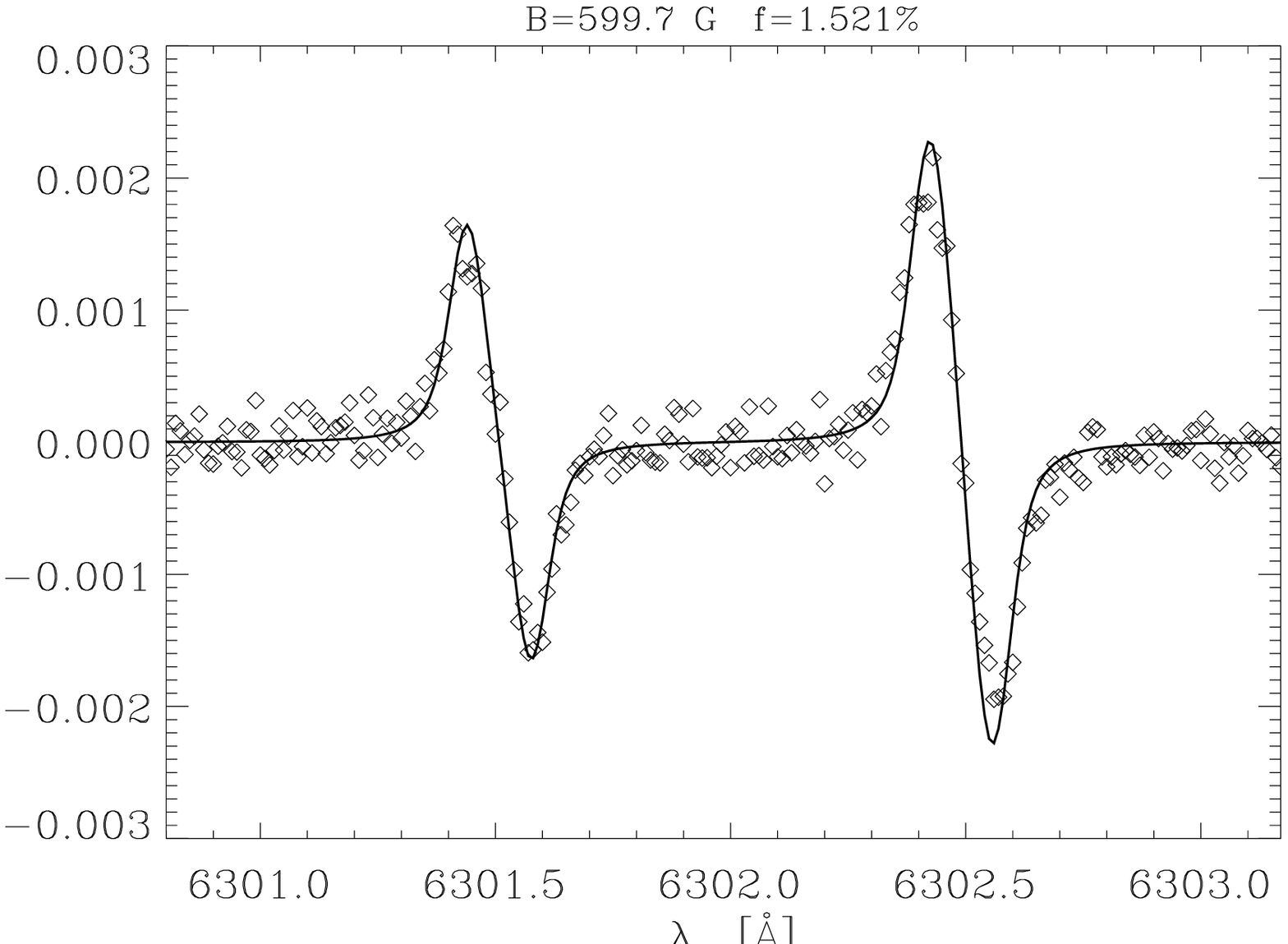}%
\includegraphics[width=0.5\hsize,clip]{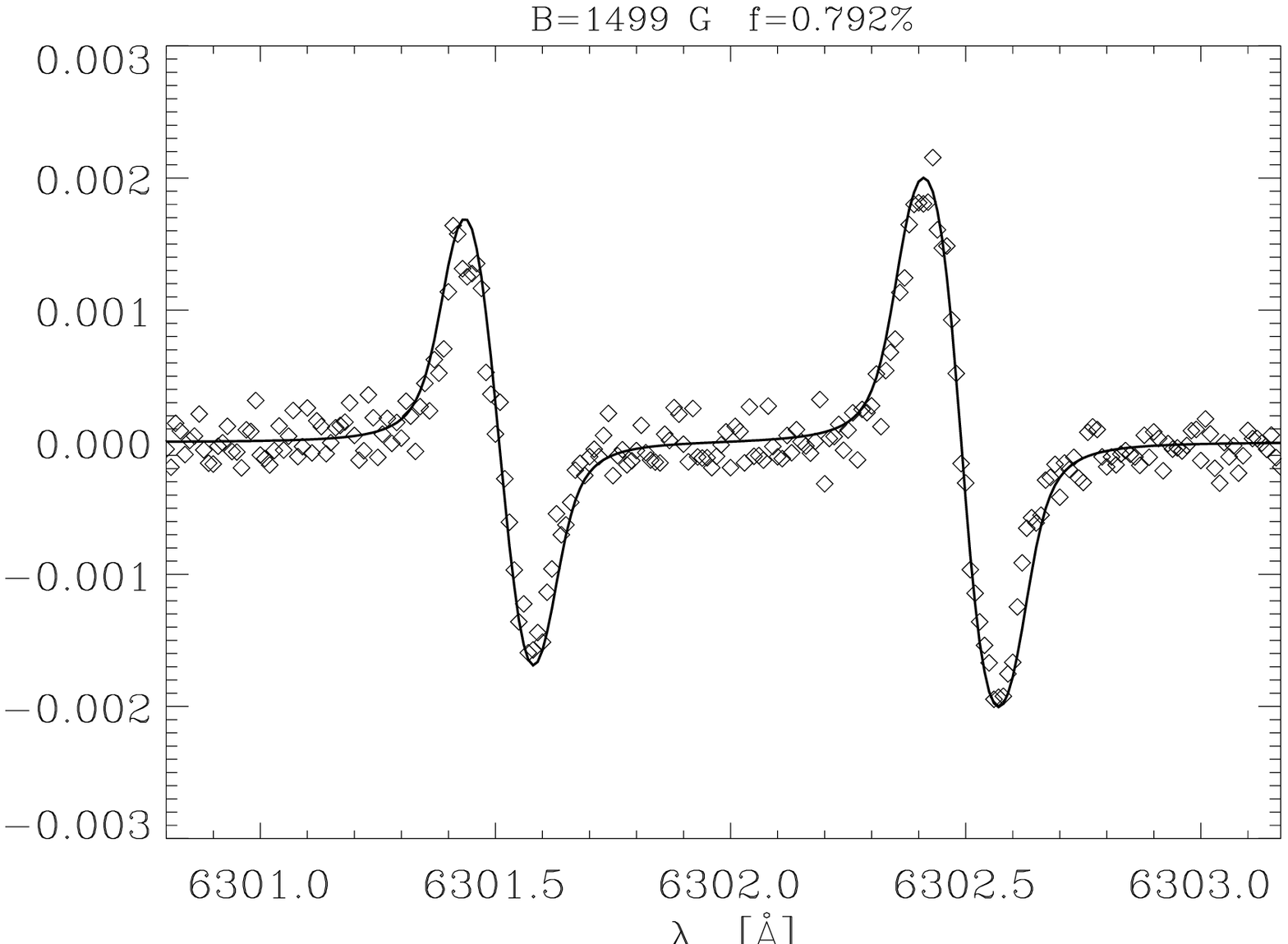}
\caption{Synthetic profiles with added noise (diamonds) together with two different synthetic profiles
corresponding to models presenting magnetic field
strengths differing by 900 G (solid lines). Since both models fit the profiles below 1$\sigma$, there is no objective reason
to favor one of them.}
\label{fig:fits}
\end{figure*}

According to Fig. \ref{fig:qs1}, fields above 500 G and below 1800 G fit the synthetic profile with added noise with a precision
smaller than 1$\sigma$. When this constraint is relaxed to 2$\sigma$, the fields can be even larger or smaller.
Using the three plots of the upper panel
of Fig. \ref{fig:qs1}, it is possible to detect a large number of combinations where fits inside the 68\%
confidence level can be obtained with sub-kG and kG fields. As an example, we show in Fig. \ref{fig:fits}
a fit to the synthetic Stokes profiles with added noise with a field of 600 G and with a field of 1500 G. There is 
no objective reason to prefer one fit over the other under a
1$\sigma$ uncertainty, as consistent with the results presented in Fig. \ref{fig:qs1}. Note that this 
result was pointed out for the first time by \cite{martinez_gonzalez06}.

\end{appendix}

\end{document}